# Assessing trade-offs among electrification and grid decarbonization in a clean energy transition: Application to New York State

Terence Conlon[a†*], Michael Waite[a†**], Yuezi Wu[a***], and Vijay Modi[a****]

[a]Department of Mechanical Engineering, Columbia University
220 S.W. Mudd Building, 500 West 120th Street, New York, NY 10027, USA

[†] Corresponding author
* tmc2180@columbia.edu
** mbw2113@columbia.edu
*** yw3054@columbia.edu
**** modi@columbia.edu

## Abstract

A modeling framework is presented to investigate trade-offs among decarbonization from increased low-carbon electricity generation and electrification of heating and vehicles. The model is broadly applicable but relies on high-fidelity parameterization of existing infrastructure and anticipated electrified loads; this study applies it to New York State where detailed data is available. Trade-offs are investigated between end use electrification and renewable energy deployment in terms of supply costs, generation and storage capacities, renewable resource mix, and system operation. Results indicate that equivalent emissions reductions can be achieved at lower costs to the grid by prioritizing electrification with 40-70% low-carbon electricity supply instead of aiming for complete grid decarbonization. With 60% electrification and 50% low-carbon electricity, approximately 1/3 emissions reductions can be achieved at current supply costs; with only 20% electrification, 90% low-carbon electricity is required to achieve the same emissions reductions, resulting in 43% higher grid costs. In addition, three primary cost drivers are identified for a system undergoing decarbonization: (1) decreasing per-unit costs of existing infrastructure with increasing electrified demand, (2) higher in-state generation costs from low-carbon sources relative to gas-based and hydropower generation, and (3) increasing integration costs at high percentages of low-carbon electricity.

## Keywords

Decarbonization; Heating and vehicle electrification; Low-carbon electricity; Energy system modeling; New York State.

# Nomenclature

*Fixed variables and parameters*

| | |
|---|---|
| $A_{P_x,j}$ | capital annualization rate for annualization period *P*, technology *x*, and interest rate *j* [years$^{-1}$] |
| $C_{existing-cap}$ | total cost of existing transmission and generation capacity over entire analysis period [$] |
| $C_{generation}$ | total generation cost over entire analysis period [$] |
| $C_{new-cap}$ | total new capacity cost over entire analysis period [$] |
| $c_{bio,i}$ | biofuel generated electricity price at node *i* [$/MWh] |
| $c_{existing-ramp}$ | existing fossil fuel-based generation ramping cost [$/MW-h] |
| $c_{ff,i}$ | fossil fuel price at node *i* [$/MMBTU] |
| $c_{hydro,i}$ | hydropower generated electricity price at node *i* [$/MWh] |
| $c_{imp,i}$ | imported electricity price at node *i* [$/MWh] |
| $c_{new-ramp}$ | new fossil fuel-based generation ramping cost [$/MW-h] |
| $c_{nuc,i}$ | nuclear generated electricity price at node *i* [$/MWh] |
| $CAP_{batt-e,i}$ | battery storage energy capital cost at node *i* [$/MWh] |
| $CAP_{batt-p,i}$ | battery storage power capital cost at node *i* [$/MW] |
| $CAP_{ff,i}$ | new fossil fuel-based generation capital cost at node *i* [$/MW] |
| $CAP_{on,i}$ | onshore wind power capital cost at node *i* [$/MW] |
| $CAP_{off,i}$ | offshore wind power capital cost at node *i* [$/MW] |
| $CAP_{us-solar,i}$ | utility-scale solar generation capital cost at node *i* [$/MW] |
| $CAP_{tx,ii'}$ | capital cost of upgraded transmission from node *i* to adjacent node *i'* [$/MW-mi] |
| $D_{elec,i}^t$ | existing electricity demand at node *i* and timestep *t* [MWh] |
| $D_{heat,i}^t$ | electrified heating demand at node *i* and timestep *t* [MWh] |
| $D_{heat,i}^{t,full}$ | full electrified heating demand at node *i* and timestep *t* [MWh] |
| $D_{veh,i}^t$ | vehicle charging demand at node *i* and timestep *t* [MWh] |
| $D_{veh,i}^{t,full}$ | full electric vehicle charging demand at node *i* and timestep *t* [MWh] |
| $d_{ii'}$ | distance between node *i* and adjacent node *i'* [mi] |
| $EX_{cap,i}$ | annual cost of maintaining existing generation capacity at node *i* [$/MW-yr] |
| $EX_{tx,i}$ | annual cost of existing transmission at node *i* [$/MWh-yr] |
| $F$ | quantity of fuel consumed [MJ] |
| $H_{fix,i}^t$ | fixed hydropower electricity generation at node *i* and timestep *t* [MWh] |
| $H_{flex,i}^{max}$ | flexible hydropower maximum electricity generation at node *i* [MWh] |
| $I$ | set of all nodes in study region |
| i | single node in the study region |
| i' | node adjacent to *i* |
| j | interest rate |
| l | transmission loss rate |
| $N_i^t$ | nuclear-generated electricity at node *i* [MWh] |
| $n_{years}$ | number of years in the analysis [years] |
| $omf_{ff}$ | new fossil fuel-based generation fixed operations and management cost [$/MW-yr] |
| $omf_{on}$ | onshore wind power fixed operations and management cost [$/MW-yr] |
| $omf_{off}$ | offshore wind power fixed operations and management cost [$/MW-yr] |
| $omf_{us-solar}$ | utility-scale solar power fixed operations and management cost [$/MW-yr] |
| $omf_{tx,ii'}$ | fixed operations and management cost of upgraded transmission from node *i* to adjacent node *i'* [$/MW-yr] |
| $omv_{ff}$ | new fossil fuel-based generation variable operations and management cost [$/MWh] |
| $P$ | annualization period [years] |
| $t$ | hourly time step |
| $T$ | total number of hourly time steps in analysis |
| $U_{tx-flow,i}^{existing}$ | annual existing intranodal transmission flow at node *i* [MWh] |
| $W_{off,i}^{t}$ | potential offshore wind-generated electricity at node *i* and timestep *t* [MWh$_{generation}$/MW$_{installed}$] |
| $W_{on,i}^{t}$ | potential onshore wind-generated electricity at node *i* and timestep *t* [MWh$_{generation}$/MW$_{installed}$] |
| $W_{btm-solar,i}^{t}$ | potential behind-the-meter solar-generated electricity at node *i* and timestep *t* [MWh$_{generation}$/MW$_{installed}$] |
| $W_{us-solar,i}^{t}$ | potential utility-scale solar-generated electricity at node *i* and timestep *t* [MWh$_{generation}$/MW$_{installed}$] |
| $X_{btm-solar,i}$ | capacity of behind-the-meter solar generation (existing and newly simulated) at node *i* [MW] |
| $X_{cap,i}^{existing}$ | capacity of existing generation with associated maintenance costs at node *i* [MW] |
| $X_{off,i}^{existing}$ | capacity of existing offshore wind generation at node *i* [MW] |
| $X_{on,i}^{existing}$ | capacity of existing onshore wind generation at node *i* [MW] |
| $X_{us-solar,i}^{existing}$ | capacity of existing utility-scale solar generation at node *i* [MW] |
| $\eta_{ff-existing}$ | fossil fuel-based generation efficiency of existing capacity |
| $\eta_{ff-new}$ | fossil fuel-based generation efficiency of new capacity |
| $\varepsilon$ | emissions [CO$_2$e] |
| $\theta$ | emissions rate [CO$_2$e/unit energy] |

*Decision variables*
*All variables are constrained to be greater than or equal to 0.*

| | |
|---|---|
| $H_{flex,i}^{t}$ | flexible hydropower electricity generation at node *i* and timestep *t* [MWh] |
| $G_{existing,i}^{t}$ | fossil fuel-based generation from existing capacity at node *i* and timestep *t* [MWh] |
| $G^t_{existing-diff,i}$ | absolute value of the difference in fossil fuel-based generation from existing capacity at node *i* between time steps *t* and *t-1* [MWh] |
| $G^t_{new,i}$ | fossil fuel-based generation from new capacity at node *i* and timestep *t* [MWh] |
| $G^t_{new-diff,i}$ | absolute value of the difference in fossil fuel-based generation from new capacity at node *i* between time steps *t* and *t-1* [MWh] |
| $L^t_i$ | biofuel generation at node *i* and timestep *t* [MWh] |
| $V^t_i$ | imported electricity at node *i* and timestep *t* [MWh] |
| $X_{batt-e,i}$ | battery storage energy capacity installed at node *i* [MWh] |
| $X_{batt-p,i}$ | battery storage power capacity installed at node *i* [MW] |
| $X_{ff,i}$ | capacity of fossil fuel-based generation installed at node *i* [MW] |
| $X_{off,i}$ | capacity of offshore wind generation installed at node *i* [MW] |
| $X_{on,i}$ | capacity of onshore wind generation installed at node *i* [MW] |
| $X_{us-solar,i}$ | capacity of utility-scale solar generation installed at node *i* [MW] |
| $X_{tx,ii'}$ | capacity of new transmission from node *i* to adjacent node *i'* [MW] |
| $Z^t_{ii'}$ | electricity transmitted from node *i* to adjacent node *i'* at timestep *t* [MWh] |
| $\gamma^t_{batt,i}$ | increase in battery storage state of charge at node *i* and timestep *t* [MWh] |
| $\delta^t_{batt,i}$ | decrease in battery storage state of charge at node *i* and timestep *t* [MWh] |

*Scenario configuration parameters*

| | |
|---|---|
| $LCP$ | low-carbon electricity generation percent: Fraction of total demand that must be met by low-carbon energy (combined nuclear, wind, water, and solar power) |
| $p_{heat,i}$ | fraction of full heating electrification demand simulated at node *i* |
| $p_{veh,i}$ | fraction of full vehicle electrification demand simulated at node *i* |
| $\omega$ | percent reduction in total greenhouse gas emissions |

*Subscripts and superscripts*
(Note: Some fixed variables and parameters defined above are used in subscripts and superscripts. These terms are not redefined here.)

| | |
|---|---|
| *batt* | battery storage |
| *bio* | biofuel |
| *btm* | behind-the-meter |
| *diff* | difference |
| *elec* | electricity |
| *fix* | fixed |
| *ff* | fossil-fuel |
| *flex* | flexible |
| *heat* | heating |
| *imp* | imports |
| *ind* | industrial sector |
| *off* | offshore wind |
| *on* | onshore wind |
| *other* | out-of-scope |
| *p2e* | power-to-energy |
| *tot* | total |
| *transp* | transportation sector |
| *tx* | transmission |
| *us* | utility scale |
| *veh* | vehicle |

*Acronyms*

| | |
|---|---|
| *SECTR* | System Electrification and Capacity TRansition |
| *SECTR-NY* | System Electrification and Capacity TRansition – applied to New York State |
| *HVE* | heating and vehicle electrification |
| *GHG* | greenhouse gas |
| *LCOE* | levelized cost of electricity |
| *CEM* | capacity expansion model |
| *RTO* | regional transmission organization |
| *ISO* | independent system operator |
| *NYS* | New York State |
| *NREL* | National Renewable Energy Laboratory |
| *VRE* | variable renewable energy |

# 1. Introduction

The United States is at a clean energy crossroads. Economically, per-unit costs of new solar and wind generation have become lower than coal and gas generation in parts of the country [1]. Policy-wise, several states have recently passed major climate legislation [2]. Public opinion mirrors these changes: A growing consensus acknowledges that a clean energy transition would have numerous social [3] and economic benefits [4]. As a result, support for sweeping federal action has reached new heights [5]. Even so, the cost-effectiveness of this transition will be influenced by region-specific nuances of legacy infrastructure, energy sources, and constraints [6]. This paper proposes an open-source framework that offers a means to evaluate decarbonizing the electricity grid while considering electrification of heating and vehicles. The framework is then to New York State (NYS) to highlight trade-offs among dominant decarbonization options emblematic of a region with a well-defined electricity system and a variety of climates, renewable energy resources, and existing fossil fuel end use needs.

There is widespread consensus that coupling electrification of heating and vehicles with renewable energy expansion is the best approach to reducing energy-related greenhouse gas (GHG) emissions [7]. In fact, it is infeasible to meet deep decarbonization targets without both cleaning the grid and replacing current fossil fuel transportation and heating technologies with low-carbon alternatives [8]. However, less well understood are how prioritizing fossil fuel end use electrification or the percentage of electricity from low-carbon sources influences the cost-effectiveness of emissions reductions, electrification's potential benefits to the electricity system, and how transitioning existing heating and transportation infrastructure impacts hourly energy system operation.

Many energy system models seek to determine economically optimal technology mixes for future electricity scenarios, including those set in NYS [9]. Modeling unit commitment and dispatch [10] at the scale of individual generators [11] under varying degrees of foresight [12] can provide detailed operational understanding for a fully defined system. Capacity expansion models (CEMs) generally aggregate generators with similar characteristics in order to avoid the significant computational requirements of high spatial and temporal resolution models with capacities as decision variables [13]. The improved tractability of CEMs (often called "macro-energy system models"[14] when applied to regional systems) allows them to incorporate a larger number of system characteristics [15]. CEMs have expanded to include additional technological options, demonstrating that higher fidelity to existing systems results in more accurate capacity expansion scenarios [16]. By modeling resource stochasticity, other CEMs find that optimal system design changes under uncertainty [17]. Moreover, the inclusion of environmental considerations shifts the deployment of renewable generation capacity compared to CEMs that do not account for land-use limitations [18]. CEMs that simulate interconnected energy systems such as transportation [19] and heating [20] have modeled sector-wide clean energy transitions, showing that the interplay of different energy demands is critical in understanding decarbonization pathways. Nevertheless, because characterizing actual systems can be time-consuming (if sufficient information and data is even available), CEMs often do not contain high-fidelity parameterizations of all existing system conditions [20]. These shortcomings are particularly problematic for regional energy systems (e.g. at the Regional Transmission Organization (RTO) or Independent System Operator (ISO) scale) with unique existing infrastructure and resource mixes that are likely to affect deep decarbonization efforts, as well as intra-regional heterogeneity that may not be captured in larger-scale models [21].

While CEMs have previously been used to investigate the impact of electrified loads on least-cost model decisions, there remain opportunities for improvements in methods and applications. A group of CEM-based studies by the U.S. National Renewable Energy Laboratory (NREL) explores the effects of electrification and decarbonization on model-selected energy infrastructure capacities [22], electricity cost [23], emissions [24], variable renewable electricity (VRE) integration [25], and electricity demand curves [26] in the continental US. These NREL studies use representative time slices in place of continuous time series to solve models with high spatial resolution, but this approach precludes thorough investigation of system operation. Similarly, a recent study on achieving net-zero emissions in the continental U.S. through expanded low-carbon electricity and end use electrification simulates power sector operations at an hourly resolution for 41 representative days [21]; as with the NREL studies, representative time slices prevent a full accounting for system operation over a continuous time period. Other studies include continuous supply and demand time series to evaluate power flow for discrete scenarios (i.e. with fixed infrastructure capacities rather than optimal capacity expansion decision-making) to evaluate the effects of electrification on VRE integration [27]. Another study of this type applies a grid model introduced in [28] to evaluate the effects of electrified heating demand in California on both GHG emissions and grid resource capacity needs. Here, resource mixes are exogenously defined, and electricity costs in future electrification scenarios are not presented [29].

Recent studies of NYS have found that deep decarbonization is feasible using existing technologies, and that different pathways exist to a carbon neutral future [30]. One such report issued by New York's Climate Action Council concludes that substantial progress on heating and vehicle electrification is required by 2030, and that nearly 100 GW of renewable generation capacity is required for full energy sector decarbonization by 2050 [31]. Related work uses a capacity expansion model and representative timeseries to show that battery storage will be required to ensure electricity reliability during a low-carbon transition [32]. However, these studies also list areas for future research, including incorporation of an updated GHG emissions assumptions accounting [30].

A gap in the literature thus remains: An evaluation of both cost-optimal capacity expansion and system operation for a well-characterized existing regional energy system, under various combinations of electrification and low-carbon electricity adoption rates, using multiple years of real data, with improved emissions assumptions. This paper addresses this gap by introducing an open-source System Electrification and Capacity TRansition (SECTR) modeling framework. To determine optimal system characteristics, SECTR computes the lowest total cost of electricity generation, transmission, and storage resource mix for specified combinations of: (a) low-carbon electricity supply percentage, (b) building end use and vehicle electrification, and (c) percent GHG emissions reduction. SECTR is designed to replicate existing system characteristics: spatially heterogeneous hourly electricity demands, generation technologies, and capital and operating costs; inter-nodal transmission limits; energy storage; temperature-dependent electric vehicle charging demands; and electrified heating demand time series [33]. Agriculture and industrial emissions are included in GHG computations, but SECTR does not endogenously model changes in those sectors. In this paper, the SECTR framework is applied to New York State's energy system (SECTR-NY). Lastly, for the SECTR-NY application, this paper includes an emissions accounting that improves upon the accounting contained in current NYS reports, as it incorporates methane leakage and adopts the longer duration GHG warming potentials specified by a recent state climate law.

# 2. Methodology

Section 2 contains a description of the SECTR model general formulation, and the motivation for its application to New York State. All modeling information not specified in Section 2 is contained in Section S2 of the Supplementary Materials.

## 2.1 System Electrification and Capacity Transition model general formulation

A SECTR model study region is defined by individual nodes, *i,* representing geographical sub-areas within the larger region of interest. Along with existing electricity demand, each node contains electrified heating[i] and vehicle charging loads at each timestep, *t*, within the overall time period simulated, *T.* To determine the least-cost infrastructure mix in future model scenarios, decision variables are assigned node-specific costs. SECTR uses a characterization of the region's energy-related GHG emissions as both a reference quantity for GHG emissions reduction computations and to compute the emissions impact of reduced fossil fuel usage associated with heating and vehicle electrification; the model does not consider improved efficiency or growth of fossil fuel end uses.

SECTR evaluates different low-carbon electricity supply and end use electrification scenarios by computing the total cost of new and existing infrastructure capacity and maintenance, fuels, and resource operation to estimate the total annual cost of electricity generation and transmission; these returned costs do not include delivery expenses (primarily distribution system costs). The modeling framework does not include the cost of replacing current fossil fuel-based building systems and vehicles or electricity distribution system costs; as such, SECTR cost computations can be considered those that typically constitute the "supply" portion of a utility customer's bill.

The remainder of Section 2.1 contains a subset of the SECTR governing equations that establish the model configuration, along with additional equations that define how costs and emissions are calculated. Due to space constraints, Section S2 of the Supplementary Materials presents the remainder of the SECTR governing equations, including those constraining fossil fuel generation, wind capacity, solar capacity, internodal transmission, battery storage, nuclear generation, hydropower generation, biofuel generation, interregional exports, and additional generation capacity costs.

*Objective function*

SECTR's objective function minimizes the total annual electricity system supply cost based on specification of two of the following three configuration parameters: (1) minimum percent of in-state electricity generated from low-carbon resources, $LCP$; 2) minimum percent electrification of current fossil fuel-based heating, $p_{heat}$, and vehicle electrification, $p_{veh}$; and (3) minimum GHG emissions reduction requirement, $\omega$. Eqs. (1-4) describe the objective function, where $C_{new-cap}$ is the total cost of new capacity, $C_{generation}$ is the total cost of generation, and $C_{existing-cap}$ is the total cost of maintaining existing capacity:

$$obj = minimize(C_{new-cap} + C_{generation} + C_{existing-cap}) \tag{1}$$

$$\begin{aligned}C_{new-cap} = n_{years} \\ * \sum_{i \in I} \Big[ &(A_{P_{on},j} * CAP_{on,i} + omf_{on}) * X_{on,i} + (A_{P_{off},j} * CAP_{off,i} + omf_{off}) \\ &* X_{off,i} + (A_{P_{us-solar},j} * CAP_{us-solar,i} + omf_{us-solar}) * X_{us-solar,i} \\ &+ (A_{P_{batt},j} * CAP_{batt-e,i}) * X_{batt-e,i} + (A_{P_{batt},j} * CAP_{batt-p,i}) * X_{batt-p,i} \\ &+ (A_{P_{ff},j} * CAP_{ff,i} + omf_{ff}) * X_{ff,i} \\ &+ \sum_{i'} (A_{P_{tx},j} * CAP_{tx,ii'} * d_{ii'} + omf_{tx,ii'}) * X_{tx,ii'} \Big]\end{aligned} \tag{2}$$

$$\begin{aligned}C_{generation} = \sum_{i \in I} \sum_{t \in T} \Big[ &c_{hydro,i} * (H^t_{fixed,i} + H^t_{flex,i}) + c_{nuc,i} * N^t_i + c_{bio,i} * L^t_i + c_{imp,i} * V^t_i \\ &+ 3.412 * c_{ff,i} * \left( \frac{G^t_{existing,i}}{\eta_{ff-existing}} + \frac{G^t_{new,i}}{\eta_{ff-new}} \right) + omv_{ff} * G^t_{new,i} + c_{existing-ramp} \\ &* G^t_{existing-diff,i} + c_{new-ramp} * G^t_{new-diff,i} \Big]\end{aligned} \tag{3}$$

$$C_{existing-cap} = n_{years} * \sum_{i \in I} \left[ EX_{cap,i} * X^{existing}_{cap,i} + EX_{tx,i} * U^{existing}_{tx-flow,i} \right] \tag{4}$$

---

[i] Note that SECTR incorporates the ability to model shifts of any fossil fuel-based building end use, which generally depend on heat in some form: In US residences, 93% of natural gas, 86% of propane, and 98% of fuel oil consumption is used for either space or water heating [44]; in commercial buildings, 78% of natural gas and 70% of fuel oil consumption is used for space or water heating [45]. As such, "heating" is used for short.

*Levelized cost of electricity calculations*

The levelized cost of electricity (LCOE) is calculated per Eq. (5):

$$LCOE = \frac{C_{new-cap} + C_{generation} + C_{existing-cap}}{\sum_{t \in T} \sum_{i \in I} [D^t_{elec,i} + D^t_{heat,i} + D^t_{veh,i} - X_{btm-solar,i} * W^t_{btm-solar,i}]} \tag{5}$$

Note that the LCOE is simply the total electricity supply cost divided by the total electricity demand, after subtracting contributions from behind-the-meter (BTM) solar generation. LCOE is used as a general comparative metric between scenarios.

*Capital cost annualization*

For a given technology, *x*, the annualization rate ($A_{P_x,j}$) associated with the capacity cost, *CAP$_x$*, is computed from a technology-specific annualization period ($P_x$) and a 5% interest rate (*j*), per Eq. (6).

$$A_{P_x,j} = \frac{j*(1+j)^{P_x}}{((1+j)^{P_x}-1)}$$

(6)

*Heating and vehicle electrification*

Hourly demands for electrified heating, $D_{heat,i}^t$, are based on the nodal percentage of heating electrification, $p_{heat,i}$, and user-provided nodal electricity demands for full heating electrification, $D_{heat,i}^{t,full}$, per Eq. (7).

$$D_{heat,i}^t = p_{heat,i} * D_{heat,i}^{t,full}$$

(7)

Electric vehicle demand at each time step, $D_{veh,i}^t$, is based on the nodal percentage of vehicle electrification, $p_{veh,i}$, and user-provided nodal electricity demands for full vehicle electrification, $D_{veh,i}^{t,full}$, per Eq. (8).

$$D_{veh,i}^t = p_{veh,i} * D_{veh,i}^{t,full}$$

(8)

*Energy balance constraint*

The nodal energy balance is constrained by the following inequality, with all variables defined in the Nomenclature:

$$\begin{aligned}(X_{on,i} + X_{on,i}^{existing}) * W_{on,i}^t + (X_{off,i} + X_{off,i}^{existing}) * W_{off,i}^t + (X_{us-solar,i} + X_{us-solar,i}^{existing}) \\ * W_{us-solar,i}^t + X_{btm-solar,i} * W_{btm-solar,i}^t + H_{flex,i}^t + H_{fixed,i}^t + N_i^t \\ + G_{existing,i}^t + G_{new,i}^t + L_i^t + V_i^t - \gamma_{batt,i}^t + \delta_{batt,i}^t \\ + \sum_{i'}[(1-l)*Z_{i'i}^t - Z_{ii'}^t] \geq D_{elec,i}^t + D_{heat,i}^t + D_{veh,i}^t\end{aligned}$$

(9)

The low-carbon electricity generation curtailment is computed from the slack in this constraint at each node.

*Low-carbon electricity generation targets*

For certain SECTR configurations, the user selects a low-carbon percent (LCP) – a minimum percentage of in-state electricity supply from onshore and offshore wind, hydropower, solar, and nuclear power after subtracting out contributions from BTM generation; the electricity generated from fossil fuels and biofuels over the full simulation period is thus constrained per Eq. (10).

$$\sum_{t\in T}\sum_{i\in I}(G^t_{existing,i} + G^t_{new,i} + L^t_i) \leq (1 - LCP) *$$
$$\sum_{t\in T}\sum_{i\in I}[D^t_{elec,i} + D^t_{heat,i} + D^t_{veh,i} - V^t_i - X_{btm-solar,i} * W^t_{btm-solar,i}]$$

(10)

*Emission reduction calculations and assumptions*

In-region electricity generation emissions are calculated with emissions rate of fossil fuel-based generation, $\theta_{ff}$, and generation from existing, $G^t_{existing,i}$, and new, $G^t_{new,i}$, fossil fuel plants, after accounting for their respective efficiencies, $\eta_{ff-existing}$ and $\eta_{ff-new}$. Emissions from imported electricity are determined by the product of the emissions rate of imports, $\theta_{imp,i}$, and the quantity of imports, $V^t_i$. Together, emissions from in-region generated electricity and imports are summed over all nodes *i* and timesteps *t* to compute total electricity related GHG emissions, $\varepsilon_{elec}$, for each scenario, per Eq. (11).

$$\varepsilon_{elec} = \sum_{t\in I}\sum_{i\in I}\left[\theta_{ff} * \left(\frac{G^t_{existing,i}}{\eta_{ff-existing}} + \frac{G^t_{new,i}}{\eta_{ff-new}}\right) + \theta_{imp,i} * V^t_i\right]$$

(11)

GHG emissions of remaining fossil fuel heating, $\varepsilon_{heat}$, are equal to product of the complement of the heating electrification fraction simulated, $p_{heat,i}$; the blended emissions rate for heating, $\theta_{heat}$; and the total quantity of heating fuel consumed $F_{heat,tot,i}$. This quantity is summed over all nodes *i* and is computed per Eq. (12):

$$\varepsilon_{heat} = \sum_{i\in I}(1 - p_{heat,i}) * \theta_{heat} * F_{heat,tot,i}$$

(12)

GHG emissions of non-electrified vehicles, $\varepsilon_{veh}$, are calculated per Eq. (13). This accounting is analogous to that for heating emissions, using the fraction of vehicle electrification simulated, $p_{veh,i}$; the blended emissions rate for vehicles, $\theta_{veh}$; and the total quantity of vehicle fuel consumed, $F_{veh,tot,i}$. Total transportation sector emissions also include existing transportation emissions outside the scope of the current analysis, $\varepsilon_{transp,other}$, per Eq. (14):

$$\varepsilon_{veh} = \sum_{i \in I}(1 - p_{veh,i}) * \theta_{veh} * F_{veh,tot,i}$$

(13)

$$\varepsilon_{transp} = \varepsilon_{veh} + \varepsilon_{transp,other}$$

(14)

Industrial sector emissions from energy consumption, $\varepsilon_{ind}$, are added to compute total GHG emissions. Emissions from the incineration of waste are excluded from the specific formulations of future energy scenarios. To compute the overall percent reduction in GHG emissions, $\omega$, SECTR compares total computed emissions to the user-provided reference quantity, $\varepsilon_{reference}$ per Eq. (15).

$$\omega = \frac{\varepsilon_{reference} - (\varepsilon_{elec} + \varepsilon_{heat} + \varepsilon_{transp} + \varepsilon_{ind})}{\varepsilon_{reference}}$$

(15)

Fig. 1 presents a flowchart that summarizes the main steps for a user – broadly defined as anyone defining or executing a SECTR configuration – to instantiate and solve SECTR model scenarios. In short, after defining the fixed variables and parameters (see *Nomenclature*), specifying two of the three scenario configuration parameters – low-carbon electricity percent, $LCP$; heating and vehicle electrification (HVE) rates $p_{heat}$ and $p_{veh}$; and GHG reduction, $\omega$ – allows SECTR to determine the cost-optimal energy system design for a future decarbonization scenario.

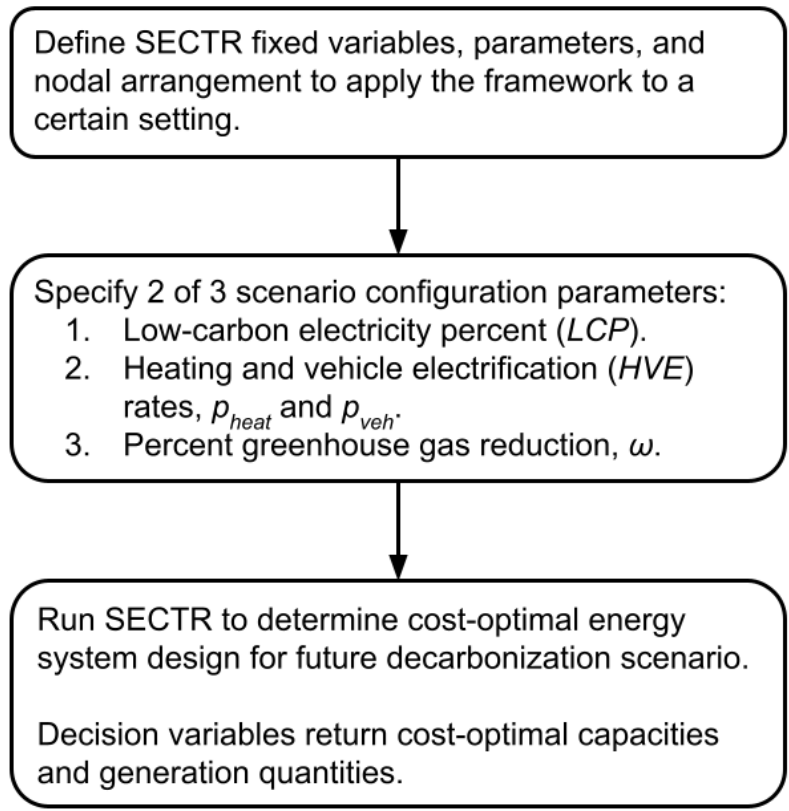

Fig 1: Flowchart for instantiating and solving SECTR general formulation model scenarios.

## 2.2 Application to New York State

This paper applies the SECTR framework to New York State (SECTR-NY), which provides a useful study area for several reasons, including:

- A 2019 law [34] mandating significant, quantifiable decarbonization targets in the years 2030, 2040, and 2050.
- A single electricity supply system operator and market administrator – the New York Independent System Operator (NYISO) – covering the extent of New York State.
- Well-defined transmission interfaces, both internal (between NYISO zones) and external (imports/exports between NYISO and other load areas).
- Diverse and geographically heterogeneous loads and potential renewable resources.
- Definable effects of population and built environment density on current system costs and documented costs of new infrastructure capacity.
- Extensive data availability for the current electricity system and statewide GHG emissions.

The Supplementary Materials contain a full parameterization of SECTR-NY, including descriptions of all data sources used and developed model data. Four nodes are defined for NYS by grouping NYISO zones based on the state's major transmission interfaces. The existing system is generally defined by the most recent reference data available; however, load and weather time series data for 2007-2012 are used in the model formulation because the reference model data for hourly wind and solar resource potential are available for only those years. Monthly characteristics of electricity supply and demand time series over the six modeled years are shown in Fig. 2.

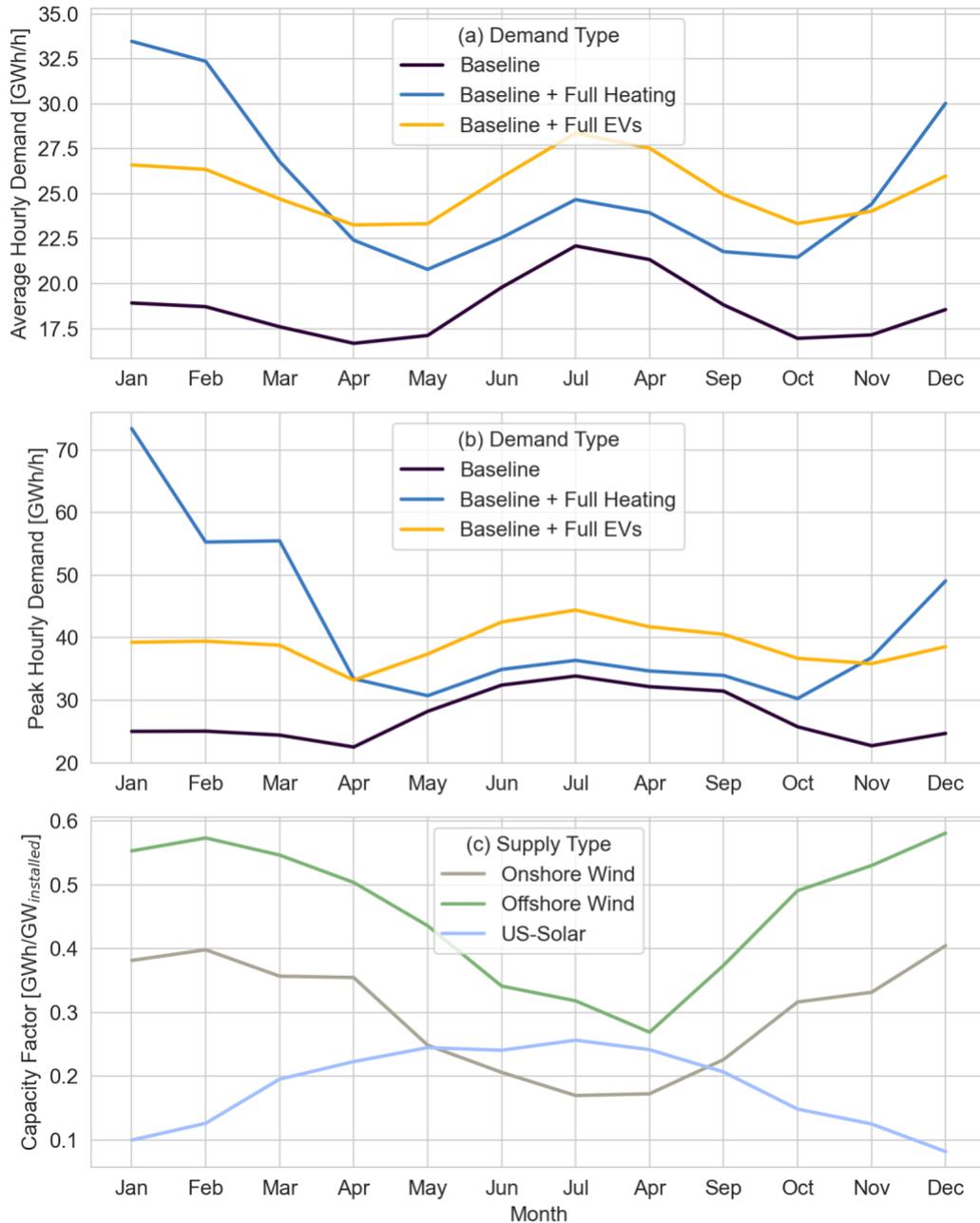

*Fig. 2: (a) monthly averages of hourly electricity demand, (b) monthly peak of hourly electricity demand, and (c) monthly capacity factors for wind and solar resources in NYS.*

# 3. Results

Section 3.1 establishes and distinguishes between a "Current" model configuration that mirrors existing NYS system characteristics, and a "Baseline" configuration for decarbonization scenario comparison. Section 3.2 presents the results of SECTR-NY Baseline configuration simulations for different combinations of in-state low-carbon electricity generation percentages (LCP) and heating and vehicle electrification rates (HVE). Section 3.3 compares SECTR-NY results to those published in recent NYS studies on decarbonization pathways. All results are presented for SECTR-NY simulations solved over the entire 6-year time period modeled; all specified generation and demand quantities are presented as hourly averages in Wh/h over the full 6-year simulation period.

## 3.1 Current system validation and Baseline configuration

The Baseline configuration deviates from the Current system configuration in three ways summarized in Table 1: The Baseline configuration excludes existing nuclear power at Node 1, includes an additional 5 GW of solar BTM capacity corresponding to a simulation year of 2030, and simulates an additional planned 1.25 GW of hydropower import capacity into New York City (NYC). For comparative purposes, Table 1 also includes a "Baseline with Nuclear" scenario. All Table 1 scenarios exclude any additional HVE beyond current electric heating and vehicles.

*Table 1: 'Current', 'Baseline with Nuclear', and 'Baseline' system configuration comparisons.*

| | Configuration Parameters | | | Specified System Characteristics[a,b] | | | | Model-returned System Characteristics[b] | | | | |
|---|---|---|---|---|---|---|---|---|---|---|---|---|
| Configuration | % GHG[c] | % HVE[d] | % LCP[e] | Instate Hydro [GW] | Nuclear [GW] | BTM Solar [GW] | Hydro Imports [GW] | Onshore Wind [GW] | Utility-Scale Solar [GW] | Battery [GWh] | Wind and Solar LCOE [$/MWh] | Total LCOE [$/MWh] |
| Current | 3.6 | 0 | 38.2 | 5.3 | 3.5 | 1.6 | 1.5 | 2.0[f] | 0.1[f] | 0.2[f] | 69.7 | 65.3 |
| Baseline w. Nuclear | -2.0 | 0 | 42.4 | 5.3 | 3.5 | 6.6 | 2.8 | 2.0 | 0.1 | 1.1 | 69.3 | 68.6 |
| Baseline | -1.6 | 0 | 40 | 5.3 | 0 | 6.6 | 2.8 | 9.1 | 2.6 | 2.0 | 67.8 | 72.1 |

[a] See Supplementary Methods for existing system characteristics and the text of this section for any modifications for the specific configuration.
[b] Besides LCOE values, all system characteristics presented indicate capacities.
[c] '% GHG' refers to the percent change in greenhouse gas emissions compared to the 1990 reference quantity. A positive value indicates a computed increase in emissions, a negative value indicates a reduction.
[d] '% HVE' refers to the percent of additional heating and vehicle electrification simulated; some heating electrification (and a very small amount of vehicle electrification) currently exists in NYS.

e '% LCP' refers to the percent of in-state electricity supply from low-carbon sources.
f Indicates model capacities that are constrained to existing capacity in the 'current' configuration.

The model-computed Current configuration LCOE of $65.3/MWh compares favorably to the actual system. An actual NYS electricity supply cost of $69.1/MWh is estimated, based on 2019 NYS generation and transmission costs [35], electricity sales [35], and total zonal electricity demands; this actual cost would include ancillary service and NYISO operation costs of approximately $2/MWh [36] that are not included in SECTR-NY. Despite the difference between these two values, the close alignment in computed costs supports SECTR-NY's applicability to the NYS system and its suitability for further analyses.

The Current configuration computes an LCP of 38.2% and a 3.6% increase in GHG emissions compared to the 1990 reference quantity. Total emissions increase because $CO_2$ reductions from natural gas displacing coal and fuel oil combustion are offset by GHG increases from larger transportation energy demands, methane leakage associated with natural gas production and transmission, and the retirement of a large nuclear power plant; these effects are more pronounced due to the use of the 20-year GWP value for methane in place of 100-year GWP value. Moreover, the calculated LCP of 38.2% is lower than the 2019 fraction of NYS electricity demand met by low-carbon sources (62.3%) for two reasons: 1) per the language of the CLCPA, LCP only considers in-state generation, and does not account for substantial hydropower imports from Canada; and 2) SECTR-NY does not include nuclear generation from Indian Point, as this facility was fully closed on April 30, 2021[ii].

The Baseline with Nuclear configuration – adding BTM PV and NYC hydropower imports to the Current configuration – computes a 2% GHG reduction and $68.6/MWh LCOE; the $3.3/MWh higher LCOE is due to the higher cost of hydropower imported into NYC and the reduction of regional demands due to solar BTM (i.e., existing system capacity costs are distributed over less load). Removing all nuclear capacity establishes the Baseline configuration; a 40% LCP is set for round number comparison in subsequent sections that is close to the current 38.2%. Approximately 10 GW of solar and wind capacity are installed to replace the nuclear generation, resulting in a slightly lower reduction in GHG (-1.6%) and a slightly higher LCOE ($72.1/MWh). Given the reasonable deviations from the current system model, the Baseline configuration is adopted for future scenario evaluations.

## 3.2 Analysis of low-carbon electricity and end use electrification scenarios

For a series of SECTR-NY simulations with different combinations of LCPs and HVEs[iii], relationships among LCOE, GHG emissions, HVE, LCP, and renewable energy capacity are shown in Fig. 3. Here, computed LCOEs represent the total costs for supply (primarily generation, storage, and transmission), excluding delivery costs (primarily distribution system costs). HVE rates refer to

---

[ii] The "Current" and "Baseline with Nuclear" configurations do include generation from NYS nuclear plants besides Indian Point, as these plants remain operational as of this paper's publication.
[iii] In the scenarios presented, heating and vehicle electrification rates are equal.

new heating and vehicle electrification, as some heating (and a small share of vehicles) currently uses electricity. Note that the 40% LCP and 0% HVE scenario presented in Table 1 is located in the bottom-left of the figure; for comparison beyond NYS, 39.7% of US electricity generation was from low-carbon sources in 2020 [37].

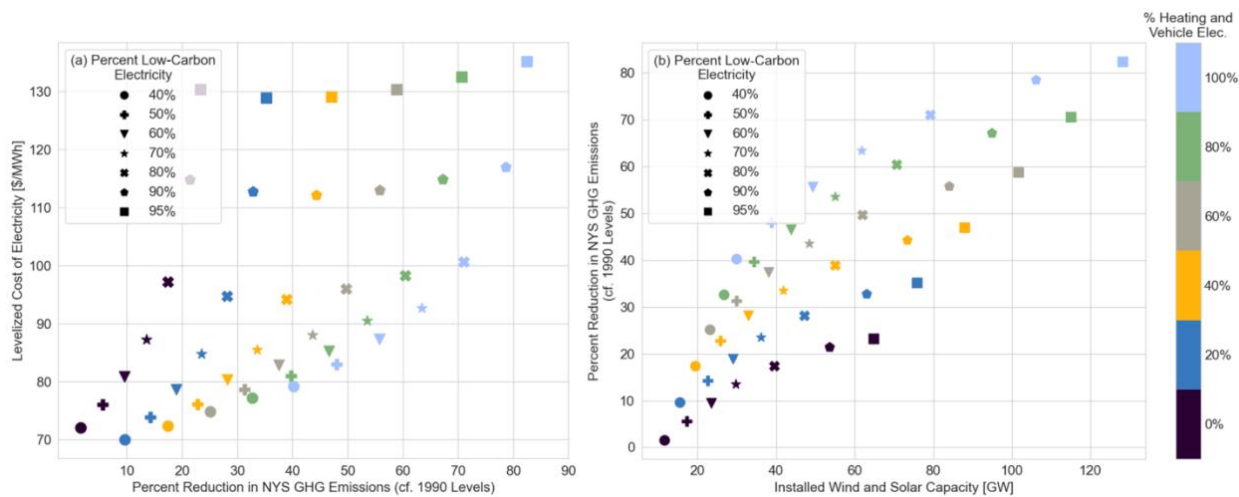

*Fig. 3:* (a) *LCOE vs. percent emissions reduction; (b) percent emissions reduction vs. installed wind and solar capacity. All emissions reductions are compared to 1990 levels. Marker shape indicates percent low-carbon electricity (LCP), and marker color indicates heating and vehicle electrification (HVE). All points represent independently solved SECTR-NY decarbonization scenarios with specified LCP + HVEs. For scenarios shown, all low-carbon electricity generation is from wind, solar, and hydropower.*

Fig. 3(a) shows how computed grid supply LCOE (strictly that of the electricity utilized) rises sharply with increasing LCP for a specified HVE, whereas for a specified LCP, higher HVEs cause limited growth in LCOE. Fig. 3(b) provides a partial explanation, showing that high HVE scenarios achieve the same GHG reductions with lower installed wind and solar capacities. The results suggest that emissions reductions can be achieved with a shallower initial rise in LCOE by prioritizing electrification of heating and vehicles in conjunction with deployment of solar and wind, as opposed to the latter by itself. Added loads from HVE can even slightly reduce LCOE up to a point (20-40% HVE, depending on LCP), as the additional electricity demand decreases the per-unit cost of existing infrastructure. (The same trend holds when the system includes an average of 3 GWh/h of nuclear generation in Node 1, albeit at LCOEs approximately 10% lower; see Supplementary Figure S2.)

It is worth noting the straightforward impact of HVE on GHG emissions: In NYS, a current average emissions rate for fossil fuel-based heating of 148 kgCO2e/MMBtu$_t$ (i.e. per unit heat delivered) is computed based a recent heating model [38] and GHG emissions rate assumptions described in the Supplementary Methodology; with electrified heating and 40% low-carbon electricity supply[iv] in SECTR-NY, this reduces to 44 kgCO2e/MMBtu$_t$. Similar reductions occur for vehicle electrification: A current average emissions rate for fossil fuel vehicles of 543 gCO$_2$e/mi (per

---

[iv] 40% LCP mirrors the current NYS fuel mix.

vehicle mile traveled) is computed, and 241 gCO$_2$e/mi for electric vehicles with 40% LCP in SECTR-NY. Therefore, even with the remaining 60% of grid power being supplied by gas-based generation, substantial reductions in overall emissions from electrification are computed.

Consider two changes in system characteristics starting at the 40% LCP and 0% HVE point of Fig. 3(a). Approximately 10% GHG emissions reductions could be achieved without additional electrification and with 60% LCP at an LCOE of $80.9/MWh; this scenario represents a 3.1 GWh/h increase in average wind and solar supply. A similar emissions reduction could be achieved with a 20% HVE and no LCP increase at a cost of $70.0/MWh; the average wind and solar supply increases by 1.1 GWh/h to maintain 40% LCP with the electrification-driven increase of 2.7 GWh/h average demand. Consider now two scenarios in Fig. 3(b) with approximately 30 GW wind and solar capacity: The scenario with 50% LCP and 60% HVE has computed GHG emissions reductions of 31%, more than double the 14% reduction in the scenario containing 70% LCP and 0% HVE. Here, the computed LCOE for the first scenario ($78.7/MWh) is nearly $10/MWh less than the second scenario ($87.2/MWh).

These various trade-offs are demonstrated with four scenarios that all contain approximately 1/3 reductions in GHG, but via different combinations of LCP and HVE. For the lowest LCP scenario shown in Table 2 (Scenario 1), GHG reductions require a high HVE that increases average load and peak load, the latter requiring larger amounts of gas turbine capacity. Comparatively, Scenario 3 contains 33 GW less gas generation capacity, accompanied by a drop in average gas generation from 15.3 GWh/h to 6.0 GWh/h. Here, higher LCP scenarios avoid increases in gas capacity and generation through additional renewable generation and battery capacity, a tradeoff that increases supply costs by $10/MWh.

*Table 2: Select scenarios achieving emissions reductions of approximately 1/3 compared to the 1990 reference quantity.*

| Scenario | % GHG[a] | % HVE[b] | % LCP[c] | Avg. Load [GWh/h] | Wind and Solar Cap. [GW][d] | Battery Cap. [GW] | Gas Cap. [GW][e] | Avg. Gas Gen. [GWh/h][f] | LCOE [$/MWh] |
|---|---|---|---|---|---|---|---|---|---|
| 1 | -32.9 | 80 | 40 | 29.4 | 26.6 | 4.7 | 63.0 | 15.3 | 77.2 |
| 2 | -31.3 | 60 | 50 | 26.7 | 29.8 | 4.2 | 48.9 | 11.4 | 78.7 |
| 3 | -33.6 | 40 | 70 | 24.0 | 41.8 | 6.9 | 29.9 | 6.0 | 85.5 |
| 4 | -32.8 | 20 | 90 | 21.3 | 63.0 | 15.0 | 27.0 | 1.8 | 112.8 |

[a] '% GHG' refers to the percent change in greenhouse gas emissions compared to the 1990 reference quantity. Negative values indicate reductions.
[b] '% HVE' refers to the percent of additional heating and vehicle electrification simulated; some heating electrification (and a very small amount of vehicle electrification) currently exists in NYS.
[c] '% LCP' refers to the percent of in-state electricity supply from low-carbon sources.

[d] 'Wind and Solar Cap.' refers to installed onshore wind, offshore wind, and utility-scale solar capacity.
[e] 'Gas Cap.' contains 27.0 GW existing gas-based generation capacity and model selected new gas turbines.
[f] 'Avg. Gas Gen.' refers to the average generation over the entire 6-year simulation period from existing gas-based generation and model-selected new gas turbines.

The synergy of renewable energy generation and electrification is further explained by looking at "excess low-carbon generation": Potential electricity generation from model-selected wind and solar capacities exceeding demand. Excess low-carbon generation exists as an hourly time series of either 0 MWh (when total low-carbon generation is less than the demand) or a positive value equal to the amount of low-carbon electricity generation that exceeds demand. In model simulations, excess low carbon generation must be either 1) stored for later use, or 2) curtailed. Fig. 4(a) shows that despite significant growth in renewable energy capacity with increasing HVE, excess low-carbon electricity generation remains below 6% as long as LCP does not exceed 70%; at LCP of 50% or less, excess generation is below 1%. Fig. 4(b) shows the relationship between excess low-carbon generation and LCOE for the same scenarios in Fig. 4(a).

By combining the effects discussed thus far, three primary LCOE drivers are identified: (1) decreasing per-unit costs of existing infrastructure with increasing demand from HVE, (2) higher generation costs from wind and solar power relative to existing resources, and (3) increasing integration costs when large amounts of wind and solar power produce electricity in excess of demand. Fig. 4(b) shows a general linear trend of integration costs (curtailment and battery storage) increasing LCOE at higher percents excess low-carbon generation, but also how the effects of the three cost drivers change over the entire range of LCPs and HVEs simulated. At LCPs at or below 60%, the primary cost tradeoffs discussed earlier are observed: Higher LCOEs from more wind and solar are partially mitigated by higher utilization of existing infrastructure with HVE. In the 70-80% LCP range, a transition begins in which some spread in excess low-carbon generation affects LCOE, but the first two LCOE drivers prevail. Beyond 80%, the integration cost-driven linear relationship between increasing excess low-carbon generation and computed LCOE dominate.

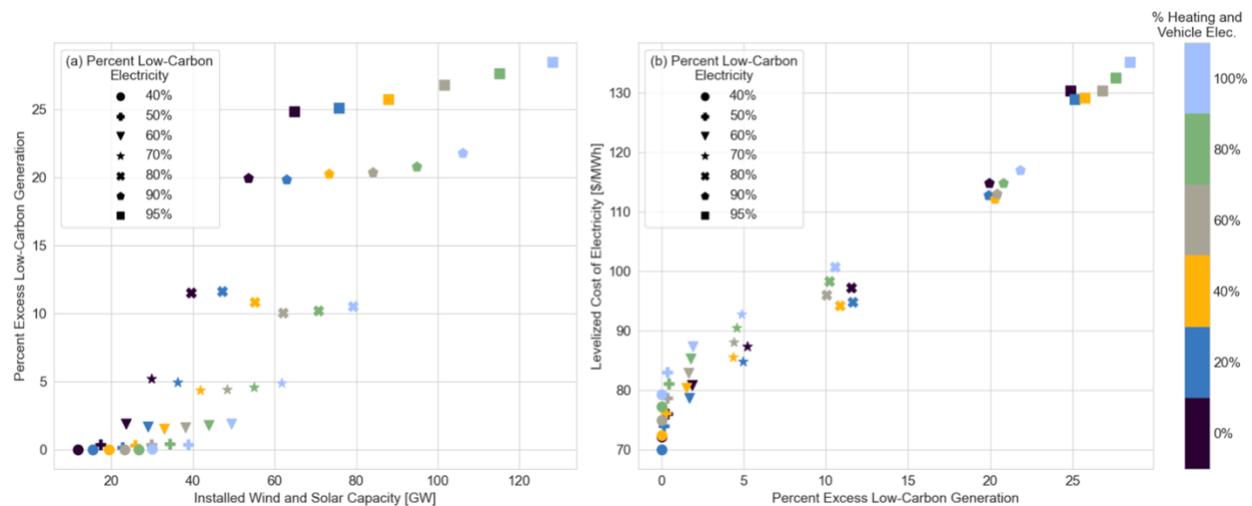

*Fig. 4: (a) Average percent excess low-carbon generation for the entire 6-year simulation period vs. installed wind and solar capacity; (b) LCOE vs. percent excess low-carbon generation. Results*

*are shown for the same independent decarbonization scenarios in Fig. 3, whereby the low-carbon electricity percent and the rate of heating and vehicle electrification are set, and SECTR-NY determines the least-cost energy system.*

The results presented thus far show how electrification accompanied by a significant buildout of renewable energy can keep LCOE low. On the other hand, a focus on large LCP fractions beyond 70% represents a major cost escalation. Competing drivers and trade-offs are next examined among scenarios with increasing HVE while maintaining LCP at 60% (Fig. 5(a-d)) vs. scenarios where HVE is 40% and LCP is progressively increased (Fig. 5(e-h)). (The trends observed here hold for other combinations of HVE and LCP; see Supplementary Figures S7-S8.) Fig. 5(a-d) demonstrates the stable buildout of generation capacity and consistency of system behavior and costs as electrification increases. In order to meet the increased demand, low-carbon generation, gas generation, and battery capacity all increase with electrification, per Fig. 5(a); gas generation undergoes the largest capacity increase – from 27.0 GW to 67.2 GW at 100% HVE – in order to meet higher electrification-induced demand peaks. Here, additional gas capacity is selected due to its low cost relative to the model's other dispatchable generation option, battery storage. With additional policy-based constraints in place, such as a limit on additional gas turbine capacity or demand-side strategies to mitigate peak heating loads, much less new gas capacity would be built out. Electricity generation trends (Fig. 5(b)) largely mirror the expansion in generation capacity, with the ratio of solar to wind generation (combined onshore and offshore) staying consistent from 0.31 at 0% HVE to 0.34 at 100% HVE, although with an increasing amount of wind generation coming from offshore capacity. Fig. 5(d) reveals the reason for consistency in system behavior: Despite increasing average uncurtailed low-carbon electricity generation from 9.5 GWh/h at 0% HVE to 17.7 GWh/h at 100% HVE, average excess low-carbon generation only increases from 177 MWh/h to 336 MWh/h. Electrification thus supports renewable energy integration by keeping the LCOEs of those supply resources low (Fig. 5(c)).

Conversely, optimal energy system characteristics change substantially with increasing LCPs. The previously noted inflection point at 70-80% LCPs is characterized by a large increase in battery capacity (Fig. 5(e)): Of the 33.4 GW of installed battery capacity at 95%, 26.1 GW is installed between 80% and 95%. As implied by Fig. 4, this buildout is due to the significant increase in excess low-carbon generation shown in Fig. 5(h). Furthermore, as battery capacity increases, battery energy throughput does not increase as much (Fig. 5(f)), resulting in battery LCOE growth from \$117/MWh at 80% LCP to \$198/MWh at 95% LCP (Fig. 5(g)). Similarly, gas-based generation capacity remains fairly steady even at very high LCPs, but the electricity generation from that capacity decreases significantly. The result is gas generation LCOE steadily increasing from \$57/MWh at 40% LCP to \$72/MWh at 70% LCP and accelerating to \$260/MWh at 95% LCOE. It is worth noting that these results partially reflect the constraints of the model; they suggest that other technologies not included in SECTR-NY due to their non-competitive costs become beneficial in pushes to eliminate emissions from electricity generation. Regardless, these technology costs coupled with the significant increase in wind and solar LCOEs due to curtailment give a strong indication of the dominance of integration costs at high LCP.

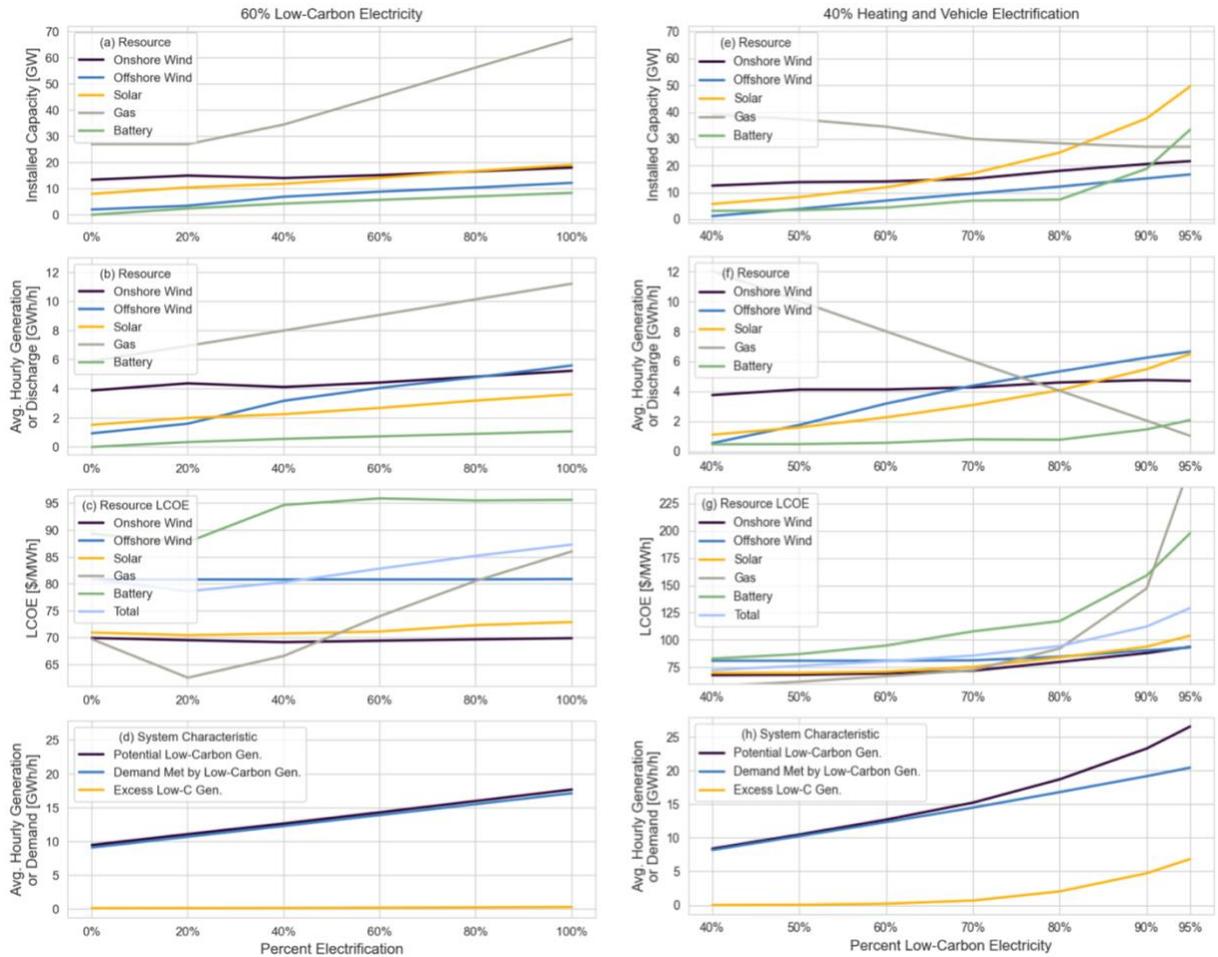

*Fig. 5: System characteristics for scenarios with (a-d) increasing HVE at 60% LCP; and (e-h) increasing LCP at 40% HVE. Subplots (a, e) present installed capacity; (b, f) present average generation by resource; (c, g) present LCOE per MWh for the generation and storage resources; and (d, h) present demand and generation quantities. In (c, g), resource LCOE for onshore wind, offshore wind, and solar refers to the LCOE of generation; LCOE for battery storage is per-MWh discharge; total LCOE contains all system costs; and in (c), gas generation LCOE at 95% LCP ($260/MWh) is cropped out to preserve y-axis resolution.*

Fig. 6 shows the monthly low-carbon electricity supply for (a) 60% LCP for HVEs of 0%, 40% and 80%, and (b) 40% HVE for LCPs of 60%, 80% and 95%. The seasonal low-carbon supply in Fig. 6(a) is nearly identical regardless of HVE and is largely in line with wind supply patterns shown in Fig. 2; this holds despite the low-carbon generation supply increasing 68% between HVEs of 0% and 80%. Accordingly, low-carbon electricity supply phenomena are shown to be essentially independent of HVE, despite very significant shifts in diurnal and seasonal demand patterns with HVE. In contrast, Fig. 6(b) shows a significant shift in seasonal low-carbon supply behavior reflecting the increased share of solar shown in Fig. 5(f). (Additional system operation characteristics were investigated on this monthly timescale to inform the findings here; given space considerations, these have been included in Supplementary Figures S3-S5.)

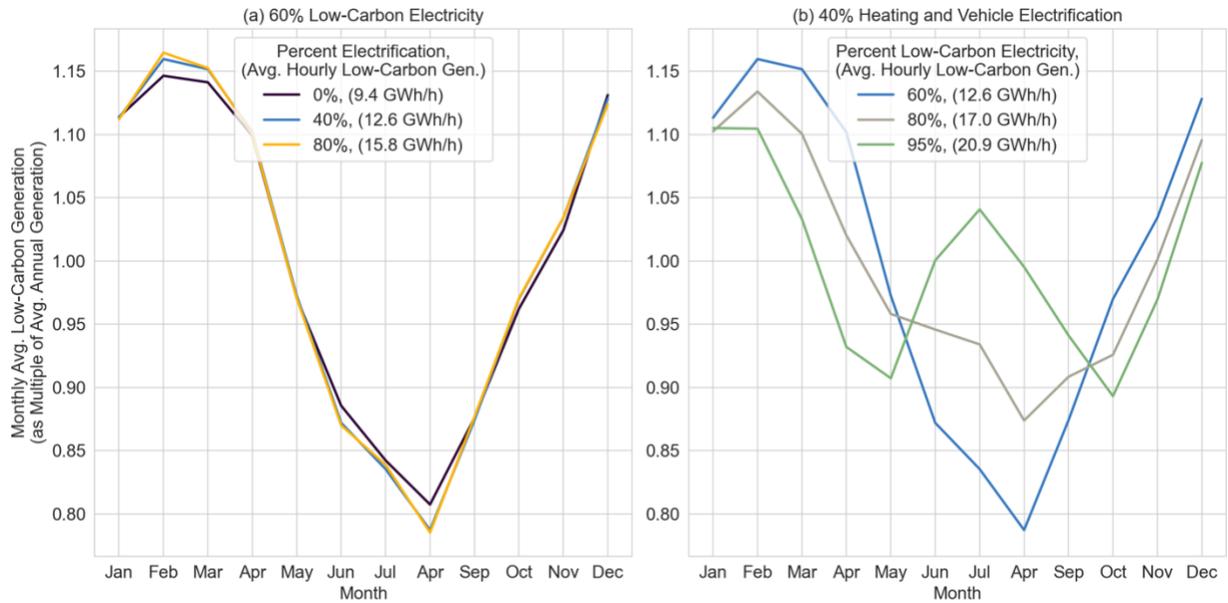

*Fig. 6: Monthly average low-carbon generation as a multiple of the average annual low-carbon generation. (a) monthly averages for 0%, 40%, 80% HVEs at 60% LCP; (b) monthly averages for 60%, 80%, and 95% LCPs at 40% HVE.*

Solar's contribution to the overall supply mix increases most dramatically beyond 80% LCP as battery storage increases: whereas 19.2 GW of solar capacity is installed between 40% and 80% LCP, 24.6 GW of capacity is installed just between 80% and 95% LCP (see Fig. 5(e)). This reflects complex dynamics in which overall system behavior may mask unique marginal behaviors of individual components: the operation of the same resource at lower LCP may be quite different with other resources present at higher LCPs. To this end, the paired buildout of solar and battery capacity at very high LCPs provides the most cost-effective method of displacing the remaining gas generation, as the daily cycling of solar generation allows for regular battery charging during the day and discharging at night even as it becomes the highest LCOE renewable resource (Fig. 5(g)). Fig. 7 shows how battery behavior and its relation to wind and solar supply changes at increasing LCPs for a given 40% HVE. (See Supplementary Figures S9-S10 for other HVEs, which show the same trends as Fig. 7.)

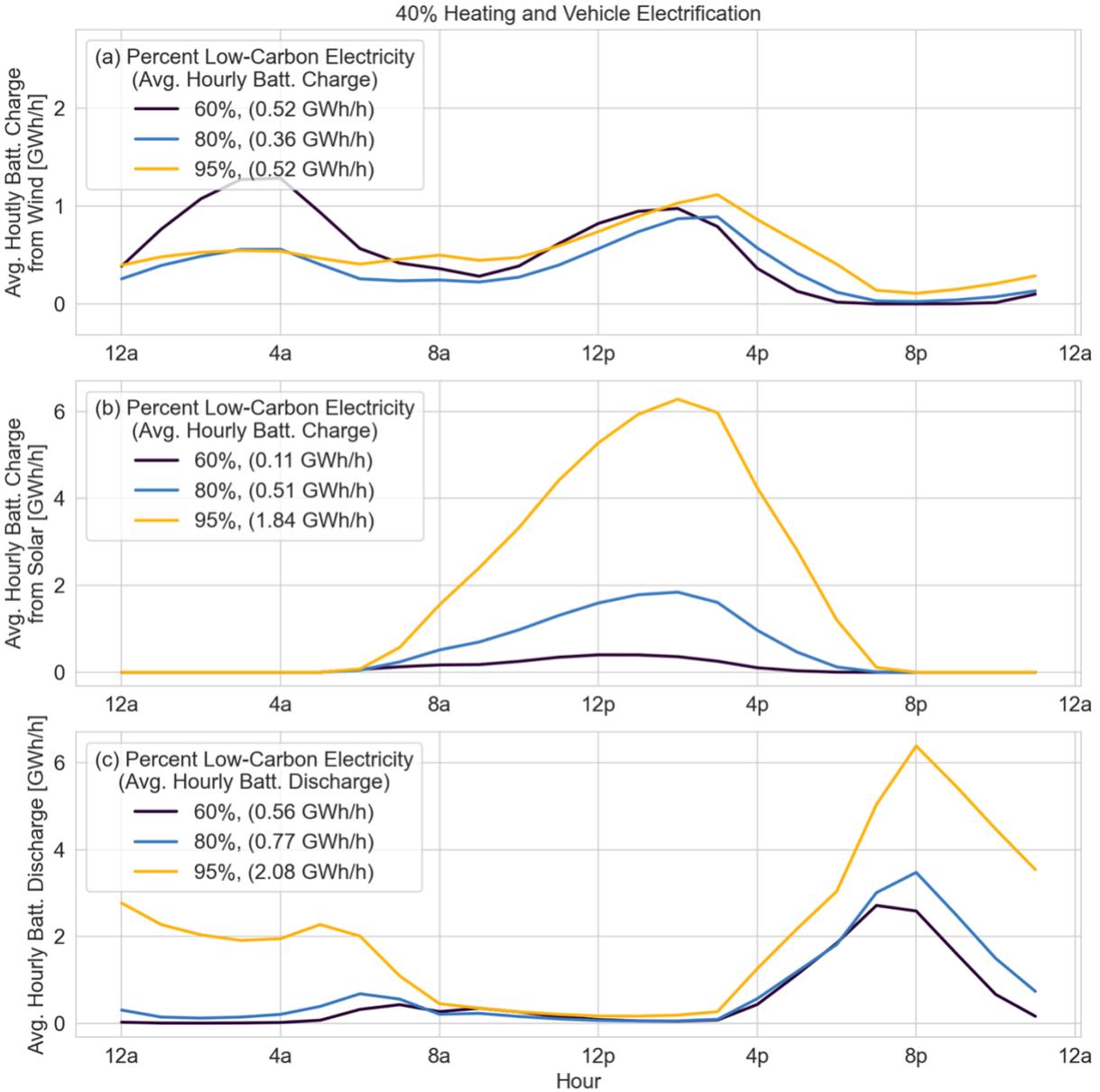

*Fig. 7: Average battery operation by hour for 60%, 80%, and 95% LCPs* over the entire 6-year simulation period. *(a) average hourly battery charging from wind (note y-axis scale is unique from (b) and (c)); (b) average hourly battery charging from solar; and (c) average battery discharge, all in GWh/h.*

At 60% LCP in Fig. 7, when total wind supply is roughly three times total solar supply, battery charging from wind is approximately 5 times higher than solar with distinct overnight and afternoon charging periods. At 80% LCP, wind's overnight charging reduces while both wind and solar charge the batteries in the afternoon; battery charging from solar becomes 1.4 times that from wind despite total wind supply being 2.4 times solar supply. Despite this shift between 60% and 80% LCP, the battery discharge remains almost entirely in the evening while total battery throughput increases by 38%. From 80% to 95% LCP, the maximum hourly average discharge in the evening doubles from 3 GWh/h to 6 GWh/h, extending throughout the night with a steady average 2-3 GWh/h supply resulting in a near tripling of the total throughput. The additional energy supply to the battery comes almost entirely from solar: While total wind supply remains 1.8 times the solar supply, battery charging from solar is 3.5 times that from wind. Here, the diurnal pattern of solar generation allows for daily battery cycling and higher battery throughput, behavior that enables the integration of more low-carbon generation.

While the average diurnal behavior shown in Fig. 7 is useful in understanding broad system behavior and the results of model decisions, decision-making is often based on complex dynamics occurring at hourly timescales over particular periods of time that set capacity and operational needs. Figs. 8 and 9 show representative weeks in the winter and summer, respectively: The upper figures (Figs. 8(a) and 9(a)) show scenarios of 80% LCP and 40% HVE, and the lower figures (Figs. 8(b) and 9(b)) show scenarios of 95% LCP and 40% HVE. Fig. 8(a) shows that the lowest LCOE low-carbon option of wind provides much of the winter energy needs at 80% LCP, due to the resource's high seasonal productivity. Conversely, there are higher needs for gas-based generation in the summer (Fig. 9(a)). In both figures, curtailment (i.e., slack in the SECTR-NY energy balance constraint) is attributed to solar and wind in proportion to their hourly generation; however, as noted in the discussion around Fig. 7, the natural pairing of solar generation and battery storage means that more wind generation is curtailed relative to solar.

As the LCP increases to 95% (Figs. 8(b) and 9(b)), the reason for coupling more solar power with battery storage is revealed: Solar generation exceeding demand during the afternoon is used to charge battery storage, which is then discharged to meet evening demand (and overnight demand, if enough stored energy is available). In Figs. 8(b) and 9(b), approximately 5% of demand met by gas generation occurs during extended hours of low wind production. Here, batteries are not as cost-effective in displacing gas generation: low wind generation potentials lasting a day or longer would require multi-day battery cycling periods, and accordingly, underutilization of storage capacity relative to its usage with solar. (For further exploration that reinforces this interpretation, Supplementary Figures S11-S12 present the same representative week and LCPs as Figs. 8-9 but at 80% HVE.)

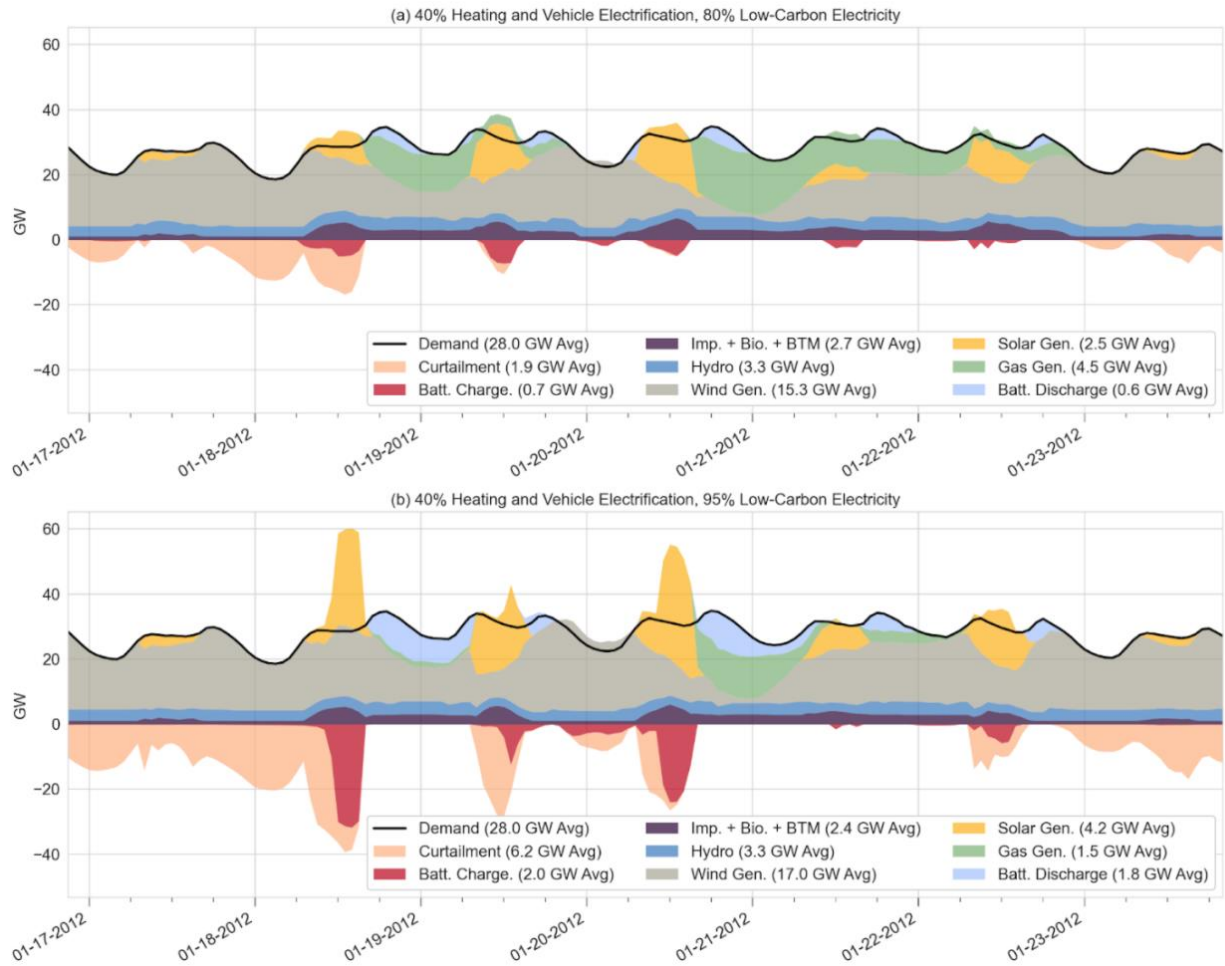

Fig. 8: Electricity generation and demand for a representative winter week with 40% HVE. (a) 80% LCP; (b) 95% LCP. 'Imp. + Bio. + BTM' represents the sum of imports, biofuel, and behind-the-meter solar generation. Average values reported in the legend are for the week shown.

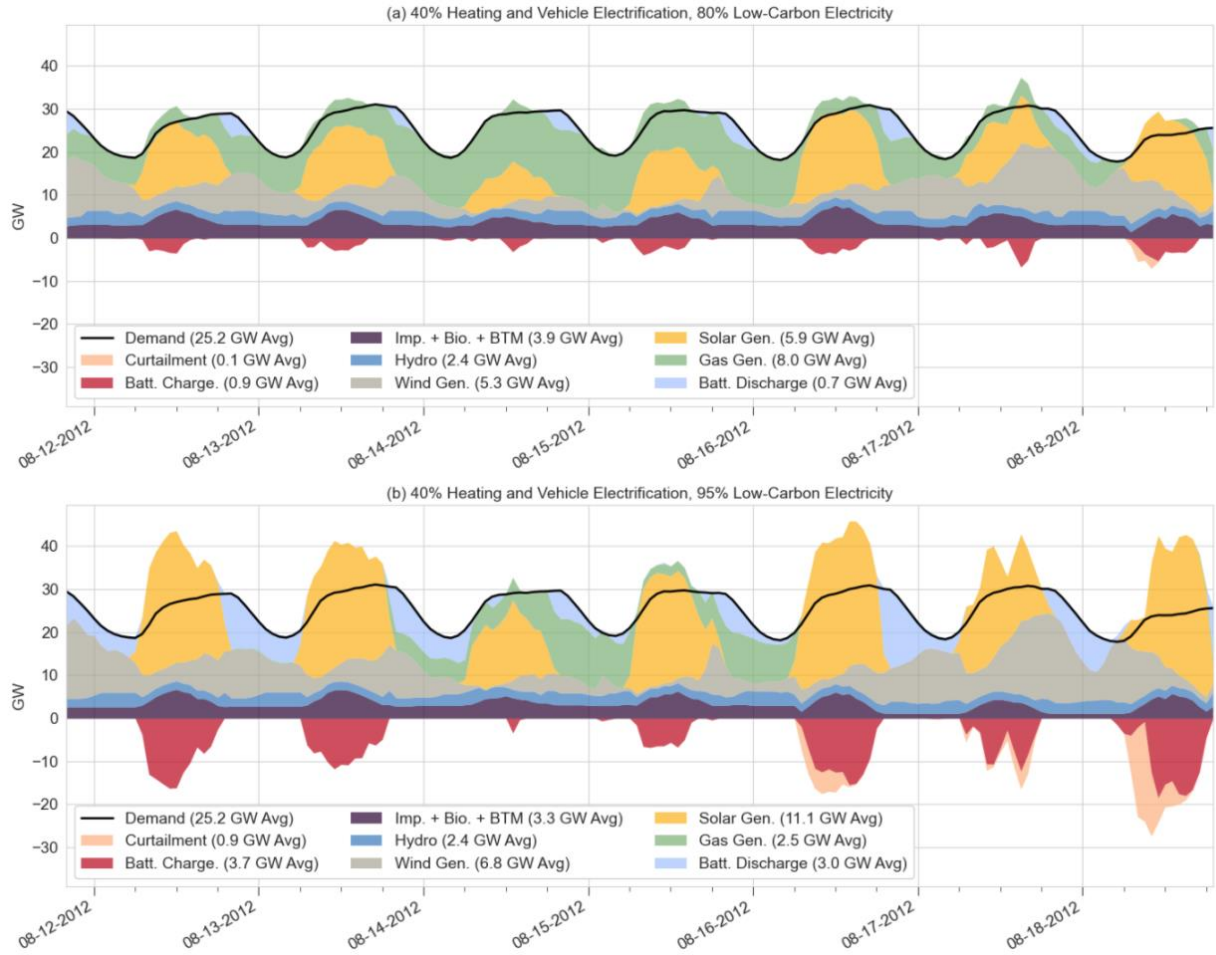

Fig. 9: Electricity generation and demand for a representative summer week with 40% HVE. (a) 80% LCP; (b) 95% LCP. 'Imp. + Bio. + BTM' represents the sum of imports, biofuel, and behind-the-meter solar generation. Average values reported in the legend are for the week shown.

## 3.3 Comparison to New York State policy studies

SECTR-NY model results are compared to initial analyses presented to the New York State Climate Action Council[v], a committee preparing a scoping plan for CLCPA, both to validate SECTR-NY outputs and to evaluate the effects of different model assumptions and input data. A comparison of select characteristics of the NYS Climate Action Council Advisory Panel (AP) 2030 scenario and two SECTR-NY scenarios is shown in Table 3. The AP 2030 scenario includes an 85% LCP and approximately 15% HVE[vi] with a computed energy-related GHG emissions reduction of 47.4% (relative to 1990, as are all GHG reductions discussed here); this scenario includes 28.4 GW of total wind and solar capacity and 3 GW battery storage capacity. For the same LCP and HVE, SECTR-NY Scenario A computes a total wind and solar capacity of 39.2 GW, 3.2 GW battery storage capacity, and GHG emissions reduction of 27.7%. There are two primary drivers for the greater SECTR-NY capacities here:

1. 14% higher average total wind and solar generation in SECTR-NY Scenario A (9.0 GW) than in AP 2030 (7.9 GW). This is due to more hydropower generation in AP 2030 than in the historical data used in SECTR-NY [39] and approximately 2.3 GW higher average statewide load in SECTR-NY Scenario A. The latter stems from a combination of SECTR-NY using historical electricity demand timeseries containing a higher existing average load (18.7 GW) than is simulated in NYS studies (18.2 GW); 15% SECTR-NY HVE likely being slightly higher than the estimate for AP 2030; AP 2030 considering combinations of population growth and efficiency savings; and SECTR-NY's more accurate representation of low-temperature effects of EHPs and EVs. (These low-temperature effects also explain the difference in fossil fuel-based generation capacity to meet the 35.4 GW peak statewide load computed in SECTR compared to the 29.6 GW statewide peak in AP 2030.)
2. 21% higher aggregate wind and solar capacity factor (CF) in AP 2030 (0.278) than in SECTR-NY Scenario A (0.230). This is primarily driven by significantly lower solar and onshore wind CFs in the latter. Model wind output in SECTR-NY is less than that of most available wind data: SECTR-NY employs a dataset that contains adjusted model data based on historical output of actual wind farms in NYS [40]. A comparison of solar data series was not performed; however, the authors believe SECTR-NY Scenario A's statewide solar CF of 0.166 represents more realistic expectations for NYS's latitude range than AP 2030's 0.194.

The difference in computed GHG reductions between AP 2030 and SECTR-NY Scenario A stems from model assumptions related to methane leakage in natural gas production and transport upstream of NYS. SECTR-NY relies on research on natural gas leakage [41,42] that estimates approximately 3.6% leakage with an associated impact on fossil fuel emissions factors [43]. AP 2030 reduces the leakage to approximately 2%, though the authors have not seen an explanation for this assumption. The implications of these assumptions can be seen in SECTR-NY Scenario B, in which more heating and vehicle electrification is needed to achieve the same percentage GHG emissions reduction as that computed for AP 2030. Here, total computed wind and solar capacity increases to 51.4 GW, 81% greater than that anticipated by the recent analyses presented to the NYS Climate Action Council.

---

[v] NYS published studies are available at the following link: https://climate.ny.gov/Climate-Resources. Technical analysis of initial results [31] and of key drivers and outputs [46] last updated in November and December of 2021 are of particular use in understanding the state's modeling methodology and simulated decarbonization pathways.
[vi] The AP considered different electrification rates for different end uses, so this estimate is not directly analogous to that of SECTR-NY presented here. See Table 3, footnote 2 for a breakdown of the different electrification rates assumed in the AP recommendations.

Table 3: Comparison of NYS Climate Action Council Advisory Panel (AP) recommendations and SECTR-NY simulation results for modeled 2030 decarbonization scenarios.

|  | Modeled Scenario | | |
|---|---|---|---|
|  | NYS AP 2030 | SECTR-NY, A | SECTR-NY, B |
| **Low-Carbon Electricity Percent (LCP)** | 85% | 85%[1] | 85%[1] |
| **Heating and Vehicle Electrification (HVE)** | 15%[2] | 15%[1] | 50% |
| **GHG Emissions Change (Compared to 1990)** | -47.4% | -27.7% | -47.4%[1] |
| **Electricity Demand Peak [GW] | Average [GWh/h]** | 29.6 | 18.4 | 35.4 | 20.7 | 52.7 | 25.4 |
| **Onshore Wind Capacity [GW] | Average Generation [GWh/h]** | 5.2 | 1.7 | 11.2 | 2.6 | 14.2 | 3.4 |
| **Offshore Wind Capacity [GW] | Average Generation [GWh/h]** | 6.2 | 2.9 | 8.4 | 3.8 | 13.2 | 5.8 |
| **Solar Capacity [GW] | Average Generation [GWh/h]** | 17.0 | 3.3 | 19.6 | 2.7 | 24.0 | 3.9 |
| **In-State Hydropower Capacity [GW] | Average Generation [GWh/h]** | 4.6 | 3.5 | 5.3 | 3.0 | 5.3 | 3.0 |
| **Hydropower Imports Capacity [GW] | Average Generation [GWh/h]** | 2.7 | 2.2 | 2.8 | 2.0 | 2.8 | 2.0 |
| **Nuclear Capacity [GW] | Average Generation [GWh/h]** | 3.4 | 3.0 | 3.5 | 3.2 | 3.5 | 3.2 |
| **Battery Capacity [GW]** | 3.0 | 3.2 | 9.9 |
| **Fossil Fuel Capacity [GW] | Average Generation [GWh/h]** | 20.8 | 2.7 | 27.0 | 2.5 | 27.6 | 3.2 |

[1] Indicates configuration parameters specified for the SECTR-NY model scenario.
[2] Approximated from the following proportions of vehicle and building stock end use equipment transitioning to electric alternatives in the AP 2030 scenario: 14% of light duty vehicles, 6% of heavy duty vehicles, 11% of residential space heating; 11% of commercial space heating, 25% of residential water heating, and 19% of commercial water heating.

# 4. Discussion

This study's results are broadly consistent with previously published research that deep greenhouse gas (GHG) emissions reductions require both a significant low-carbon electricity percentage (LCP) and increases in heating and vehicle electrification (HVE); however, an important finding is that by prioritizing heating and vehicle electrification in conjunction with renewable energy deployment rather than first focusing on LCP, emissions reductions can be achieved with lower electricity supply costs. Through comparative scenarios, the benefits of end use electrification to the electricity system are emphasized: Heating and vehicle electrification allows the same amount of renewable energy to be installed with significantly lower electricity supply costs all while producing deeper reductions in GHG emissions.

First order GHG reductions from electrification occur because of improved energy efficiency compared to the direct use of fossil fuels for heating and vehicles, even when the LCP is close to 40%, i.e. that of the existing NYS electricity grid. At this LCP, average heating emissions per unit heat delivered are 70% lower with current electric technologies than existing fossil fuel-based heating; average vehicle emissions per mile traveled are 56% lower.

For LCPs at or below 60%, higher levelized costs of electricity (LCOE) of wind and solar generation are mitigated by higher utilization of existing infrastructure with increased HVE (with LCOE even decreasing at HVEs up to 20-40%). The 70-80% LCP range represents a transition phase: Beyond 80%, integration costs (e.g., curtailment and battery storage) lead to rapidly rising LCOEs. Accordingly, three primary levelized cost of electricity (LCOE) drivers are identified from the range of LCPs and HVEs investigated: (1) per-unit costs of existing infrastructure decrease with increasing demand from HVE, (2) wind and solar power generation costs rise relative to gas-based and hydropower generation, and (3) costs of integration increase when large amounts of wind and solar power produce electricity in excess of demand.

For LCPs below 80%, wind generation meets most of the low-carbon generation requirement, as onshore wind represents the lowest LCOE renewable resource, followed by offshore wind resources near the dense load areas of New York City and Long Island. Beyond 80% LCP, paired solar generation and batteries become the most cost-effective method of displacing fossil fuel-based electricity generation. At higher LCPs, battery cycling occurs daily, making solar a more appropriate paired generation resource – at least some electricity is generated from solar daily whereas wind can drop off considerably for multi-day periods, particularly in the summer.

The marginal costs of lowering emissions from the limited set of electricity supply technologies considered here (wind, solar, battery and gas turbines) become high enough at LCPs larger than 80% to suggest that other nascent technologies (e.g., hydrogen storage) may play a role in achieving full energy sector decarbonization. Moreover, targeted deployment of other demand-side technologies not modeled – such as upgraded building envelopes, thermal storage and ground-source heat pumps – could further reduce supply costs by reducing heating-driven system peaks. Demand-side flexibility measures – like dual-fuel capabilities and grid-interactive

controls – may also mitigate integration costs and reduce dispatchable capacity requirements. Lastly, breakthroughs in energy and emissions intensive industrial sectors could partially scale down emissions reductions needed in the residential, commercial, transportation, and electricity sectors.

A comparison of model results described in this paper to initial analyses presented to the New York State Climate Action Council ("NYS study") validated SECTR-NY outputs, but also highlighted important factors in assessing the planning implications of such models. While SECTR-NY and the NYS study compute similar energy resource capacities for a scenario in line with the State's Year 2030 targets, deviations between the two can largely be attributed to differences in time series data for wind/solar potential time series and historical demand data, and to this paper's particular attention to low-temperature effects on heat pump and electric vehicle performance. Accurately modeling the potential generation from renewable resources and new electrification-driven peak demands does thus affect the resource capacity required to meet the electric load. However, the two models do diverge significantly in the calculation of GHG emissions. SECTR-NY computes lower emission reductions than the NYS study for a given combination of LCP and HVE; SECTR-NY includes upstream natural gas leakage in line with recent research and its related quantifiable GHG effects, whereas the NYS study assumes a lower leakage rate. As detailed in the paper, this distinction has significant implications for the amount of electrification needed to meet the State's GHG reduction targets.

A couple of caveats surrounding this paper's methodology and results are also worth mentioning. Foremost, SECTR does not model the electricity distribution network. As there will be a need to upgrade distribution to incorporate end-use heating and vehicle electrification, future work should investigate the scale, location, and costs of this reinforced capacity. Second, all SECTR generation is considered to be lumped. While this assumption substantially increases model tractability, it masks operating practices at the individual generator level where decisions are made. Third, LCPs are imposed on the amount of instate electricity generation, and do not account for the carbon content of any imported electricity. Should state regulations change to allow clean, imported electricity to satisfy low-carbon generation targets, the SECTR general formulation will need to be adjusted. Lastly, this paper presents results for a single set of cost assumptions. Should these assumptions prove inaccurate, rerunning the presented decarbonization scenarios will be required.

As the SECTR framework is an open-source, computationally efficient, capacity transition and system operation framework, the energy systems research community can adapt it in a number of ways for future work. One possibility is parameterizing SECTR for other RTO/ISO settings to explore comparative lowest cost decarbonization pathways. Moreover, within an RTO/ISO, researchers can investigate the impact of further interconnections to external generation. Lastly, researchers can build upon the SECTR framework by addressing the caveats mentioned above, such as by adding location specific costs for upgraded distribution capacity.

# 5. Conclusions

This paper introduces an open-source System Electrification and Capacity Transition (SECTR) modeling framework; the framework is then applied to the New York State (NYS) regional energy system (SECTR-NY). By characterizing existing system capacities, loads, and pricing structures, SECTR-NY reasonably approximates current electricity supply costs, establishing a reliable baseline from which to investigate different combinations of low-carbon electricity percentages (LCP) and heating and vehicle electrification rates (HVE).

Methodologically, SECTR addresses several shortcomings of traditional capacity expansion models (CEMs), including characterization of existing energy infrastructure systems, multi-year simulations with weather-dependent time series inputs, and spatially resolved end-use electrification effects. In parameterizing the model for NYS, the model incorporates improved emissions accounting assumptions specified by recent climate legislation but previously unimplemented in state decarbonization studies. This study demonstrates that overall energy emissions reductions can be achieved at lower electricity costs by prioritizing heating and vehicle electrification ahead of complete grid decarbonization; the former approach still requires a major buildout of wind and solar power, but at lower percentage penetration into the grid because of higher demands from more electrification. Moreover, three main electricity supply cost drivers are established for a decarbonizing energy system: (1) decreasing per-unit supply costs of existing infrastructure with increasing electrification (i.e. with higher demand); (2) higher wind and solar power supply costs relative to current hydropower and fossil fuel-based generation; and (3) increasing costs of integration (due to curtailment and energy storage) as solar and wind supply in excess of demand increase with LCP.

# 6. Data Availability

All code and data used for the SECTR-NY model formulation can be found in the following GitHub repository: https://github.com/SEL-Columbia/sectr-ny.

# 7. Acknowledgements

Partial support for this effort was provided by the National Science Foundation. T.C. received support through INFEWS NSF Award Number 1639214; M.B.W. received support through SRN NSF Award Number 1444745.

V.M. and M.B.W received support through Breakthrough Energy Grant Number BE CU20-3670.

# 7. Appendix A

Table 4 contains a full listing of all nodal cost assumptions in SECTR-NY. The Supplementary Materials provides a full accounting of how these assumptions were reached. Internodal transmission upgrade and O&M costs are presented in Supplementary Table S2.

Table 4: Cost assumptions used in SECTR-NY.

| Quantity | Unit | Node 1 | Node 2 | Node 3 | Node 4 | Notes |
|---|---|---|---|---|---|---|
| Onshore Wind Capacity Cost, High | $/kW | 1992 | 1992 | N/A | N/A | See SM, page 18 |
| Onshore Wind Capacity Cost, Low | $/kW | 1698 | 1698 | N/A | N/A | See SM, page 18 |
| Offshore Wind Capacity Cost, High | $/kW | N/A | N/A | 3583 | 3583 | See SM, page 18 |
| Offshore Wind Capacity Cost, Low | $/kW | N/A | N/A | 2256 | 2256 | See SM, page 18 |
| Utility-Scale Solar Capacity Cost, High | $/kW | 1341 | 1341 | 1593 | 1593 | See SM, page 19 |
| Utility-Scale Solar Capacity Cost, Low | $/kW | 1006 | 1006 | 1195 | 1195 | See SM, page 19 |
| Battery Storage Energy Cost, High | $/kWh | 208 | 208 | 208 | 208 | See SM, page 21 |
| Battery Storage Energy Cost, High | $/kWh | 144 | 144 | 144 | 144 | See SM, page 21 |
| Hydrogen Storage Energy Cost | $/kWh | 0.35 | 8.29 | 8.29 | 8.29 | See SM, page 22 |
| Hydrogen Storage Power Cost | $/kW | 3013 | 3013 | 4036 | 4036 | See SM, page 22 |
| New Fossil Fuel-Based Generation Capacity Cost | $/kW | 772 | 772 | 1034 | 1034 | See SM, page 17 |
| Hydropower Generation Cost | $/MWh | 18.47 | 28.02 | N/A | N/A | See SM, page 23 |
| Nuclear Generation Cost | $/MWh | 37.94 | N/A | 26.82 | N/A | See SM, page 22 |
| Biofuel Generation Cost | $/MWh | 20.66 | 27.41 | 27.05 | 32.39 | See SM, page 24 |
| Imported Electricity Cost | $/MWh | 22.13 | N/A | 70 | N/A | See SM, page 25 |
| Wholesale Natural Gas Price | $/MMBTU | 2.89 | 4.04 | 3.67 | 3.62 | See SM, page 17 |
| Existing Fossil Fuel-Based Generation Ramping Cost | $/MW-h | 79 | 79 | 79 | 79 | See SM, page 17 |
| New Fossil Fuel-Based Generation Ramping Cost | $/MW-h | 69 | 69 | 69 | 69 | See SM, page 17 |
| New Fossil Fuel-Based Generation Fixed O&M Cost | $/kW-yr | 6.97 | 6.97 | 6.97 | 6.97 | See SM, page 17 |
| Onshore Wind Capacity Fixed O&M Cost | $/kW-yr | 18.1 | 18.1 | N/A | N/A | See SM, page 18 |
| Offshore Wind Capacity Fixed O&M Cost | $/kW-yr | N/A | N/A | 38 | 38 | See SM, page 18 |
| Utility-Scale Solar Capacity Fixed O&M Cost | $/kW-yr | 10.4 | 10.4 | 10.4 | 10.4 | See SM, page 19 |
| Hydrogen Storage Fixed O&M Cost | $/kW-yr | 48.87 | 48.87 | 48.87 | 48.87 | See SM, page 22 |
| New Fossil Fuel Based Generation Variable O&M Cost | $/MWh | 4.48 | 4.48 | 4.48 | 4.48 | See SM, page 17 |
| Existing Generation Capacity Maintenance Cost | $/kW-yr | 27.64 | 53.44 | 101.303 | 104.6 | See SM, page 17 |
| Existing Transmission Capacity Maintenance Cost | $/MWh | 16.9 | 16.9 | 27.3 | 27.3 | See SM, page 17 |

Table 5 contains a full listing of existing nodal capacities modeled in SECTR-NY. The Supplementary Materials provides a full accounting of how these values were reached. Internodal existing transmission capacities are presented in Supplementary Table S2.

Table 5: Existing capacities modeled in SECTR-NY.

| Capacity Type | Unit | Node 1 | Node 2 | Node 3 | Node 4 | Notes |
|---|---|---|---|---|---|---|
| Onshore Wind | MW | 1985 | 0 | 0 | 0 | See SM, page 18 |
| Offshore Wind | MW | 0 | 0 | 0 | 0 | See SM, page 18 |
| Utility Scale Solar | MW | 0 | 0 | 0 | 56.5 | See SM, page 19 |
| Behind-the-Meter Solar | MW | 562 | 523 | 293 | 259 | See SM, page 20 |
| Gas-Fueled | MW | 3934.2 | 8622.5 | 10249.9 | 4192.7 | See SM, page 17 |
| Hydropower | MW | 4717.4 | 608.7 | 0 | 0 | See SM, page 23 |
| Nuclear | MW | 3536.8 | 0 | 2311 | 0 | See SM, page 22 |
| Biofuel | MW | 258 | 45 | 59.7 | 142.2 | See SM, page 24 |
| Interregional Import Limits | MW | 1500 | 0 | 1250 | 0 | See SM, page 25 |
| Battery Storage, Energy | MWh | 5.2 | 80 | 0 | 65 | See SM, page 21 |
| Battery Storage, Power | MW | 3 | 20 | 0 | 10 | See SM, page 21 |

# 7. References

[1] Clack CTM, Choukulkar A, Cote B, McKee SA. Technical Report: Economic & Clean Energy Benefits of Establishing a Competitive Wholesale Electricity Market in the Southeast United States 2020:153. https://vibrantcleanenergy.com/wp-content/uploads/2020/08/SERTO_WISdomP_VCE-EI.pdf.

[2] Clean Air Task Force. State and utility climate change targets shift to carbon reductions 2019:1–9. https://www.catf.us/wp-content/uploads/2019/05/State-and-Utility-Climate-Change-Targets.pdf.

[3] Scovronick N, Budolfson M, Dennig F, Errickson F, Fleurbaey M, Peng W, et al. The impact of human health co-benefits on evaluations of global climate policy. Nat Commun 2019;10:1–12. https://doi.org/10.1038/s41467-019-09499-x.

[4] Fitzroy F. A Green New Deal: The Economic Benefits of Energy Transition. Substantia 2019;3(2) Suppl:55–67. https://doi.org/10.13128/Substantia-276.

[5] Tyson A, Kennedy B. Two-Thirds of Americans Think Government Should Do More on Climate. Pew Res Cent 2020. https://www.pewresearch.org/science/2020/06/23/two-thirds-of-americans-think-government-should-do-more-on-climate/.

[6] Guelpa E, Bischi A, Verda V, Chertkov M, Lund H. Towards future infrastructures for sustainable multi-energy systems: A review. Energy 2019;184:2–21. https://doi.org/10.1016/j.energy.2019.05.057.

[7] Grubler A, Wilson C, Bento N, Boza-Kiss B, Krey V, McCollum DL, et al. A low energy demand scenario for meeting the 1.5 °C target and sustainable development goals without negative emission technologies. Nat Energy 2018;3:515–27. https://doi.org/10.1038/s41560-018-0172-6.

[8] Bistline JET. Roadmaps to net-zero emissions systems: Emerging insights and modeling challenges. Joule 2021;5:2551–63. https://doi.org/10.1016/j.joule.2021.09.012.

[9] Conlon T, Waite M, Modi V. Assessing new transmission and energy storage in achieving increasing renewable generation targets in a regional grid. Appl Energy 2019;250:1085–98. https://doi.org/10.1016/j.apenergy.2019.05.066.

[10] Barth R, Brand H, Meibom P, Weber C. A stochastic unit-commitment model for the evaluation of the impacts of integration of large amounts of intermittent wind power. 2006 9th Int Conf Probabilistic Methods Appl to Power Syst PMAPS 2006. https://doi.org/10.1109/PMAPS.2006.360195.

[11] Kern JD, Patino-Echeverri D, Characklis GW. An integrated reservoir-power system model for evaluating the impacts of wind integration on hydropower resources. Renew Energy 2014. https://doi.org/10.1016/j.renene.2014.06.014.

[12] An Y, Zeng B. Exploring the modeling capacity of two-stage robust optimization: Variants of robust unit commitment model. IEEE Trans Power Syst 2015;30:109–22. https://doi.org/10.1109/TPWRS.2014.2320880.

[13] Jenkins J, Sepulveda N. Enhanced Decision Support for a Changing Electricity Landscape: the GenX Configurable Electricity Resource Capacity Expansion Model. MIT Energy Initiat Work Pap 2017:1–40. https://energy.mit.edu/wp-content/uploads/2017/10/Enhanced-Decision-Support-for-a-Changing-Electricity-Landscape.pdf.

[14] Levi PJ, Kurland SD, Carbajales-Dale M, Weyant JP, Brandt AR, Benson SM. Macro-Energy Systems: Toward a New Discipline. Joule 2019;3:2282–6. https://doi.org/10.1016/j.joule.2019.07.017.

[15] Denholm P, Arent DJ, Baldwin SF, Bilello DE, Brinkman GL, Cochran JM, et al. The challenges of achieving a 100% renewable electricity system in the United States. Joule 2021;5:1331–52. https://doi.org/10.1016/j.joule.2021.03.028.

[16] Howells M, Rogner H, Strachan N, Heaps C, Huntington H, Kypreos S, et al. OSeMOSYS: The Open Source Energy Modeling System. An introduction to its ethos, structure and development. Energy Policy 2011;39:5850–70. https://doi.org/10.1016/j.enpol.2011.06.033.

[17] Gil E, Aravena I, Cardenas R. Generation Capacity Expansion Planning Under Hydro Uncertainty Using Stochastic Mixed Integer Programming and Scenario Reduction. IEEE Trans Power Syst 2015;30:1838–47. https://doi.org/10.1109/PESGM.2015.7285838.

[18] Short W, Sullivan P, Mai T, Mowers M, Uriarte C, Blair N, et al. Regional Energy Deployment System (ReEDS) - Technical Report. NREL 2011. https://www.nrel.gov/docs/fy12osti/46534.pdf.

[19] Karlsson K, Meibom P. Optimal investment paths for future renewable based energy systems-Using the optimisation model Balmorel. Int J Hydrogen Energy 2008;33:1777–87. https://doi.org/10.1016/j.ijhydene.2008.01.031.

[20] DeCarolis JF, Jaramillo P, Johnson JX, McCollum DL, Trutnevyte E, Daniels DC, et al. Leveraging Open-Source Tools for Collaborative Macro-energy System Modeling Efforts. Joule 2020;4:2523–6. https://doi.org/10.1016/j.joule.2020.11.002.

[21] Jenkins JD, Mayfield EN, Larson ED, Pacala SW, Greig C. Mission net-zero America: The nation-building path to a prosperous, net-zero emissions economy. Joule 2021;5:2755–61. https://doi.org/10.1016/j.joule.2021.10.016.

[22] Murphy C, Mai T, Sun Y, Jadun P, Muratori M, Nelson B, et al. Electrification Futures Study: Scenarios of Power System Evolution and Infrastructure Development for the United States. Natl Renew Energy Lab NREL/TP-6A20-72330 2021;1.

[23] Murphy C, Mai T, Sun Y, Jadun P, Donohoo-Vallett P, Muratori M, et al. High electrification futures: Impacts to the U.S. bulk power system. Electr J 2020;33:106878. https://doi.org/10.1016/j.tej.2020.106878.

[24] Steinberg D, Dave Bielen, Eichman J, Eurek K, Logan J, Mai T, et al. Electrification and Decarbonization: Exploring U.S. Energy Use and Greenhouse Gas Emissions in Scenarios with Widespread Electrification and Power Sector Decarbonization. Natl Renew Energy Lab 2017:43. https://doi.org/doi:10.2172/1372620.

[25] Zhou E, Mai T. Electrification Futures Study: Operational Analysis of U.S. Power Systems with Increased Electrification and Demand-Side Flexibility n.d. https://doi.org/10.2172/1785329.

[26] Mai T, Steinberg D, Logan J, Bielen D, Eurek K, McMillan C. An electrified future: Initial scenarios and future research for U.S. Energy and electricity systems. IEEE Power Energy Mag 2018;16:34–47. https://doi.org/10.1109/MPE.2018.2820445.

[27] Bellocchi S, Manno M, Noussan M, Prina MG, Vellini M. Electrification of transport and residential heating sectors in support of renewable penetration: Scenarios for the Italian energy system. Energy 2020;196:117062. https://doi.org/10.1016/j.energy.2020.117062.

[28] Eichman JD, Mueller F, Tarroja B, Schell LS, Samuelsen S. Exploration of the integration of renewable resources into California's electric system using the Holistic Grid Resource Integration and Deployment (HiGRID) tool. Energy 2013;50:353–63. https://doi.org/10.1016/j.energy.2012.11.024.

[29] Tarroja B, Chiang F, AghaKouchak A, Samuelsen S, Raghavan S V., Wei M, et al. Translating climate change and heating system electrification impacts on building energy use to future greenhouse gas emissions and electric grid capacity requirements in California. Appl Energy 2018;225:522–34. https://doi.org/10.1016/j.apenergy.2018.05.003.

[30] Environmental Economics and Energy. Pathways to Deep Decarbonization in New York State 2020. https://climate.ny.gov/-/media/Project/Climate/Files/2020-06-24-NYS-Decarbonization-Pathways-CAC-Presentation.ashx.

[31] New York State Climate Action Council. Technical Advisory Group, Integration Analysis -- Initial Results Presentation (updated November 21, 2021) 2021. https://climate.ny.gov/-/media/Project/Climate/Files/2021-11-18-Integration-Analysis-Initial-Results-Presentation.ashx.

[32] Giarola S, Molar-Cruz A, Vaillancourt K, Bahn O, Sarmiento L, Hawkes A, et al. The role of energy storage in the uptake of renewable energy: A model comparison approach. Energy Policy 2021;151:112159. https://doi.org/10.1016/j.enpol.2021.112159.

[33] Waite M, Modi V. Electricity Load Implications of Space Heating Decarbonization Pathways. Joule 2020;4:376–94. https://doi.org/10.1016/j.joule.2019.11.011.

[34] New York State Senate. Climate Leadership and Community Protection Act — Final Bill Text 2019. https://www.nysenate.gov/legislation/bills/2019/s6599.

[35] US Energy Information Administration (EIA). Annual Energy Outlook 2020 with projections to 2050. 2020. https://www.eia.gov/outlooks/aeo/pdf/aeo2020.pdf.

[36] Patton DB, LeeVanSchaick P, Chen J, Naga RP. 2019 State of the Market Report for the New York ISO Markets. Potomac Ecnonomics 2019:1–24. https://www.nyiso.com/documents/20142/2223763/NYISO-2019-SOM-Report-Full-Report-5-19-2020-final.pdf/.

[37] US Energy Information Administration (EIA). Electricity Data Browser 2021. https://www.eia.gov/electricity/data/browser/.

[38] Waite M, Modi V. Existing and projected infrastructure capacities motivate alternatives to all-electric heating decarbonization. Joule-D-19-00676 2019.

[39] US Energy Information Administration (EIA). Form EIA-923 detailed data with previous form data (EIA-906/920) 2020. https://www.eia.gov/electricity/data/eia923/.

[40] Waite M, Modi V. Modeling wind power curtailment with increased capacity in a regional electricity grid supplying a dense urban demand. Appl Energy 2016. https://doi.org/10.1016/j.apenergy.2016.08.078.

[41] Alvarez RA, Zavala-Araiza D, Lyon DR, Allen DT, Barkley ZR, Brandt AR, et al. Assessment of methane emissions from the U.S. oil and gas supply chain. Science (80- ) 2018;361:186–8. https://doi.org/10.1126/science.aar7204.

[42] Howarth RW, Santoro R, Ingraffea A. Methane and the greenhouse-gas footprint of natural gas from shale formations. Clim Change 2011;106:679–90. https://doi.org/10.1007/s10584-011-0061-5.

[43] Howarth RW. Methane Emissions and Greenhouse Gas Accounting: A Case Study of a New Approach Pioneered by the State of New York 2019:14. https://documents.dps.ny.gov/public/Common/ViewDoc.aspx?DocRefId=%7B3498AB82-B671-451E-A556-A917A61F939A%7D.

[44] US Energy Information Administration (EIA). Annual household site end-use consumption by fuel in the U.S. 2018:8. https://www.eia.gov/consumption/residential/data/2015/c&e/pdf/ce4.1.pdf.

[45] US Energy Information Administration (EIA). Commercial Building Energy Consumption Survey (CBECS) 2018. https://www.eia.gov/consumption/commercial/data/2018/.

[46] New York State Climate Action Council. Technical Advisory Group, Integration Analysis -- Key Drivers and Outputs (updated December 30, 2021) 2021. https://climate.ny.gov/-/media/Project/Climate/Files/IA-Tech-Supplement-Annex-2-Key-Drivers-Outputs.ashx.

# Assessing trade-offs among electrification and grid decarbonization in a clean energy transition: Application to New York State

# Supplementary Materials

Terence Conlon[a†*], Michael Waite[a†**], Yuezi Wu[a***], and Vijay Modi[a****]

[a]Department of Mechanical Engineering, Columbia University
220 S.W. Mudd Building, 500 West 120th Street, New York, NY 10027, USA

[†] Corresponding author
* tmc2180@columbia.edu
** mbw2113@columbia.edu
*** yw3054@columbia.edu
**** modi@columbia.edu

Table of Contents

# S1 Supplementary Nomenclature
*Additional fixed variables and parameters*

| | |
|---|---|
| $C_{fuel}$ | total fuel cost for fossil fuel-based generation over entire analysis period [$] |
| $CAP_{h2-e,i}$ | hydrogen storage energy capital cost at node *i* [$/MWh] |
| $CAP_{h2-p,i}$ | hydrogen storage power capital cost at node *i* [$/MW] |
| $D^t_{veh-fix,i}$ | vehicle fixed charging demand at node *i* [MWh] |
| $E^{daily}_{veh,i}$ | daily vehicle charging demand at node *i* [MWh] |
| $E^{daily}_{veh-fix,i}$ | daily vehicle fixed charging demand at node *i* [MWh] |
| $E^{daily}_{veh-flex,i}$ | daily vehicle flexible charging demand at node *i* [MWh] |
| $E^{daily}_{veh-tot,i}$ | total daily vehicle charging demand at node *i* (fixed plus flexible) [MWh] |
| $f_{c-s}$ | cubic spline function |
| $h_{veh-start}$ | electric vehicle charging start time |
| $h_{veh-end}$ | electric vehicle charging end time |
| $h_{veh-min}$ | minimum number of hours required for full daily electric vehicle charging [hours] |
| $H^{daily}_{flex,i}$ | daily flexible hydropower generation at node *i* [MWh] |
| $H^{monthly}_{fix,i}$ | monthly fixed hydropower generation at node *i* [MWh] |
| $H^{monthly}_{flex,i}$ | monthly flexible hydropower generation at node *i* [MWh] |
| $H^{monthly}_{tot,i}$ | monthly total hydropower electricity generation at node *i* (fixed plus flexible) [MWh] |
| $L^{daily}_i$ | daily biofuel generation at node *i* [MWh] |
| $L^{max}_i$ | biofuel maximum generation at node *i* [MWh] |
| $m$ | day index |
| $omf_{h2}$ | hydrogen storage fixed operations and management cost [$/MW-yr] |
| $V^{max}_i$ | maximum hourly electricity import limit at node *i* [MWh] |
| $X^{existing}_{batt-e,i}$ | capacity of existing battery energy at node *i* [MWh] |
| $X^{existing}_{batt-p,i}$ | capacity of existing battery power at node *i* [MW] |
| $X^{existing}_{bio,i}$ | capacity of existing biofuel generation at node *i* [MW] |
| $X^{existing}_{ff,i}$ | capacity of existing fossil fuel-based generation at node *i* [MW] |
| $X^{existing}_{hydro,i}$ | capacity of existing hydropower generation at node *i* [MW] |
| $X^{existing}_{nuc,i}$ | capacity of existing nuclear generation at node i [MW] |
| $X^{existing}_{tx,ii'}$ | capacity of existing transmission between node *i* and adjacent node *i'* [MW] |
| $y$ | fraction |
| $\eta_{batt}$ | one-way battery storage efficiency |
| $\eta_{h2}$ | one-way hydrogen storage efficiency |
| $\eta_{veh}$ | electric vehicle charging efficiency |
| $\kappa$ | storage self-discharge |
| $\sigma$ | fossil fuel-based generation reserve requirement |

$\varphi_{p2e-batt-min}$ minimum possible battery storage power-to-energy ratio
$\varphi_{p2e-batt-max}$ maximum possible battery storage power-to-energy ratio
$\varphi_{p2e-h2-min}$ minimum possible hydrogen storage power-to-energy ratio
$\varphi_{p2e-h2-max}$ maximum possible hydrogen storage power-to-energy ratio

*Additional decision variables*
*All variables are constrained to be greater than or equal to 0.*

$D_{veh-flex,i}^{t}$     hourly vehicle flexible charging demand at node *i* [MWh]
$E_{batt,i}^{t}$     aggregate battery storage state of charge at node *i* and timestep *t* [MWh]

$E_{h2,i}^{t}$     aggregate hydrogen storage state of charge at node *i* [MWh]
$X_{h2-e,i}$     hydrogen storage energy capacity installed at node *i* [MWh]
$X_{h2-p,i}$     hydrogen storage power capacity installed at node *i* [MW]
$\gamma_{h2,i}^{t}$     increase in hydrogen storage state of charge at node *i* [MWh]
$\delta_{h2,i}^{t}$     decrease in hydrogen storage state of charge at node *i* [MWh]

*Additional scenario configuration parameters*

RGT     renewable electricity generation target: Fraction of total demand that must be met by renewable energy (combined wind, water, and solar power)

*Additional subscripts and superscripts*

gas     motor gasoline
h2     hydrogen storage
max     maximum
ng     natural gas

# S2 Supplementary Methodology

Section S2 presents the remainder of methodology for the System Electrification and Capacity TRansition (SECTR) framework, and how the framework is applied to the New York State (NYS) energy system (SECTR-NY).

## S2.1 Remainder of general formulation governing equations

The following sections contain the governing equations for the SECTR general formulation not specified in Section 2.1 of the main text.

*Characterization of fossil fuel generation*

Fossil fuel-based electricity generation from existing, $X_{ff,i}^{existing}$, and new, $X_{ff,i}$, capacity is modeled. In scenarios where $X_{ff,i}$ is selected, all new generation is provided by simple cycle gas turbines, because of the very low load factors of peak load increases with heating and vehicle electrification [1]. Existing fossil fuel-based generation efficiency, $\eta_{ff-existing}$, is determined from historical data; new gas turbine efficiency, $\eta_{ff-new}$, is based on advanced combustion turbines [2]. Fossil fuel generation costs are computed per Eq. (S1).

$$C_{fuel} = \sum_{t \in T} \sum_{i \in I} 3.412 * c_{ff,i} * \left( \frac{G_{existing,i}^t}{\eta_{ff-existing}} + \frac{G_{new,i}^t}{\eta_{ff-new}} \right)$$

(S1)

A capacity reserve margin on $X_{ff,i}^{existing}$ and $X_{ff,i}$ is also imposed:

$$X_{ff,i}^{existing} \geq (1 + \sigma) * G_{existing,i}^t$$

(S2)

$$X_{ff,i} \geq (1 + \sigma) * G_{new,i}^t$$

(S3)

To avoid significant increases in computation time, fossil fuel-based generation start-up costs are linearized as ramping costs, $c_{ff-ramp}$, on a per-MW per-hour basis ($/MW-h); this quantity is applied to $G_{new-diff,i}^t$ and $G_{existing-diff,i}^t$, variables which represent the absolute value of the hourly change in gas generation (Eqs. (S4-S5)). Ramping limitations are not imposed on the gas generators [3].

$$G_{existing-diff,i}^t = | G_{existing,i}^t - G_{existing,i}^{t-1} |$$

(S4)

$$G_{new-diff,i}^t = | G_{new,i}^t - G_{new,i}^{t-1} |$$

(S5)

*Wind capacity*

Both new onshore, $X_{on,i}$, and offshore, $X_{off,i}$, wind capacities are simulated, and are limited by resource availability and maximum capacity available at each node (onshore, Eq. (S6)) or within the study region (offshore, Eq. (S7)):

$$X_{on,i}^{existing} + X_{on,i} \leq X_{on,i}^{max}$$

(S6)

$$\sum_{i \in I} (X_{off,i}^{existing} + X_{off,i}) \leq X_{off}^{max}$$

(S7)

*Solar capacity*

Node-specific BTM solar capacity, $X_{btm-solar,i}$, produces fixed generation at each node equal to the product of user-imposed capacity and the supplied generation potential time series, $W_{btm-solar,i}^t$. BTM solar is treated as must-run.

Utility-scale solar capacity is constrained per Eq. (S8):

$$X_{us-solar,i}^{existing} + X_{us-solar,i} \leq X_{us-solar,i}^{max}$$

(S8)

*Internodal transmission*

The cost of maintaining existing transmission capacity is based on user inputs for historical transmission costs and flows. Costs of new transmission capacity are defined for each internodal interface. Transmission losses of 3% between adjacent nodes are assumed, and a nominal cost of transmission ($0.01/MWh) is applied. Eq. (S9) limits internodal transmission flow, $Z_{ii'}^t$, to the combined capacity of existing, $X_{tx,ii'}^{existing}$, and new, $X_{tx,ii'}$, transmission:

$$Z_{ii'}^t \leq X_{tx,ii'}^{existing} + X_{tx,ii'}$$

(S9)

*Battery storage*

Energy storage is based on lithium-ion batteries and is modeled as bulk storage at each node. Modeled batteries are constrained to a power-to-energy ratio, $\varphi_{p2e-batt}$, and a single efficiency, $\eta_{batt}$, applied on both charge and discharge. A nominal \$0.01/MWh cost is attached to battery charge, $\gamma_{batt,i}^t$, and discharge, $\delta_{batt,i}^t$; storage self-discharge, $\kappa$, is also included. Battery storage constraints are presented in Eqs. (S10-S14).

$$\frac{\delta_{batt,i}^t}{\eta_{batt}} - \eta_{batt} * \gamma_{batt,i}^t = (1-\kappa) * E_{batt,i}^T - E_{batt,i}^t, \qquad \forall t = 0$$
(S10a)

$$\frac{\delta_{batt,i}^t}{\eta_{batt}} - \eta_{batt} * \gamma_{batt,i}^t = (1-\kappa) * E_{batt,i}^{t-1} - E_{batt,i}^t, \qquad \forall t > 0$$
(S10b)

$$E_{batt,i}^t \leq X_{batt-e,i} + X_{batt-e,i}^{existing}$$
(S11)

$$\gamma_{batt,i}^t \leq X_{batt-p,i} + X_{batt-p,i}^{existing}$$
(S12)

$$\delta_{batt,i}^t \leq X_{batt-p,i} + X_{batt-p,i}^{existing}$$
(S13)

$$\varphi_{p2e-batt-min} * (X_{batt-p,i} + X_{batt-p,i}^{existing}) \leq X_{batt-e,i} + X_{batt-e,i}^{existing}$$
$$\leq \varphi_{p2e-batt-max} * (X_{batt-p,i} + X_{batt-p,i}^{existing})$$
(S14)

In the SECTR formulation, storage self-discharge and nominal storage charge and discharge costs are included to limit the number of unique model solutions, thereby allowing the model to find an optimal solution more quickly. In the case where excess low-carbon generation is available over a period of hours, storage self-discharge reduces the number of ways to fully charge the storage to a single, unique schedule. As storage technologies undergo self-discharge in reality, the self-discharge parameter better allows SECTR to simulate likely battery operation. Moreover, when excess low-carbon generation is available and battery storage is fully charged, without nominal storage charge and discharge costs, nothing prevents the model from discharging the batteries, curtailing that energy, and then using the excess generation to recharge the batteries. Nominal charge and discharge costs prevent this type of unnecessary operation.

*Nuclear generation*

Nodal nuclear generation, $N_i^t$ is modeled as constant based on a user input value and is treated as must-run.

*Hydropower generation*

SECTR includes modules for both fixed and flexible hydropower operation per [4]. Monthly hydropower generation is split into fixed, $H_{fix,i}^{monthly}$, and flexible, $H_{flex,i}^{monthly}$ quantities based on the nodal fraction of hydropower to be considered fixed, $y_{fix,i}$, as shown in Eqs. (S15-S16); both monthly generation quantities are fit with cubic splines, $f_{c-s}$, per Eqs. (S17-S18):

$$H_{fix,i}^{monthly} = y_{fix,i} * H_{tot,i}^{monthly}$$
(S15)

$$H_{flex,i}^{monthly} = (1 - y_{fix,i}) * H_{tot,i}^{monthly}$$
(S16)

$$H_{fix,i}^t = f_{c-s}(H_{fix,i}^{monthly})$$
(S17)

$$H_{flex,i}^{daily} = f_{c-s}(H_{flex,i}^{monthly})$$
(S18)

While fixed hydropower generation time series, $H_{fix,i}^t$, are treated as must-run, flexible hydropower generation, $H_{flex,i}^t$, can vary throughout the day to meet a daily nodal total, $H_{flex,i}^{daily}$, per Eqs. (S19-S20).

$$\sum_{t=1+24m}^{24*(m+1)} H_{flex,i}^t = H_{flex,i}^{daily}, \qquad m = 0..\frac{T}{24} - 1$$
(S19)

$$H_{flex,i}^t \leq H_{flex,i}^{max}$$
(S20)

*Biofuel generation*

Biofuel generation, $L_i^t$, is assumed to have flexible operation, and can meet up to a set amount of daily generation, $L_i^{daily}$, without exceeding a nodal limit, $L_i^{max}$, at any time step per Eqs. (S21-S22):

$$\sum_{t=1+24m}^{24*(m+1)} L_i^t \leq L_i^{daily}, \qquad m = 0..\frac{T}{24} - 1$$

(S21)

$$L_i^t \leq L_i^{max}$$

(S22)

*Interregional imports*

Electricity imports into the study region, $V_i^t$, are allowed at each node. All interregional imports are subject to a maximum limit, $V_i^{max}$, per Eq. (S23).

$$V_i^t \leq V_i^{max}$$

(S23)

*Existing generation capacity costs*

A fixed cost, $EX_{cap,i}$, is applied to eligible existing generation capacity, $X_{cap,i}^{existing}$, per Eq. (4) in the main text. All existing hydropower, nuclear, fossil-fuel, and biofuel capacity is included in this approach, per Eq. (S24).

$$X_{cap,i}^{existing} = X_{hydro,i}^{existing} + X_{nuc,i}^{existing} + X_{gt,i}^{existing} + X_{bio,i}^{existing}$$

(S24)

### S2.2 Model Framework Additional Modeling Capabilities

The SECTR framework has additional modeling capabilities not used in any of the SECTR-NY results presented in the Main Text. These capabilities are detailed in the following paragraphs.

*Objective function*

With the inclusion of hydrogen storage energy and power capacity as SECTR decision variables, the total cost of new capacity is presented in Supplementary Eq. (S25):

$$C_{new-cap} = n_{years}$$
$$* \sum_{i \in I} \left[ (A_{P_{on},j} * CAP_{on,i} + omf_{on}) * X_{on,i} + (A_{P_{off},j} * CAP_{off,i} + omf_{off}) \right.$$
$$* X_{off,i} + (A_{P_{us-solar},j} * CAP_{us-solar,i} + omf_{us-solar}) * X_{us-solar,i}$$
$$+ (A_{P_{batt},j} * CAP_{batt-e,i}) * X_{batt-e,i} + (A_{P_{batt},j} * CAP_{batt-p,i}) * X_{batt-p,i}$$
$$+ (A_{P_{h2},j} * CAP_{h2-e,i} + omf_{h2}) * X_{h2-e,i} + (A_{P_{h2},j} * CAP_{h2-p,i}) * X_{h2-p,i}$$
$$+ (A_{P_{ff},j} * CAP_{ff,i} + omf_{ff}) * X_{ff,i}$$
$$\left. + \sum_{i'} (A_{P_{tx},j} * CAP_{tx,ii'} * d_{ii'} + omf_{tx,ii'}) * X_{tx,ii'} \right]$$

(S25)

The second SECTR objective function minimizes the levelized cost of electricity (LCOE) according to Supplementary Eq. (S26), where LCOE is defined in Eq. (5) of the main text. When this second objective function is applied, the user specifies a greenhouse gas (GHG) emission reduction, and SECTR determines the combination of low-carbon electricity percent (LCP) and heating and vehicle electrification rate (HVE) that allows for the lowest LCOE.

$$obj_2 = minimize(LCOE)$$

(S26)

Energy balance constraint

With the inclusion of hydrogen storage charge and discharge capabilities, nodal energy balance is constrained per Supplementary Eq. (S27):

$$(X_{on,i} + X_{on,i}^{existing}) * W_{on,i}^t + (X_{off,i} + X_{off,i}^{existing}) * W_{off,i}^t + (X_{us-solar,i} + X_{us-solar,i}^{existing})$$
$$* W_{us-solar,i}^t + X_{btm-solar,i} * W_{btm-solar,i}^t + H_{flex,i}^t + H_{fixed,i}^t + N_i^t$$
$$+ G_{existing,i}^t + G_{new,i}^t + L_i^t + V_i^t - \gamma_{batt,i}^t + \delta_{batt,i}^t - \gamma_{h2,i}^t + \delta_{h2,i}^t$$
$$+ \sum_{i'} [(1 - l) * Z_{i'i}^t - Z_{ii'}^t] \geq D_{elec,i}^t + D_{heat,i}^t + D_{veh,i}^t$$

(S27)

*Renewable electricity generation targets*

In SECTR simulations, users can also select a renewable generation target (RGT) – a minimum percentage of electricity from onshore and offshore wind, hydropower, and solar. Accordingly, the maximum allowable electricity generated from fossil fuels, biofuels, and nuclear power over the full simulation period is constrained per Supplementary Eq. (S28).

$$\sum_{t \in T} \sum_{i \in I} \left( G_{existing,i}^t + G_{new,i}^t + L_i^t + N_i^t \right) \leq (1 - RGT) * \sum_{t \in T} \sum_{i \in I} \left[ D_{elec,i}^t + D_{heat,i}^t + D_{ev,i} - V_i^t - X_{btm-solar,i} * W_{btm-solar,i}^t \right]$$

(S28)

*Flexible charging of electrified vehicle demand*

SECTR includes another formulation for electric vehicle charging in which $D_{veh,i}^t$ can be computed as the sum of a fixed electric vehicle demand, $D_{veh-fix,i}^t$, and a flexible electric vehicle demand, $D_{veh-flex,i}^t$, per Supplementary Eq. (S29):

$$D_{veh,i}^t = D_{veh-fix,i}^t + D_{veh-flex,i}^t$$

(S29)

This formulation uses a daily nodal vehicle electricity requirement, $E_{veh,i}^{daily}$, calculated as the product of the nodal percentage of vehicle electrification (user-defined or computed, depending on model configuration), $p_{veh,i}$, and user-provided daily nodal electricity requirement for full vehicle electrification, $E_{veh,i}^{daily,full}$, per Supplementary Eq. (S30).

$$E_{veh,i}^{daily} = p_{veh,i} * E_{veh,i}^{daily,full}$$

(S30)

Here, SECTR allows flexibility in meeting daily vehicle electrification energy requirements. Users can split daily vehicle electricity energy demand, $E_{veh,i}^{daily}$, into flexible, $E_{veh-flex,i}^{daily}$, and fixed, $E_{veh-fix,i}^{daily}$, portions based on a provided fraction of daily vehicle electricity requirement allowed to be flexible, $y_{veh-flex}$, as shown in Supplementary Eqs. (S31-S32).

$$E_{veh-flex,i}^{daily} = y_{veh-flex} * E_{veh,i}^{daily}$$

(S31)

$$E_{veh-fix,i}^{daily} = (1 - y_{veh-flex}) * E_{veh,i}^{daily}$$

(S32)

In determining hourly flexible vehicle charging demand, $D_{veh-flex,i}^t$, SECTR requires that the user provide a timestep for the hour at which daily charging can start, $h_{veh-start}$ and a timestep indicating the last hour at which charging is allowed, $h_{veh-end}$. The standard SECTR formulation establishes a lower limit of 4 hours, $h_{veh-min}$, for full daily flexible EV charging. The flexible vehicle charging and power constraints are shown below:

$$\sum_{t=h_{veh-start}}^{h_{veh-end}} D_{veh-flex,i}^t = \frac{E_{veh-flex,i}^{daily}}{\eta_{veh}}, \qquad for\ m = 0..\frac{T}{24} - 1$$

(S33)

$$D_{veh-flex,i}^t \leq \frac{E_{veh-flex,i}^{daily}}{h_{veh-min}}$$

(S34)

To determine the hourly fixed vehicle charging demand, $D_{veh-fix,i}^t$, the daily fixed vehicle charging load is split equally across the same charging period. The fixed charging constraint is shown in Supplementary Eq. (S35).

$$D_{veh-fix,i}^t = \frac{E_{veh-fix,i}^{daily}}{\eta_{veh}*(h_{veh-end}-h_{veh-start}+1)}, for\ t = (h_{veh-start} + 24m)..(h_{veh-end} + 24m),$$
$$for\ m = 0..\frac{T}{24} - 1$$

(S35)

*Hydrogen storage*

Long-term energy storage capabilities are modeled based on potential future system costs of grid-scale power-to-gas (P2G) with hydrogen (H2) gas: H2 produced by electrolysis, $\gamma_{h2,i}^t$; availability of a low-cost gas storage reservoir, $E_{h2,i}^t$; and electricity generated by H2 combustion in a gas turbine, $\delta_{h2,i}^t$. Nodal per-unit power capacity, $CAP_{h2-p,i}$, and energy capacity, $CAP_{h2-e,i}$, cost components are assigned. Hydrogen storage efficiency, $\eta_{h2}$, is applied on both charge and discharge. A self-discharge rate, $\kappa$, is also included.

SECTR places no constraints on the hydrogen storage power-to-energy ratio. Hydrogen storage energy balance, $E_{h2,i}^t$; power capacity, $X_{h2-p,i}$; energy capacity, $X_{h2-e,i}$; charging, $\gamma_{h2,i}^t$; and discharging, $\delta_{h2,i}^t$, constraints are shown in Supplementary Eqs. (S36-S39).

$$\frac{\delta_{h2,i}^t}{\eta_{h2}} - \eta_{h2}*\gamma_{h2,i}^t = (1-\kappa)*E_{h2,i}^T - E_{h2,i}^t, \qquad \forall t = 0$$

(S36a)

$$\frac{\delta_{h2,i}^t}{\eta_{h2}} - \eta_{h2}*\gamma_{h2,i}^t = (1-\kappa)*E_{h2,i}^{t-1} - E_{h2,i}^t, \qquad \forall t > 0$$

$$\text{(S36b)}$$

$$E_{h2,i}^t \leq X_{h2-e,i}$$

(S37)

$$\gamma_{h2,i}^t \leq X_{h2-p,i}$$

(S38)

$$\delta_{h2,i}^t \leq X_{h2-p,i}$$

(S39)

Identical to the treatment of battery storage, hydrogen storage self-discharge and nominal charging and discharging costs are included to limit the number of unique model solutions for a given SECTR configuration.

S2.3 Application of the System Electrification and Capacity TRansition framework to New York State

The subsections below detail the SECTR-NY parameterization, including descriptions of all data sources used and model data development. In SECTR-NY, New York State (NYS) is split into four nodes based on the major transmission interfaces of the New York Independent System Operator (NYISO) control area; these nodes are shown in Supplementary Figure S1.

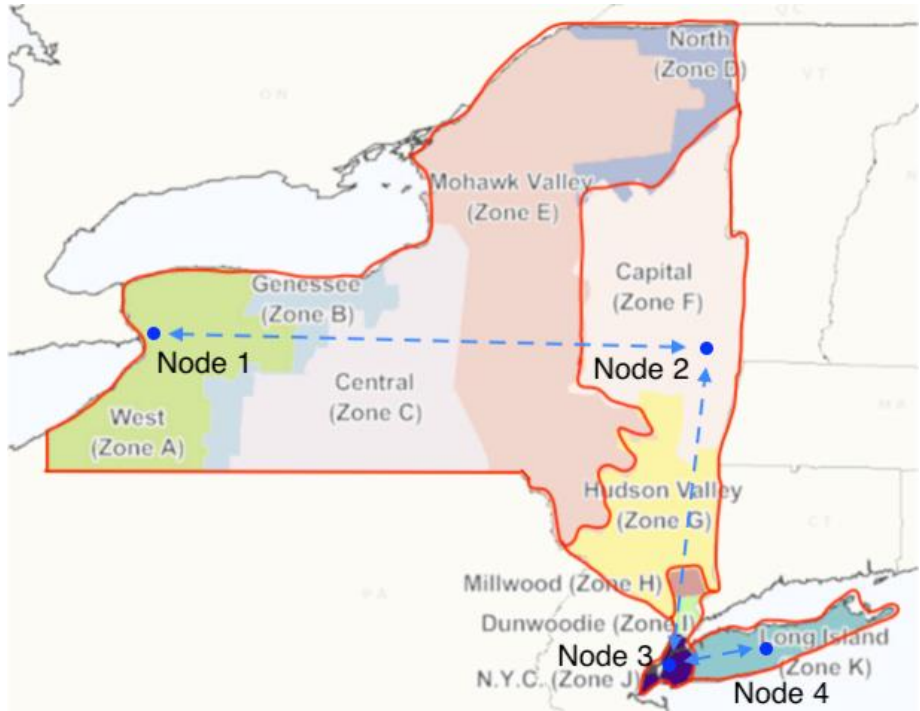

*Supplementary Figure S1: NYISO control area load zones split into model nodes. Node boundaries and connections by authors; underlying image taken from* [5].

In all simulations, low cost estimates are adopted for the technologies with multiple estimates available. All new generation technologies are annualized with a 20-year annualization period; all storage technologies are annualized with a 10-year annualization period. All model constraints presented in the Main Text that contain variables with nodal indexing are applied over all nodes in the study region; constraints which contain variables with temporal indexing are applied over all timesteps in the study period.

*Nodal electricity demands*

The existing electricity demand used is the 2007-2012 demand in each NYISO load zone [6], aggregated at each node per Supplementary Figure S1; the average existing statewide demand is 18,655 MWh/h. Supplementary Table S1 shows average and peak electricity demands at each node. Current electricity demands include some amount of electricity usage for heating and very limited use for passenger vehicles. Here, new electricity demands from converting current fossil fuel end uses in buildings and on-road vehicles to electric technologies are also considered. (As discussed in the Main Text, fossil fuel end uses in buildings are thermal and dominated by space heating, "heating" is used for short.)

Nodal electricity demands for heating fossil fuel conversion to electric heat pumps (EHPs) are based on a nationwide building heating model described in detail in a recently published paper [7] and applied to 2007-2012 temperature data [8]. To convert fossil fuel demands to thermal loads, current average fossil fuel equipment efficiencies of 82% for space heating and 58% for DHW are assumed based on average values for "Installed Base" equipment from the US Energy Information Administration (EIA) [9]. The temperature-dependent coefficient of performance (COP) of new EHPs is based on the 90$^{th}$ percentile performance of EHPs in a regularly updated database of "cold climate" EHPs [10] and modeled per [8]. The COP of domestic hot water (DHW) EHPs was assumed to be a constant 2.32 based on the highest field-validated product performance from an National Renewable Energy Laboratory (NREL) study [11]. Full heating electrification results in a computed statewide average additional electricity load of 7573 MWh/h; however, the conversion of existing electric resistance heating to EHPs is also considered, which reduces statewide average heating electricity demand to 6716 MWh/h. Regional and statewide computed average and peak electrified heating values are shown in Supplementary Table S1.

To parameterize potential electric vehicle charging demand, the total 2018 volumetric sales of gasoline and diesel to New York transportation customers [12] are converted to miles driven using an assumed 21.0 miles per gallon (mpg). The latter assumption is based on an average vehicle age of 11.8 years in 2019 per the Bureau of Transportation Statistics [13] and the corresponding average "Real World" fuel economy of 2008 model year vehicles per the US Environmental Protection Agency (EPA) [14]. The nodal distribution of the fuel sales is assumed to be equal to the distribution of 2016 county level gasoline sales aggregated to the nodal level [15]. This mileage is then converted into daily temperature-dependent EV charging profiles using NREL's EVI-Pro model API [16] assuming 1/3 100-mile range EVs and 2/3 250-mile range EVs (based on a fixed ratio of the NREL model); weekends and weekdays are treated identically, using a 5:2 weighted average of weekday and weekend profiles for each day[vii]. This approach results in a computed average statewide EV demand of 6769 MWh/h. Because of the many assumptions involved and the closeness of this value to the net additional potential demand from heating electrification, the EV demand series is scaled to an equivalent 6716 MWh/h average demand to facilitate more direct comparison between the two. Regional and statewide computed average and peak electrified vehicle values are also shown in Supplementary Table S1.

---

[vii] The NREL tool requires selections among fixed options for various inputs, the following of which were selected: 80% sedans, 20% SUVs; middle option of 80% for home charging preference; middle option of 75% for home charging access; equal usage of Level 1 and Level 2 home charging; 80% of work charging using Level 2 chargers; and minimum delay in charging at both home and work locations.

*Supplementary Table S1: Existing and potential new nodal electricity demands.*

| Node | Existing Electricity Demand [MWh/h][a] | | Computed Potential Net New Heating Electricity Demand [MWh/h][b] | | Computed Potential New Electric Vehicle Demand [MWh/h][b] | |
|---|---|---|---|---|---|---|
| | Average | Peak | Average | Peak | Average | Peak |
| 1 | 6383 | 10,467 | 2178 | 20,982 | 2458 | 5471 |
| 2 | 2495 | 4795 | 1059 | 11,347 | 1182 | 2641 |
| 3 | 7211 | 13,623 | 2376 | 13,303 | 1667 | 3642 |
| 4 | 2567 | 5933 | 1103 | 6601 | 1409 | 3081 |
| Statewide | 18,655 | 33,876 | 6716 | 51,088 | 6716 | 14,836 |

[a] NYISO [6].
[b] See the text of this section.

*Internodal transmission*

In SECTR-NY, both existing internodal transmission limits and costs are characterized. Existing transmission limits assumptions shown in Supplementary Table S2 are those assumed by NYISO for the year 2021 in recent system reliability simulations [17].

*Supplementary Table S2: SECTR internodal existing transmission limits and costs of existing and new transmission.*

| Interface | Miles[a] | Existing Limits [MW][b] | | New Transmission | | |
|---|---|---|---|---|---|---|
| | | West to East | East to West | Cost of New Transmission Capacity [$/MW-mi][c] | | New Transmission O&M costs [$/MW-yr][d] |
| | | | | $/MW-mi | $/GW | |
| 1: Node 1 to 2 | 300 | 5000 | 3400 | 2400 | 720 | 2806 |
| 2: Node 2 to 3 | 150 | 7000 | 7000 | 4800 | 720 | 2357 |
| 3: Node 3 to 4 | 60 | 1613 | 220 | 12,000 | 720 | 277 |

[a] Distance between nodes taken as the distance between the representative cities of Buffalo, Albany, New York City, and Brentwood, per Google Maps.
[b] NYISO [17].
[c] See the text of this section.
[d] NREL [18].

Projecting costs of specific large-scale transmission upgrades is difficult. To evaluate the effect of transmission prices on future energy scenarios, public information on the costs of recent and proposed transmission projects in NYS was reviewed, as well as cost assumptions used in other studies of the region. References used in this assessment include: For Interface 1 (Node 1 to 2), Supplementary Table S2 shows the approximate average of $1400/MW-mi for simulated aboveground High Voltage Direct Current (HVDC) [19]; and $3614/MW-mi for underground HVDC in a NYISO study of the region [20]. For Interface 2 (Node 2 to 3), the Supplementary Table

S2 value is approximately ¾ of the cost of $6567/MW-mi for a recent NYS underground HVDC transmission installation [21] (adjusted downward due unique challenges surrounding this project). For Interface 3 (Node 3 to 4), a transmission upgrade cost of $12,000/MW-mi is assumed based on a previous underground HVDC transmission project between New Jersey and Long Island [22]. With the above per-(MW-mi) costs of upgraded transmission and the assumed distances between each node's representative city, per-GW costs of new transmission are equal at every interface.

The NREL Jobs and Economic Development Impact (JEDI) Transmission Line Model [18] is used to compute internodal O&M costs for new transmission; new transmission capacity between Nodes 1 and 2 assumes O&M costs for 500 kVAC lines, capacity between Nodes 2 and 3 assumes O&M costs for 345 kVAC lines, and capacity between Nodes 3 and 4 assumes O&M costs for HVDC reinforcements.

The annual cost of maintaining existing transmission capacity is assumed to be the total costs recovered through electricity sales based on EIA data [23]: Based on the 2019 transmission contribution to electricity unit costs ($16.9/MWh at Nodes 1 and 2; $27.3/MWh at Nodes 3 and 4) and 2019 total electricity sales (69.683 TWh at Nodes 1 and 2; 75.52 TWh at Nodes 3 and 4), total annual cost for existing transmission was computed to be approximately $3.239B.

*Characterization of fossil fuel-based electricity generation*

SECTR uses a simplified characterization of the existing NYS fossil fuel electricity generation fleet and new generation capacity at each node without modeling individual generators; relevant assumed values described in this section are summarized in Supplementary Table S3. As natural gas provides 96% of fossil fuel-based electricity generation in NYS [24] and generators that burn natural gas (alone or as part of dual fuel capabilities) produce 99% of NYS fossil fuel-based electricity generation [25], only existing gas-fueled electricity generation capacity (including dual fuel generators) are considered, equal to the nameplate capacity operational at the end of 2019 per NYISO [26]. The assumed cost of existing electricity generation capacity – all existing generation modeled, including natural gas, hydropower, biofuel and nuclear – at each node is derived from capacity market costs used in a recent New York State Energy Research and Development Authority (NYSERDA) study[viii]. Generator start-up costs are assumed to be $79/MW-h, the value for combined cycle gas turbines (CCGT) in a recent NREL study [3]. An electricity generation efficiency of 42.8% is assumed for existing natural gas generation based on NYS electric power sector total natural gas consumption [27] and natural gas-based electricity

---

[viii] The reference study [30] contains capacity market costs for New York City (NYC), Long Island (LI), Lower Hudson Valley (LHV) and Rest of State (ROS). Here, Node 1 is assumed to be 100% ROS; Node 2 to be 50% LHV and 50% ROS per the approximate actual capacity distribution [26]; Node 3 to be 87% NYC and 13% LHV per the reference study; and Node 4 to be 100% LI.

generation [28] for 2019. Modeled natural gas prices for electricity generation at each node are derived from regional natural gas avoided costs in a recent NYSERDA study[ix].

New gas-fueled generation costs are adopted based on industrial frame gas turbines (GTs) per EIA's 2020 Annual Energy Outlook [2]. These GTs have node-specific capital costs, statewide fixed and variable operations and maintenance costs, and constant 34.4% efficiency. New generator start-up costs are assumed to be $69/MW-h, the value for GTs in a recent NREL study [3]. Natural gas prices for new generators are assumed to be the same as those for existing generators at each node. Existing and new natural gas-based electricity generation capacity are constrained to be a minimum 1.189 times larger than peak generation, based on NYISO's 18.9% statewide capacity reserve margin for the 2020-2021 capability year [29].

*Supplementary Table S3: Nodal gas-fueled electricity generation assumptions.*

| Node | Wholesale Nat. Gas Prices [$/MMBTU][a] | Existing Gas-Fueled Generation | | | New Gas-Fueled Generation | | | |
|---|---|---|---|---|---|---|---|---|
| | | Capacity [MW][b] | Capital Cost [$/kW-yr][a] | Start-up Costs [$/MW-h][c] | Capital Cost [$/kW][d] | Fixed O&M Cost [$/kW-yr][d] | Variable O&M Cost [$/MWh][d] | Start-up Costs [$/MW-h][c] |
| 1 | 2.89 | 3934.2 | 27.640 | 79 | 772 | 6.97 | 4.48 | 69 |
| 2 | 4.04 | 8622.5 | 53.440 | 79 | 772 | 6.97 | 4.48 | 69 |
| 3 | 3.67 | 10,249.9 | 101.303 | 79 | 1034 | 6.97 | 4.48 | 69 |
| 4 | 3.62 | 4192.7 | 104.600 | 79 | 1034 | 6.97 | 4.48 | 69 |

[a] NYSERDA [30].
[b] NYISO [26].
[c] Bloom et al. [3].
[d] EIA [2].

*Wind power capacity and generation*

Existing onshore wind capacities at each node are those active by the end of 2019 [26] as shown in Supplementary Table S4.

Wind power potential capacity and power output are based on model data developed by NREL for 126,000 potential wind sites [31,32]. First, onshore wind power potential time series data were adjusted to account for consistent over-predictions based on historical output of existing sites in NYS [33]. After this adjustment, a single wind potential timeseries was produced for each of the two upstate nodes[x] by computing the capacity-weighted potential timeseries of all NREL-modeled sites in each node.

---

[ix] The reference study [30] contains natural gas avoided costs for Upstate/Western NY (UWNY), Hudson Valley (HV), and New York City and Long Island (NYC-LI). Node 1 is 100% UWNY, Node 2 is 100% HV and Node 4 is 100% NYC-LI. Node 3 is assumed to be 87% NYC-LI and 13% HV per the reference study.
[x] Onshore wind capacity is ignored for downstate nodes 3 and 4 due to space constraints and the likelihood of a large buildout of offshore wind capacity connected to these nodes.

To determine the offshore wind potential timeseries, potential timeseries for all NREL modeled wind sites within NYS maritime boundaries are collected; these timeseries are then weighted by modeled site capacity to return a single potential timeseries. This single timeseries is adjusted based on a previously published logit transform method [33] so that the new capacity factor equals the estimate from a more recent NREL wind energy resource assessment [34], after subtracting electrical and wake losses[xi]. This adjusted timeseries is applied to both downstate nodes.

High and low costs are computed for onshore (only available in upstate Nodes 1 and 2) and offshore (only available in downstate Nodes 3 and 4) wind capacity. Based on the average of costs from three recent NREL wind technology reports [35–37] and predicted cost reductions [38], a high cost of $1992/kW and a low cost of $1698/kW are assumed for onshore wind capacity. For onshore wind, fixed O&M costs of $18.10/kW-yr are applied per the 2018 Bloomberg New Energy Finance Wind Operations and Maintenance Pricing Index [39]; installations are limited to the maximum capacities given in the NREL data set [31]. Based on a review of the costs of wind energy [40], along with cost reduction estimates [38], the high cost of offshore wind capacity is set to $3583/kW; a cost curve fit to a NREL estimates of offshore wind LCOE in 2030 [41] (5% interest, 20 year lifetime) yields a low cost estimate of $2256/kW. A fixed operations and management cost of $38/kW-yr is applied for offshore wind [42], and total offshore wind installations are capped to 57.9 GW based on potential capacity in water depths less than 60m as identified by NREL [34] (See Supplementary Table S4).

*Utility-scale solar capacity and generation*

Existing utility-scale solar capacities at each node are those active by the end of 2019 [26] as shown in Supplementary Table S4.

The utility-scale solar potential generation time series for each node is determined by (1) identifying the capacity and location of all NYS potential grid-scale solar PV sites in a NREL model solar data set [43]; (2) computing hourly solar PV potential output using NREL's System Advisory Model [44], assuming single-axis tracking, tilted at latitude; (3) adjusting the system efficiency according to protocols specified by the California Energy Commission [45]; and (4) aggregating the individual site time series at each node, weighted by each site's capacity per the NREL data set.

High costs of new utility-scale solar PV capacity of $1341/kW at Nodes 1 and 2, and $1593/MW at Nodes 3 and 4 are adopted based on location-specific capital cost inputs to EIA's Annual Energy Outlook [2]. Low cost estimates are computed by applying a 25% cost reduction to high cost estimates, which is approximately the average of the cost reductions seen for onshore (15%) and offshore (37%) wind capacity, described above: $1006/kW in Nodes 1 and 2 and $1195/kW in

---

[xi] From the offshore wind resource assessment [34], the potential capacity (Appendix B) and resource energy with losses (Appendix D) in water depth less than 60m areas are collected, keeping electrical losses and wake losses but removing 6% fixed losses (Appendix J). This results in a NYS offshore wind average capacity factor of 45.9%.

Nodes 3 and 4. A statewide $10.4/kW-yr fixed O&M cost is set for new solar capacity based on a recent NREL benchmark for utility-scale tracking PV [46]. To account for space limitations, the maximum potential utility-scale solar PV capacity is determined by county and then aggregated to the nodal level, per Supplementary Table S4. For each county, the maximum capacity is based on 1) the smaller quantity of (a) existing cropland, per the 2017 USDA Census of Agriculture [47], or (b) 10% of the county's total land area; and 2) an assumed 8.5 MW/acre [48].

*Supplementary Table S4: Nodal existing and maximum wind power and utility-scale solar capacities.*

| | Existing Capacity [MW] | | Maximum Potential Capacity [MW] | | |
|---|---|---|---|---|---|
| Node | Onshore Wind[a] | Utility-scale Solar[a] | Onshore Wind[b] | Offshore Wind[c] | Utility-scale Solar[d] |
| 1 | 1985.25 | 0 | 32,402 | 0 | 212,710 |
| 2 | 0 | 0 | 4376 | 0 | 44,899 |
| 3 | 0 | 0 | 0 | 57,938 | 481 |
| 4 | 0 | 56.5 | 0 | | 2743 |

[a] NYISO [26].
[b] Draxl et al. [31].
[c] Musial et al. [34].
[d] See the text of this section.

*Behind-the-meter solar capacity and generation*

Nodal BTM solar capacity is imposed exogenously on the optimization based on a user-provided year and a nodal capacity distribution, itself determined by a NYISO-projected 9 GW solar capacity scenario [49]. Statewide BTM solar capacity is based on a logistic growth function of the general form shown in Supplementary Eq. (S40) fit to historical capacity data for the years 2000-2019 [50]:

$$\sum_{i \in I} X_{btm-solar,i} = \frac{K}{1 + Qe^{-B(year-M)*1/v}}$$

(S40)

where K = 10,982.023; Q = 1.680925e-4; B = 0.1202713, M = 1995.067; $v$ = 4.955324e-6.

Existing nodal capacity as of the end of 2019 [50] and projected distribution computed per Supplementary Eq. S40 for example years are shown in Supplementary Table S5.

*Supplementary Table S5: Nodal behind-the-meter (BTM) solar capacity*

| Node | BTM Solar capacity (MW) for given year | | | |
|---|---|---|---|---|
| | Current[a] | 2030 | 2040 | 2050 |
| 1 | 562 | 2109 | 3009 | 3348 |
| 2 | 523 | 2364 | 3372 | 3752 |
| 3 | 293 | 1096 | 1564 | 1740 |
| 4 | 259 | 1039 | 1482 | 1649 |

[a] At the end of 2019 per NYSERDA [50].

The BTM PV generation time series for each node is determined by (1) selecting a representative city for each NYISO zone from those in the NREL National Solar Radiation Database [51]; (2) computing hourly solar PV potential output using NREL's System Advisory Model [44], assuming a fixed axis, tilted at latitude; (3) adjusting the system efficiency according to protocols specified by the California Energy Commission [45]; and (4) aggregating zonal time series at each node weighted by zonal capacities in the NYISO-projected 9 GW solar capacity scenario [49].

*Energy storage*

Existing battery storage power capacity was extracted from the EIA energy mapping system [52], and existing battery storage energy capacity was determined from news reports and websites corresponding to recently installed projects[xii]; these quantities are presented in Supplementary Table S6. Although the SECTR General Formulation allows per-unit power capacity and per-unit energy capacity cost components, for the present analyses only energy capacity costs are included. High and low costs are set based on the "Mid" and "Low" cost projections for 2030 from NREL [53]: $208/kWh and $144/kWh, respectively. A power-to-energy ratio of 0.25 kW/kWh is assumed based on common 4-hour battery systems, with 94.6% charge and discharge efficiencies based on the 89.5% roundtrip efficiency of a commercially available battery storage system [54]. A 10-year lifetime [55] is adopted for modeled batteries. Batteries are also assigned a self-discharge rate of 0.1%/hr.

Supplementary Table S6: Existing nodal battery energy and power capacity.

| | Existing Battery Capacity | |
|---|---|---|
| Node | Battery Energy [MWh] | Battery Power [MW] |
| 1 | 5.2[a] | 3[a] |
| 2 | 80 | 20[b] |
| 3 | 0 | 0 |
| 4 | 65[c] | 10[c] |

[a] Battery capacities taken from [56,57].
[b] Key Capture Energy [58]; the facility is assumed to be a 4 hour battery system.
[c] Battery capacities taken from [59,60].

---

[xii] Node 1: East Pulaski BESS [56] and Lockheed Martin RMS [57]. Node 2: KCE NY 1 assumed to be 4 hour battery system [58]. Node 4: East Hampton Energy Storage Center [59] and Montauk Energy Storage Center [60].

For long-term storage, the use of hydrogen electrolysis and combustion in a gas turbine is assumed, with model-selected deployment analogous to battery storage based on cost components for both power capacity and energy capacity. A power capacity cost of $3013/kW is adopted based on a recent study [61] for Nodes 1 and 2; the same capital cost adjustment for GTs is then applied for Nodes 3 and 4, resulting in $4036/kW. For hydrogen storage capital costs, a per-unit energy cost of $0.35/kWh is set for geologic storage in Node 1 based on an NREL study (and adjusting from 2008 dollars to 2020 dollars) [62]. For other nodes, hydrogen storage is assumed to occur in carbon fiber storage tanks given the lack of geologic formations for storage and higher population density; a storage cost of $8.29/kWh is applied based on annually updated Department of Energy hydrogen storage cost analysis [63]. A fixed operations cost of $48.87/kW-yr is assumed based on an earlier study [64]. Charge and discharge efficiencies of 59.2% are adopted based on 35% roundtrip efficiency in a recent NREL analysis [65] referencing an earlier study [66]. A self-discharge rate of 0.1%/hr is set.

Despite its existence in NYS, SECTR-NY does not separately model pumped hydropower storage capabilities. Pumped hydropower storage in NYS is primarily used to provide black-start capabilities, contributing less than 2% of the state's hydropower generation [6]. Battery and hydrogen storage are therefore the only two methods of energy storage implemented.

*Nuclear power*

The nuclear power landscape in NYS is evolving, as the last operational generator of the Indian Point two-generator plant in Node 3 shuttered on April 30, 2021 [67], and nuclear generators in Node 1 have been subsidized in recent years. To investigate the impact of capacity retirements, the SECTR-NY formulation can either include or ignore these nuclear generators. Nuclear capacity is distributed across all four model nodes per NYISO [26] as shown in Supplementary Table S7 (which for clarity shows no nuclear at Nodes 2 and 4). Electricity generation is assumed to be constant throughout the simulation period and equal to the average electricity production of those generators in 2019 according to NYISO [26]. The price of nuclear electricity at each node is computed from the average 2019 day-ahead locational based marginal pricing (LBMP) [6] of each nuclear generator at each node, weighted by the 2019 total net electricity generation [26] of each of those generators. The price at Node 1 is increased to account for subsidies of the nuclear generators at that node, funded by Zero Emission Credits (ZECs). Per Supplementary Eq. (S41), the per energy unit subsidy is computed from the 2020 compliance year ZEC rate [68], NYISO's 2020 baseline demand forecast [26], and the constant output of nuclear electricity at Node 1 from Supplementary Table S7.

$$\left\{\begin{matrix} Nuclear\ Price\ Subsidy \\ at\ Node\ 1 \end{matrix}\right\} = \frac{\left\{\begin{matrix} 2020\ Compliance \\ Year\ ZEC\ Rate \end{matrix}\right\} \times \left\{\begin{matrix} NYISO\ 2020\ Baseline \\ Demand\ Forecast \end{matrix}\right\}}{\{Constant\ Electricity\ Generation\ at\ Node\ 1\}}$$

(S41)

The assumed cost of existing nuclear electricity generation capacity at each node is the same as described above under "Characterization of fossil fuel-based electricity generation."

Supplementary Table S7: Nodal existing nuclear power characteristics

| Node | Generation Capacity [MW][a] | Constant Electricity Generation [MWh/h][a] | Capacity Cost [$/kW-yr][b] | Electricity Price [$/MWh][c] |
|---|---|---|---|---|
| 1 | 3536.8 | 3207 | 27.640 | 37.94 |
| 2 | 0 | 0 | N/A | N/A |
| 3 | 2311 | 1906 | 101.303 | 26.82 |
| 4 | 0 | 0 | N/A | N/A |

[a] NYISO [26].
[b] NYSERDA [30].
[c] See the text of this section.

*Hydropower*

The methodology for creating hydropower fixed and flexible generation time series is described in a recent paper [4]. Actual monthly hydropower output by facility is collected for 2007-2012[xiii] from EIA Form 923 [69], and then is aggregated at each node. The two largest NYS hydropower facilities (both located at Node 1) operate near their maximum capacity given available stream flows; accordingly, fixed hourly time series are provided for these generators. The remaining hydropower generation and capacity in Nodes 1 and 2 are considered to be flexible with provided daily total electric energy generation requirements. Total fixed and flexible hydropower capacities are computed from the nameplate capacities operational at the end of 2019 [26]. Hourly generation is determined endogenously by the hydropower methodology detailed in the General Formulation. Hydropower-generated electricity prices are based on recent prices for such electricity in NYISO's day-ahead market[xiv]. The assumed cost of existing hydropower electricity generation capacity at each node is the same as described above under "Characterization of fossil fuel-based electricity generation." The values described here are summarized in Supplementary Table S8.

---

[xiii] Monthly generation quantities for 2007-2012 are used to align with the wind, solar, and demand time series.
[xiv] All based on 2019 hourly day-ahead LBMP [6] and weightings by total 2019 electricity production [26]: Node 1 cost is the weighted average LBMP for Moses Niagara and St. Lawrence hydropower facilities; Node 2 is the weighted average LBMP of the four highest producing hydropower facilities at that node (62% of total hydroelectricity production at that node).

Supplementary Table S8: Existing hydropower characteristics

| Node | Average Generation [MWh/h][a] | | Capacity [MW][b] | | Capacity Cost [$/kW-yr][c] | Electricity Price [$/MWh][d] |
|---|---|---|---|---|---|---|
| | Fixed | Flexible | Fixed | Flexible | | |
| 1 | 2395 | 328 | 3948 | 769.4 | 27.640 | 18.47 |
| 2 | 0 | 270 | 0 | 608.7 | 53.440 | 28.02 |
| 3 | 0 | 0 | 0 | 0 | N/A | N/A |
| 4 | 0 | 0 | 0 | 0 | N/A | N/A |

[a] EIA [69].
[b] NYISO [26].
[c] NYSERDA [30].
[d] See footnote *vii*.

*Biofuel-based electricity generation*

SECTR-NY classifies various electricity generation feedstocks as "biofuels": wood and wood waste, biogas, and solid waste. In NYS, biofuel capacity is distributed across all four model nodes per NYISO [26] as shown in Supplementary Table S9. Intraday biofuel electricity generation is flexible as described in the Main Text; maximum daily electricity generation is assumed to be constant throughout the simulation period and equal to the average daily electricity production of those generators in 2019 according to NYISO [26]. Biofuel-generated electricity prices are based on recent prices for such electricity in NYISO's day-ahead market[xv]. The assumed cost of existing biofuel-based electricity generation capacity at each node is the same as described above under "Characterization of fossil fuel-based electricity generation."

Supplementary Table S9: Nodal existing biofuel characteristics

| Node | Generation Capacity [MW][a] | Daily Electricity Generation [MWh][a] | Capacity Cost [$/kW-yr][b] | Electricity Price [$/MWh][c] |
|---|---|---|---|---|
| 1 | 258.0 | 3289.041 | 27.640 | 20.66 |
| 2 | 45.0 | 473.425 | 53.440 | 27.41 |
| 3 | 59.7 | 1046.575 | 101.303 | 27.05 |
| 4 | 142.2 | 2445.479 | 104.600 | 32.29 |

[a] NYISO [26].
[b] NYSERDA [30].
[c] See footnote *ix*.

---

[xv] All based on 2019 hourly day-ahead LBMP [6] and weightings by total 2019 electricity production [26]: Node 1 cost is the average LBMP for the four highest producing biofuel facilities at that node (58% of total biofuel electricity production at that node); Node 2 is the average of Zone F and G LBMP; Node 3 is the average LBMP for the 1 biofuel facility at that node; Node 4 is the weighted average LBMP of all four biofuel facilities at that node.

*External imports*

NYISO currently imports significant quantities of low-carbon electricity from Hydro-Quebec (HQ), a net average of 1247 MWh/h in 2019 [26]; as such, electricity imported at this interface with Node 1 is included as a decision variable constrained to the maximum interface limit specified by NYISO (1.5 GW) [6]. A cost of $22.13/MWh is attributed to this imported electricity based on average 2019 day-ahead LBMP [70] and including capacity market payments for 1114 MW capacity per NYISO [71].

NYS regulators are nearing approval for plans for the Champlain Hudson Power Express, a 1250 MW HVDC transmission line that would bring hydropower-produced electricity from Quebec to New York City [21], which is also included in recent NYC local legislation [72]. As such, additional electricity import into Node 3 is included in future energy system scenarios. The precise cost of this electricity supply is unknown; however, a price of $70/MWh is adopted based on publicly available information, personal and public conversations about the project, and various possible financing parameters[xvi]. The line is assumed to provide 1125 MWh/h continuous based on the approximate 90% capacity factor of existing upstate Hydro-Quebec import lines [6] and an understanding of the project from public and personal conversations.

Imports from other external control areas are ignored to avoid characterizing or modeling future developments in regions that currently rely largely on fossil fuel-based electricity generation.

*Greenhouse gas emissions*

As accounted by NYSERDA, NYS energy sector emissions constitute 84% of total statewide GHG emissions (measured in equivalent global warming potential of carbon dioxide, $CO_2e$) as of 2016 [73]. The remaining 16% of GHG emissions comes from industrial processes, agriculture, and waste.

In New York's Climate Leadership and Community Protection Act (CLCPA), statewide GHG emissions accounting includes GHGs produced in NYS and GHGs produced outside NYS that are associated with imported electricity and fossil fuels [74]. Supplementary Table S10 shows emissions factors for carbon dioxide ($CO_2$), methane ($CH_4$), and nitrous oxide ($N_2O$) compiled from a variety of sources; the table also includes values for carbon dioxide equivalent ($CO_2e$). $CO_2e$ is a single metric that combines the effect of multiple GHGs based on their global warming potential (GWP). CLCPA requires GWP values based on the amount of warming impact relative to $CO_2$ when integrated over a 20-year time frame. Here, respective GWPs of 86 for $CH_4$ and 264 for $N_2O$ are used, in accordance with the Intergovernmental Panel on Climate Change (IPCC) Fifth Assessment Report (AR5) [75]. $CH_4$ emissions, particularly for natural gas, are largely dependent on venting

---

[xvi] Our calculations are generally in the $65-70/MWh range based on the project website's lower bound capital cost [21], higher potential upfront costs that have been discussed publicly, various annualization periods, average HQ export revenues ($1441M on 33.7 TWh in 2019 [85]), and the approximate 90% capacity factor of existing upstate Hydro-Quebec import lines [6].

at wellheads and leakage in transmission and distribution infrastructure; understanding these effects is the subject of ongoing research, but recent efforts focused on New York State provide a reliable reference point [76].

*Supplementary Table S10: Emissions factors [g/MJ] for GHG contributors*

| Energy source | $CO_2$ | $CH_4$ | $N_2O^f$ | $CO_2e$ |
|---|---|---|---|---|
| Coal | $92^a$ | $0.185^c$ | $1.52 \cdot 10^{-3}$ | 108.31 |
| Petroleum | $73^b$ | $0.093^d$ | $5.69 \cdot 10^{-4}$ | 81.15 |
| Natural Gas | $55^a$ | $0.641^e$ | $9.48 \cdot 10^{-5}$ | 110.18 |

[a] Based on high-heating values per Hayhoe et al. [77] as documented by Howarth et al. [76].
[b] $CO_2$ emission factor for petroleum is the high-heating value from Howarth et al. [78] as reported by Howarth et al. [76].
[c] As computed by Howarth et al. [76] based on the ratio of total methane emissions during coal mining and total coal production in the U.S. in 1990 from IPCC reporting [79], with a coal heating value of 27 MJ/kg [78].
[d] Based on $CH_4$ emissions from petroleum production from the National Energy Technology Laboratory (NETL) [80] as documented by Howarth et al. [76].
[e] Computed from assumptions of: $CH_4$ emission rate of 3.6% as used for NYS in Howarth et al. [76] based on a range computed by Alvarez et al. [81] and Howarth et al. [78]; natural gas to be 93% $CH_4$ [82]; and a high-heating value of 52.2 MJ/kg for natural gas in the U.S. market [83].
[f] EPA [84].

Per the targets set in the CLCPA [74], emissions reductions relative to a 1990 reference value are computed. Reference $CO_2$ and $CH_4$ emissions for electricity, buildings, industrial, transportation are calculated by using the 1990 EIA fuel consumption estimates [12] and emission factors in Supplementary Table S10; $CO_2$ and $CH_4$ emissions for electricity imports in 1990 are taken directly from Howarth et al. [76]; $CO_2$ and $CH_4$ emissions for waste incineration and all $N_2O$ emissions in 1990 are from the NYSERDA inventory of GHG in NYS [73]. Thus an $\varepsilon_{baseline}$ of 302.770 MMtCO$_2$e/year is computed, per Supplementary Table S11. Supplementary Table S11 further delineates emissions that are fixed in the model and those that are variable: variable emissions can change as computed by the model for a given user-defined scenario and as described by Eqs. (11-15) in the Main Text.

Current NYS electricity emissions are calculated by using SECTR to model a "current scenario". The current scenario includes all existing NYS energy infrastructure parameterized and discussed above, and assumes current capacities of wind and solar power, no additional electrification of vehicle or heating demand, and no generation from the Indian Point nuclear facility. Using the natural gas emissions factors in Supplementary Table S10 and the model-returned amount of natural gas generation needed to meet the existing electricity demand, current electricity emissions of 84.889 MMtCO$_2$e/year are computed, per Supplementary Table S11. Since SECTR-NY assumes the modeled electricity sector can be fully decarbonized, these emissions are considered variable.

Total fossil fuel usage for heating, $F_{heat,tot,i}$, is computed from the heating model [7] described above; portions of this fossil fuel usage are attributed to natural gas, fuel oil, and propane based on the 2018 residential and commercial usage of these fuels [12]. Annual heating GHG emissions, $\varepsilon_{heat}$, are calculated as 110.853 MMtCO$_2$e/year averaged over the six-year computation period based on the emissions factors for natural gas and petroleum (for fuel oil and propane) in Supplementary Table S10. As SECTR-NY assumes that NYS heating demand can be fully electrified, these emissions are considered variable in Supplementary Table S11.

Transportation sector emissions are determined from the 2018 EIA fuel consumption estimates [12] that are used to calculate statewide vehicle energy demand as described above. Gasoline and diesel consumption for transportation that can be electrified, $F_{veh,tot}$, and the petroleum emissions factor in Supplementary Table S10 are used to compute 73.703 MMtCO$_2$e/year variable emissions for vehicles included in the model electrification scope, $\varepsilon_{veh}$. Aviation fuel, hydrocarbon gas liquids, jet fuel, lubricants, residual oil, and natural gas consumption for transportation listed in the same EIA dataset [12] are considered fixed and constitute transportation emissions outside the scope of the model, $\varepsilon_{transp,other}$. From this usage data and the appropriate emissions factors from Supplementary Table S10, $\varepsilon_{transp,other}$ is computed to be 21.956 MMtCO$_2$e/year, a quantity fixed in every model run.

Industrial emissions, $\varepsilon_{ind}$, are calculated from the 2018 EIA fuel consumption estimates [12] and the appropriate emissions factors from Supplementary Table S10. Here, coal, natural gas, and petroleum products result in computed total emissions of 19.365 MMtCO$_2$e/year; this quantity is fixed in every model run.

Supplementary Table S11 also displays emissions from waste incineration in New York for the current system. In SECTR-NY model scenarios, waste incineration is excluded per the CLCPA [74]; as this emissions quantity is set to 0 in all model runs, it is presented as variable.

Supplementary Table S11: Relevant aggregate greenhouse gas emissions (MMtCO$_2$e/year)

| Emissions Source | 1990 Reference Emissions | Current Emissions as Modeled | |
|---|---|---|---|
| | | Variable | Fixed |
| Electricity | 86.772 | 84.889[a] | 0 |
| Electricity Imports | 1.909 | 0 | 0[b] |
| Heating (Buildings) | 100.468 | 110.853 | 0 |
| Industrial | 32.824 | 0 | 19.365 |
| Transportation | 79.532 | 73.703 | 21.956 |
| Waste Incineration | 1.265 | 2.784[c] | 0 |
| Total | 302.770 | 272.229 | 41.321 |
| | | 313.550 | |

[a] Based on SECTR-NY model of current system as described in this section.
[b] Electricity imports are only considered from hydropower generation.
[c] This quantity represents the 2016 value from NYSERDA inventory for waste incineration [73]. In SECTR model runs, it is set to zero.

*Seasonal distribution of demand and renewable generation potentials*

Existing electricity demand, electrified demand, and the renewable generation potentials of wind and solar resources used in SECTR-NY simulations all demonstrate substantial seasonal variability. Fig. 2 in the main text contains monthly values for the mean and maximum existing electricity demand; the means and maximums of existing electricity demand combined with either electrified heating or transport; and the means of onshore wind, offshore wind, and utility-scale solar generation potentials. From the top and middle panels in Fig. 2, one observes that electrified heating increases average and peak electricity demands in the winter months: Full electrification of heating increases average load by up to 15 GWh/h and peak load by up to 52 GWh/h. In contrast, electrification of transport has smoother effect. With 100% transport electrification, average load rises by 6 to 8 GWh/h in all months of the year, with the larger increases coming during the winter due to the inverse relationship between temperature and EV charging demand. The effect on peak load is similarly consistent: Peak electricity demand increases by 12 to 15 GWh/h in all months.

Wind and solar generation potentials in NYS also display a strong seasonal dependence. Offshore and onshore wind potentials both peak in the winter months, reaching an average of 0.57 $MWh_{generation}/MW_{installed}$ and 0.40 $MWh_{generation}/MW_{installed}$, respectively; in the summer months, average generation for each decreases by approximately 50%. In contrast, utility-scale solar capacity offers peak generation potentials during the summer months, up to an average 0.26 $MWh_{generation}/MW_{installed}$, while winter months see this quantity drop to 0.10 $MWh_{generation}/MW_{installed}$.

Taking all three panels of Fig. 2 together, a clear seasonal alignment is identified between electrified heating demand and wind generation potential, indicating that electrified heating may prove effective in integrating large amounts of installed wind capacity. Moreover, summer-peaking solar generation is well-suited to meet summer loads in NYS, both those that currently exist and those that are increased by transport electrification.

# S3 Supplementary Results

Supplementary Section S3 contains additional results for the SECTR Baseline configuration; results that investigate the impact of SECTR system parameterization assumptions; and main text results for different heating and vehicle electrification rates (HVEs) and low carbon electricity percents (LCPs).

## S3.1 Additional Baseline configuration results

Supplementary Figure S2 presents an analogous plot to Fig. 3(a), but with a continuous 3.2 GWh/h of upstate nuclear generation present. Here, nuclear generation allows for approximately 10% lower LCOEs on average at the simulated scenarios, cost savings that grow larger at higher LCPs. However, the addition of nuclear generation does not change the overall shape of Fig. 3(a), and accordingly the same conclusions are reached: 1) Emissions reductions can be achieved at lower LCOEs by prioritizing electrification of heating and vehicles in conjunction with deployment of solar and wind, as opposed to the latter by itself, and 2) system costs increase substantially above 70-80% LCPs.

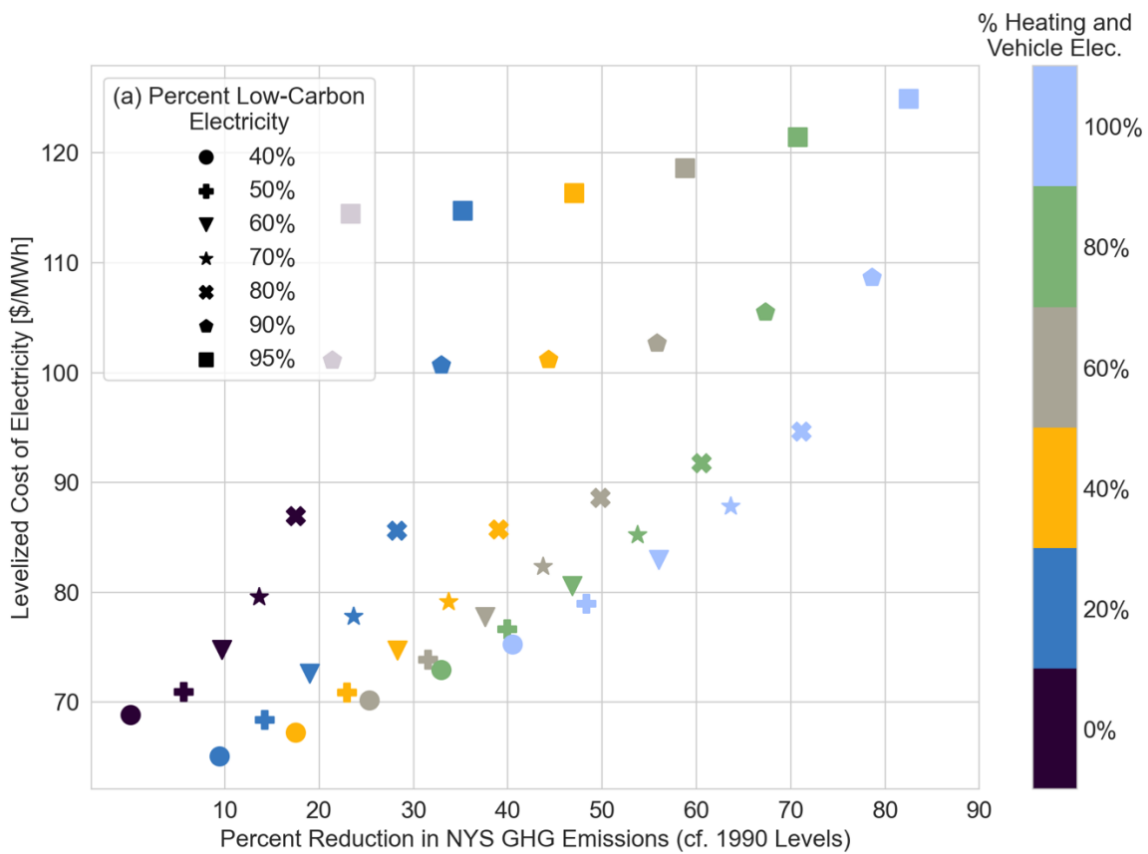

*Supplementary Figure S2: LCOE vs. percent reduction in NYS GHG emissions (compared to 1990 levels). Marker shape indicates percent low-carbon electricity (LCP), and marker color indicates heating and vehicle electrification (HVE). For scenarios shown, all low-carbon electricity generation is from wind, solar, nuclear, and hydropower.*

Next explored are the effects of either increased HVE or LCP on peak gas generation, average gas generation, low-carbon electricity generation, and battery storage throughput. In evaluating the peak gas generation characteristics, increasing electrification at a set LCP results in substantial winter peaks: Supplementary Figure S3(a) presents the monthly peak to annual average gas generation ratio at 60% LCP for 0%, 40% and 80% HVE. At 80% HVE, additional, peaky heating demand causes January gas generation peaks of 46.9 GWh/h, equal to 4.6 times the annual average, compared 15.9 GWh/h at 0% HVE with a peak-to-average ratio of 2.7. In contrast, the July peak only increases from 22.4 GWh/h at 0% HVE to 25.5 GW at 100% HVE. Supplementary Figure S3(b) shows that there are no equivalent seasonal effects to increasing the LCP at 40% electrification. However, increasing the LCP to 80% and 95% results in lower average gas generation (4.0 GWh/h and 1.0 GWh/h, respectively, compared to 8.0 GWh/h), quantities which result in substantial peak-to-average ratios (above 20 in December and January for the 95% LCP).

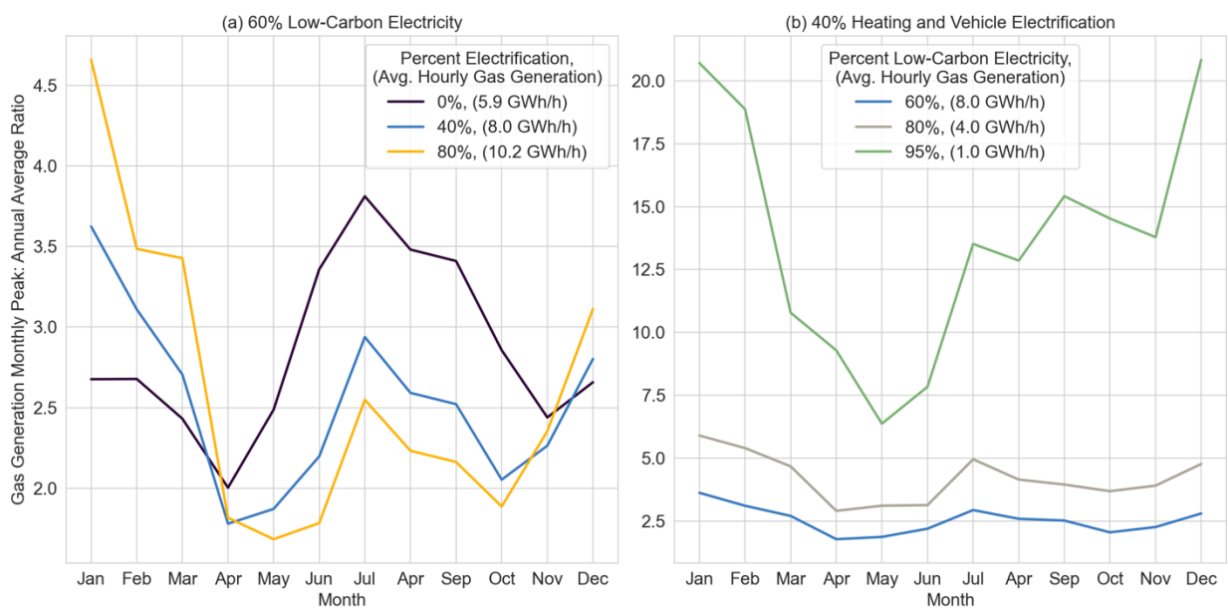

*Supplementary Figure S3: Monthly peak to annual average gas generation ratios for (a) scenarios containing 60% low carbon electricity with increasing amounts of electrification; and (b) scenarios containing 40% electrification with increasing percents low-carbon electricity.*

Increasing electrification at a set LCP has a similar seasonal shift on average gas generation, shown in Supplementary Figure S4(a). For the same 60% LCP and 0%, 40%, and 80% HVEs, increased electrification results in higher average winter gas generation – in absolute terms and relative to the annual average – and lower relative generation during the summer. In January, 0% HVE corresponds to an average 5.3 GWh/h of gas generation, or 0.9 times the annual average; 100% HVE increases this to 16.3 GWh/h, or 1.6 times the annual average. Again, this increase in average generation is attributable to the higher amounts of peaky heating demand on the system: Heating demand proves difficult to meet with low-carbon electricity and is accordingly satisfied by dispatchable gas generation. The suitability of gas generation in meeting electrified heating demand also explains the relative decreases in gas generation during summer months.

As the same LCP needs to be achieved despite increased winter gas generation, gas generation during the summer is reduced (1.2 times the annual average with 80% HVE compared to 1.8 times at 0% HVE in the month of July), as this less-peaky demand can more easily be met by a combination of solar generation and battery storage.

Supplementary Figure S4(b) demonstrates that raising the LCP from 60% to 95% increases the January gas generation from 1.3 to 2.5 times the annual average, a shift that indicates the costliness of meeting electrified heating demand with only low-carbon generation and battery storage. In contrast, gas generation in the shoulder seasons is the first to be displaced by low-carbon generation, due to 1) the high productivity of onshore wind, offshore wind, and solar resources, and 2) the lack of peaky heating demand during these months.

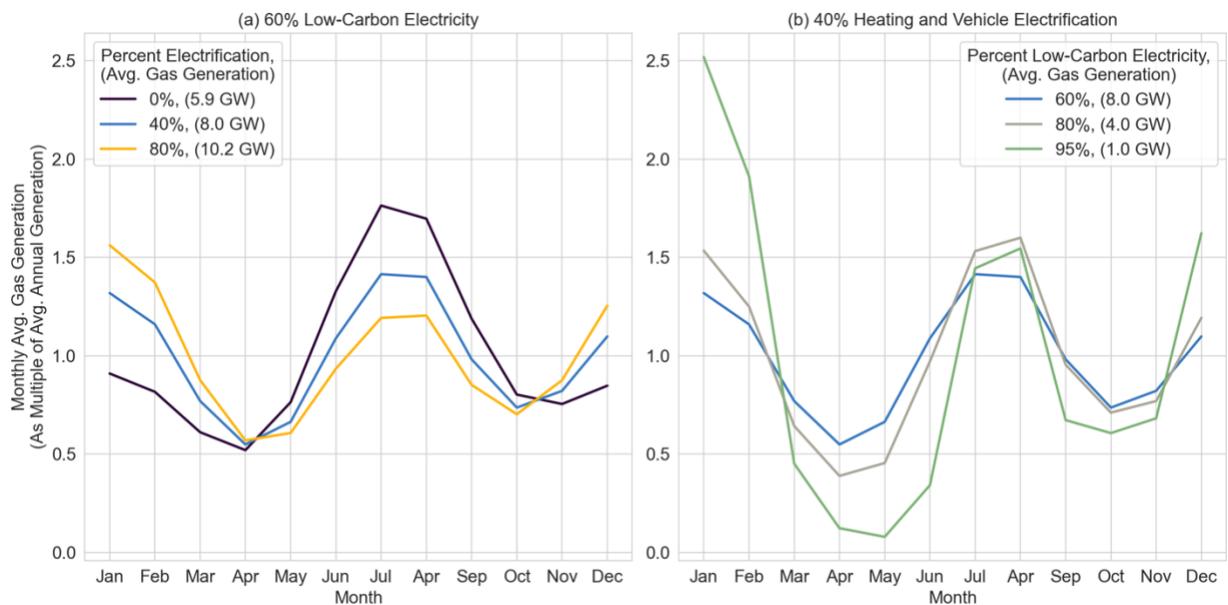

*Supplementary Figure S4: Monthly gas generation as a multiple of the annual average for scenarios containing (a) 60% low-carbon electricity with increasing amounts of electrification; and (b) 40% electrification with increasing percents low-carbon electricity.*

Evaluation of monthly battery storage behavior reinforces the findings presented in main text. Increasing electrification at 60% low-carbon electricity shifts battery throughput towards summer months when battery storage is well-paired with the daily cycles of productive solar generation, per Supplementary Figure S5(a). While this relative seasonal shift is apparent in the changing shapes of the normalized throughput curves, the absolute seasonal difference in battery throughput is not as stark: Increasing HVE from 0% to 80% only raises battery throughput by an average 1.0 GWh/h, indicating that battery output is not utilized to meet a significant portion of demand at 60% LCP. In contrast, battery throughput increases substantially in the summer months – in both absolute and relative terms – and experiences a relative drop during the shoulder seasons as LCP increases from 60% to 95% at 40% HVE (Supplementary Figure S5(b)). At 95% LCP, battery throughput reaches an average of 3.1 GWh/h in August (1.4 times the annual average), a quantity that is double the average throughput in April (1.6 GWh/h); to compare, the 60% LCP scenario contains average throughputs in August and April both roughly equal to the annual average of 0.6 GWh/h. From this figure, one concludes that pairing batteries with productive solar generation during summer months provides a cost-effect method of meeting additional load with low-carbon electricity. It is also notable that this effect is substantially greater when increasing the LCP at a given HVE, due to the greater amounts of excess low-carbon generation present in these scenarios.

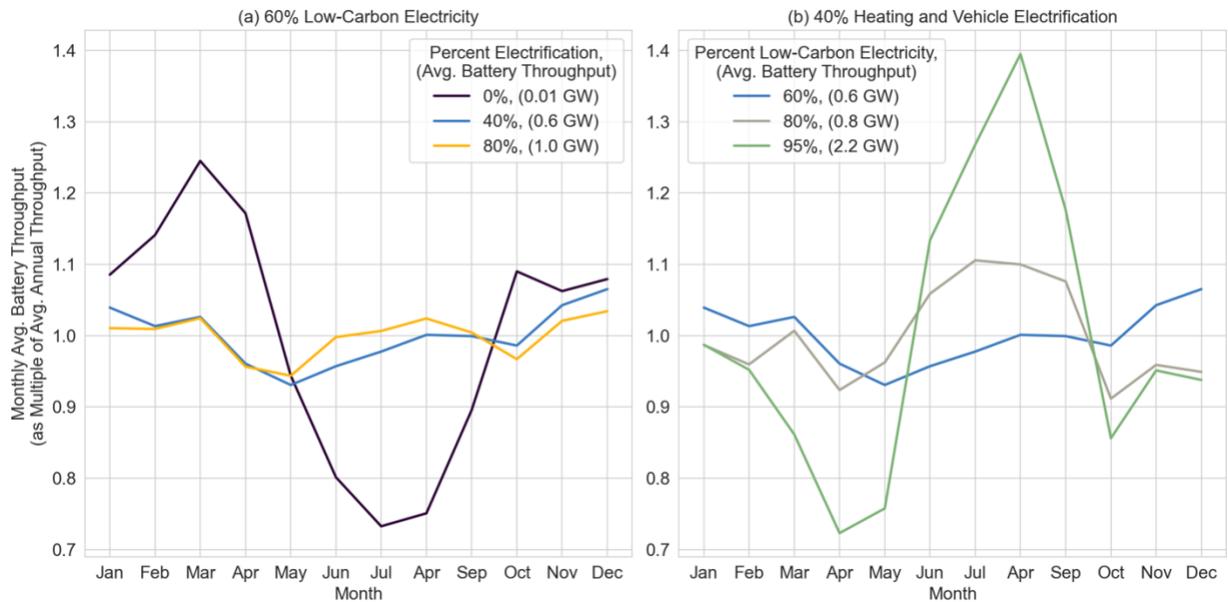

*Supplementary Figure S5: Monthly battery throughput as a multiple of average annual throughput. Results are presented for scenarios containing (a) 60% low carbon electricity with increasing amounts of electrification; and (b) 40% electrification with increasing percents low-carbon electricity.*

## S2.2 Impact of existing system parameterization

To understand the impact of SECTR baseline parameters and how different parameterizations affect model results, two additional configurations are evaluated: A 'Greenfield' configuration and a 'Greenfield with Constant Costs' configuration. The Greenfield configuration represents a type of parameterization often seen in the capacity expansion modeling literature: This configuration includes no existing solar, wind, gas, biofuel, or transmission capacity; and no existing biofuel generation. The Greenfield with Constant Costs configuration combines the greenfield parameterization with homogenous nodal costs, which are calculated via a weighted average of the costs associated with the returned capacity and generation quantities from the Greenfield configuration model solution.

All configurations are evaluated at two scenarios, one representing a combination of a high LCP and a low HVE (referred to as the high LCP scenario), and the other representing a combination of a lower LCP and a higher HVE (referred to as the low LCP scenario). In both scenarios, the GHG reduction is set to 40%. For the high LCP scenario, electrification of heating and transport is set to 40% and the LCP is determined by the model; for the low LCP scenario, LCP is set to 60% and the HVE is determined by the model. To ensure equivalent LCPs across configurations, the efficiency of new gas turbines in a Greenfield-based configuration is set to the weighted average efficiency of existing and new generation in the corresponding Baseline scenario.

*Supplementary Table S12: Select computed characteristics of Baseline and Greenfield configurations.*

| Configuration | Configuration Parameters | | | Model-returned Generation and Storage Capacities (Cap.) and Transmission (Tx.) Characteristics | | | | | | | |
|---|---|---|---|---|---|---|---|---|---|---|---|
| | % GHG[a] | % HVE[b] | % LCP[b] | Total Gas Cap. [GW] | Total Upstate Battery Cap. [GW] | Total Downstate Battery Cap. [GW] | Total Pos. Tx. Cap. [GW-mi][c] | Total Rev. Tx. Cap. [GW-mi][d] | Avg. Pos. Tx. Util. %[c] | Existing Cap. LCOE [$/MWh][e] | Total LCOE [$/MWh] |
| Baseline | -40 | 40 | 81.7 | 27.2 | 7.3 | 1.2 | 2646.8 | 2287.4 | 24.3 | 27.0 | 96.4 |
| Baseline | -40 | 64.8 | 60 | 47.2 | 2.8 | 3.9 | 2646.8 | 2083.2 | 28.1 | 23.7 | 83.5 |
| Greenfield | -40 | 40 | 81.3 | 26.1 | 8.8 | 3.5 | 2473.3 | 0.0 | 31.7 | 0.8 | 86.5 |
| Greenfield | -40 | 64.3 | 60 | 48.4 | 2.6 | 3.9 | 1371.0 | 0.0 | 41.7 | 0.7 | 75.8 |
| Greenfield w. Constant Costs | -40 | 40 | 81.3 | 25.7 | 10.9 | 2.2 | 2339.8 | 0.0 | 27.3 | 0.8 | 86.7 |
| Greenfield w. Constant Costs | -40 | 64.4 | 60 | 48.0 | 5.7 | 2.7 | 629.0 | 5.3 | 29.9 | 0.7 | 75.0 |

[a] '% GHG' refers to the percent change in greenhouse gas emissions compared to the 1990 reference values. Negative values indicate reductions.
[b] LCPs/HVEs are not identical across configurations due to slight differences in model-computed electricity imports given the specified GHG reduction and the HVE/LCP.
[c] 'Pos.' refers to "positive" upstate-to-downstate transmission directionality, i.e. from Node 1 to 2, Node 2 to 3, and Node 3 to 4.
[d] 'Rev.' refers to "reverse" downstate-to-upstate transmission directionality, i.e. from Node 4 to 3, Node 3 to 2, and Node 2 to 1.
[e] The costs of maintaining existing gas, hydropower, biofuel, and transmission capacity constitute the cost portion of 'Existing Capacity LCOE.'

Supplementary Table S12 presents a comparison of model-selected gas, battery, transmission, and LCOE characteristics. Here, both Greenfield configurations (with and without constant costs) contain LCOEs approximately 10% lower than in the fully-parameterized Baseline configuration, regardless of the combination of HVE/LCP. As the Greenfield configurations do not include any existing gas, biofuel, or transmission capacity, the fixed costs associated with maintaining this infrastructure (see 'Existing Cap. LCOE' column) drop to nearly $0/MWh, a reduction that causes the total LCOE decline. Moving from the Baseline to the Greenfield configuration, an average

60.2% decline in total installed transmission capacity is observed across both scenarios, with reverse transmission being completely eliminated; moving to the Greenfield with Constant Costs configuration causes an average 69.2% decline. Accordingly, the transmission capacity that is installed in the positive direction is utilized more frequently, a trend which is particularly pronounced in the Greenfield configuration results, due to lower amounts of installed downstate gas generation (see following Supplementary Figure S6).

For the high LCP scenarios in the two greenfield configurations, less transmission capacity and lower amounts of installed gas generation are compensated by increased battery capacity: The Greenfield configuration contains 2.8 GW additional storage capacity (a 29.4% increase), while the Greenfield with Constant Costs configuration contains 3.6 GW additional storage capacity (a 37.9% increase). This larger quantity of installed battery capacity is less prominent in the low LCP scenarios, due to their lower need for low-carbon electricity shifting; however, the low-LCP scenario in the Greenfield with Constant Cost configuration contains 1.7 GW more battery capacity than its Baseline analogue, an increase of 25.4%

Supplementary Figure S6 displays the change in gas capacity and generation characteristics across the three configurations. Here, the spatial heterogeneity of SECTR-NY results is investigated by splitting NYS into upstate and downstate regions[xvii]. Upstate NYS contains the state's onshore wind capacity, low-cost utility-scale solar, and existing low-carbon generation, while downstate NYS contains substantial electricity demand in and around New York City and offshore wind capacity. These differences in regional characteristics results in distinct system behavior on either side of the interface between Nodes 2 and 3.

Comparing results across configurations, the top row – representing the high LCP scenario – contains a 7.1 GW shift in gas capacity from downstate to upstate nodes when changing from the Baseline to the Greenfield configuration, due to the relatively higher cost of downstate gas capacity. Adopting constant nodal costs causes a smaller shift: When all new capacity has the same cost, the high LCP scenario shifts 1.8 GW gas capacity towards downstate regions compared to its equivalent Baseline configuration. In the low LCP scenarios (bottom row of Supplementary Figure S6), a consistent shift from upstate to downstate gas generation capacity is observed: The Greenfield configuration contains a shift of 5.4 GW, while the Greenfield with Constant Costs configuration contains a shift of 12.0 GW.

---

[xvii] 'Upstate' is defined as a region containing Nodes 1 and 2; 'downstate' refers to a combination of Nodes 3 and 4.

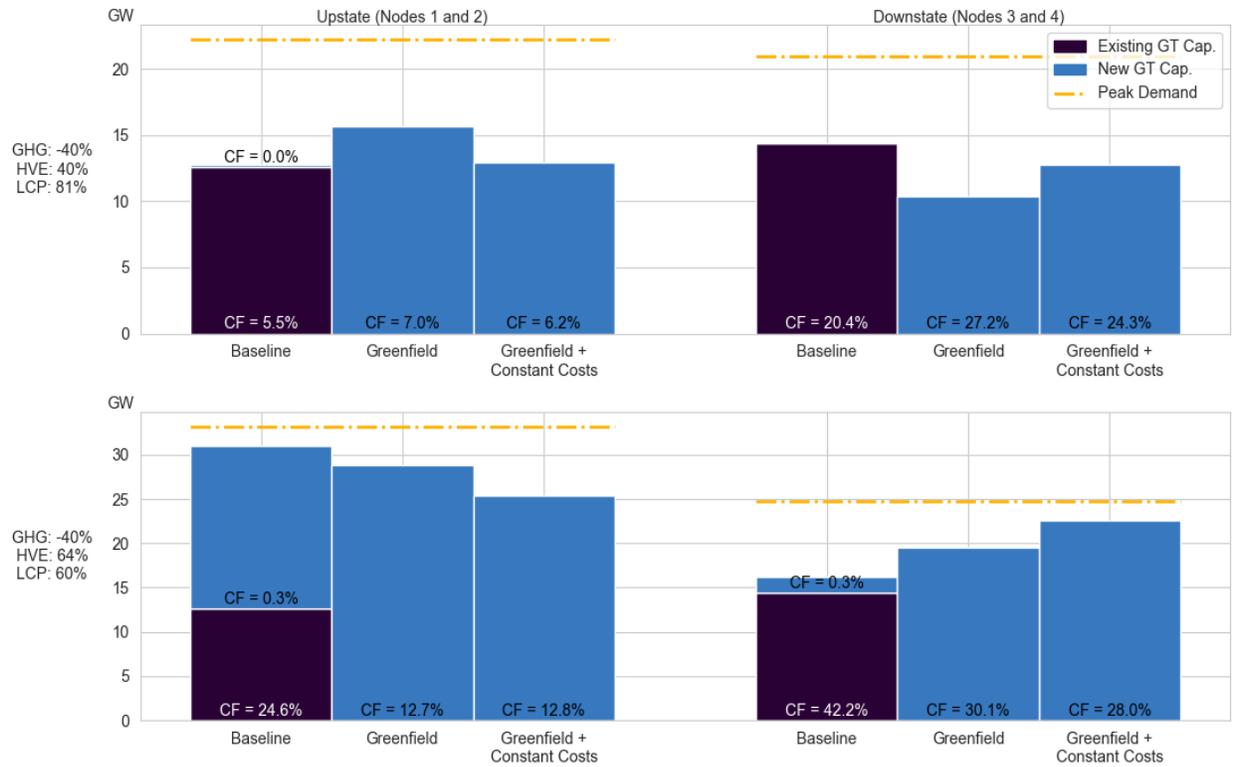

*Supplementary Figure S6: Existing and new gas capacity, distribution, and capacity factors (CF), shown with peak demand, for upstate and downstate New York State regions. The top row presents results for the high low-carbon electricity percent (LCP) scenario; the bottom row presents results for the LCP scenario.*

Both scenarios reveal the low capacity factors (CFs) of gas capacity in energy systems that achieve 40% GHG reduction, regardless of the configuration. In the top row, the high percent low-carbon electricity means that gas generation meets 19% of demand; this corresponds to capacity factors less than 7% upstate and less than 28% downstate. Here, CFs are lower upstate as this where the bulk of the renewable generation capacity is located. In comparison, gas generation CFs are higher on average for the low LCP scenario despite the larger amounts of GT capacity required to meet the additional electrified load: The looser low-carbon electricity constraint means that gas generation can satisfy approximately 40% of the demand. The outlier to this trend is the new gas capacity installed upstate in the Baseline configuration. For this scenario, 18.4 GW of new upstate capacity generates an average of 62.7 MWh/h, and 1.8 GW of new downstate capacity generates an average of 5.7 MWh/h, both corresponding to rounded CFs of 0.3%.

## S3.3 Main text figures presented at different rates of heating and vehicle electrification and different percents low-carbon electricity

Supplementary Figures S7-S8 display versions of Fig. 5 at different HVEs and LCPs; Supplementary Figures S9-S10 display versions of Fig. 7 at different HVEs; and Supplementary Figures S11-S12 display versions of Figs. 8-9 at different HVEs. These figures demonstrate that the results presented in the main text are not unique to the selected percents low-carbon electricity or electrification rates therein.

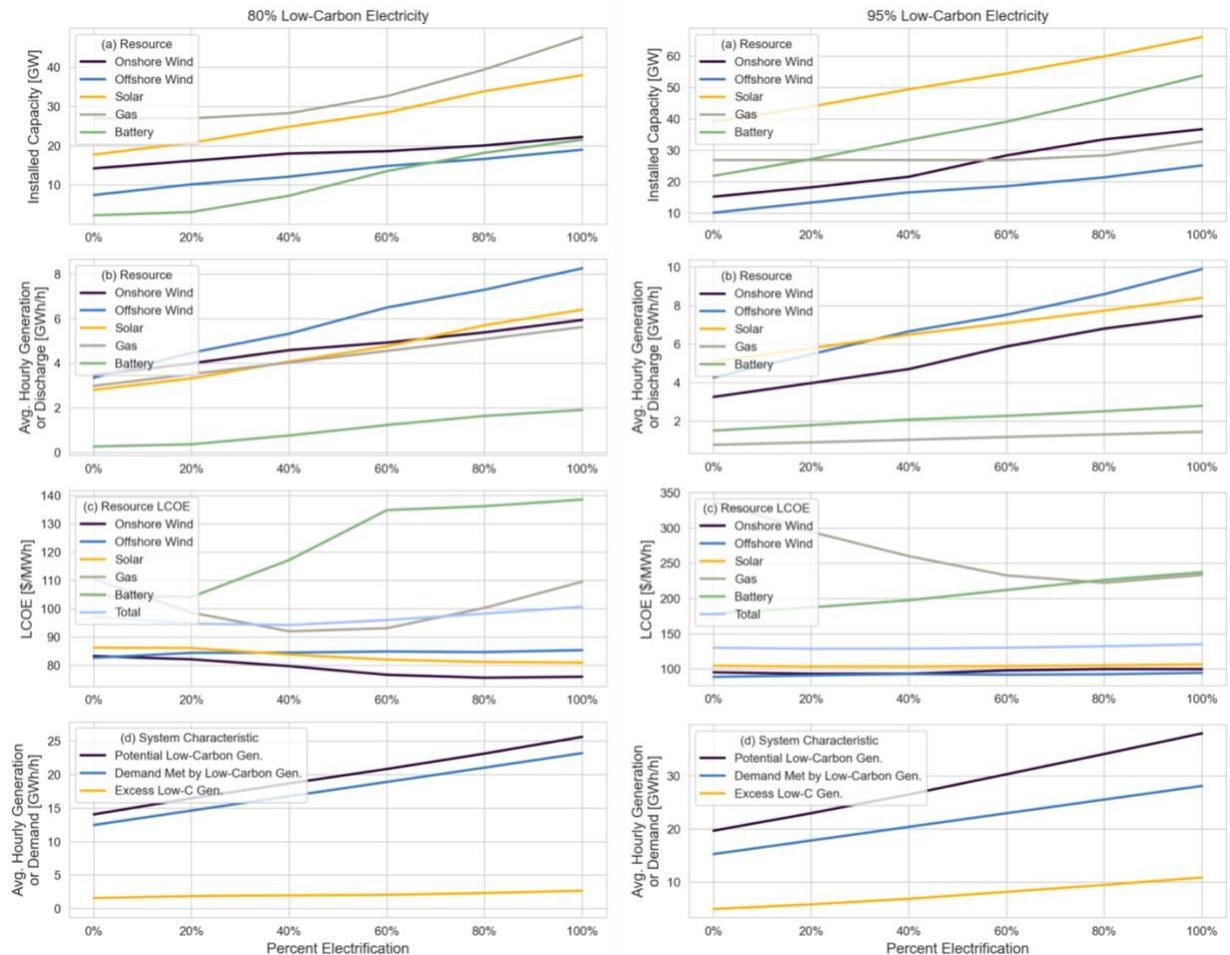

*Supplementary Figure S7: System characteristics for scenarios with (a-d) increasing HVE at 80% LCP; and (b) increasing HVE at 95% LCP. Subplots (a, e) present installed capacity; (b, f) present average generation by resource; (c, g) present LCOE per MWh for the generation and storage resources; and (d, h) present demand and generation quantities. In (c, g), resource LCOE for onshore wind, offshore wind, and solar refers to the LCOE of generation; LCOE for battery storage is per-MWh discharge. Note the different y-axis ranges for side-by-side panels.*

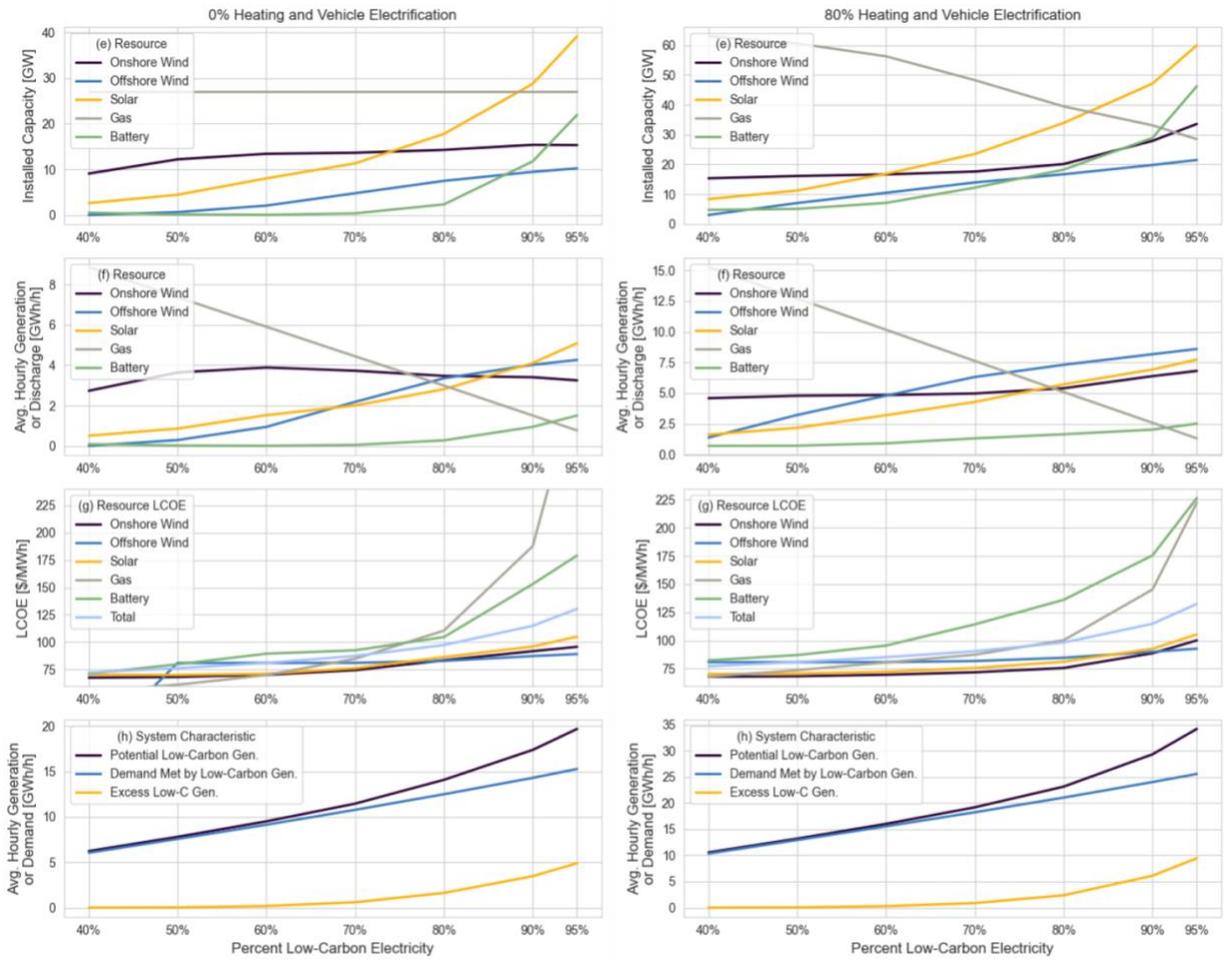

*Supplementary Figure S8: System characteristics for scenarios with (a-d) increasing LCP at 0% HVE; and (b) increasing LCP at 80% HVE. Subplots (a, e) present installed capacity; (b, f) present average generation by resource; (c, g) present LCOE per MWh for the generation and storage resources; and (d, h) present demand and generation quantities. In (c, g), resource LCOE for onshore wind, offshore wind, and solar refers to the LCOE of generation; LCOE for battery storage is per-MWh discharge; and in (c), gas generation LCOE at 95% LCP ($338/MWh) is cropped out to preserve y-axis resolution at lower LCOE values. Note the different y-axis ranges for side-by-side panels.*

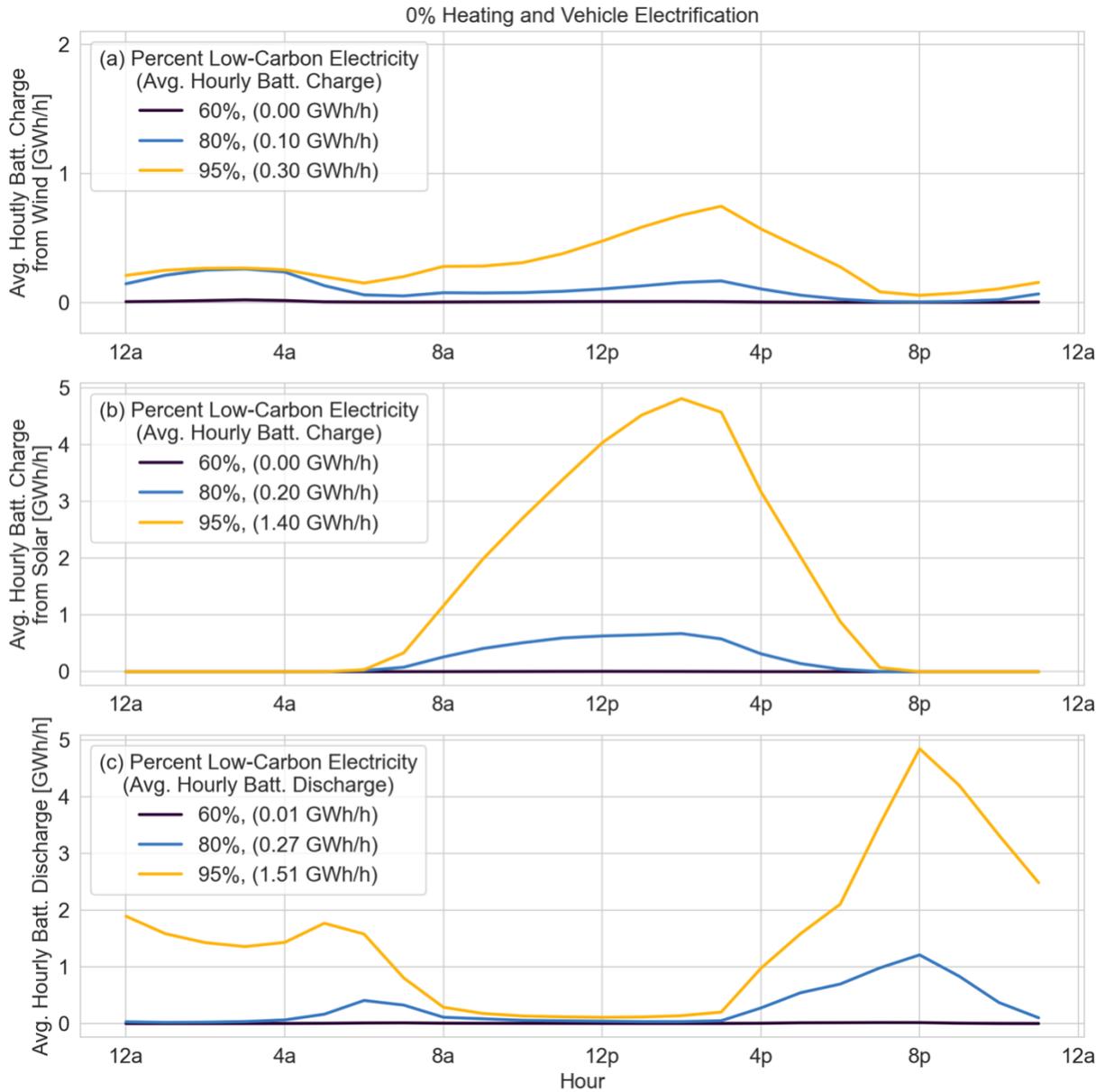

Supplementary Figure S9: *Average battery operation by hour for 60%, 80%, and 95% LCPs at 0% HVE. (a) Average hourly battery charging from wind (note y-axis scale is unique from (b) and (c)); (b) average hourly battery charging from solar; and (c) average battery discharge, all in GWh/h.*

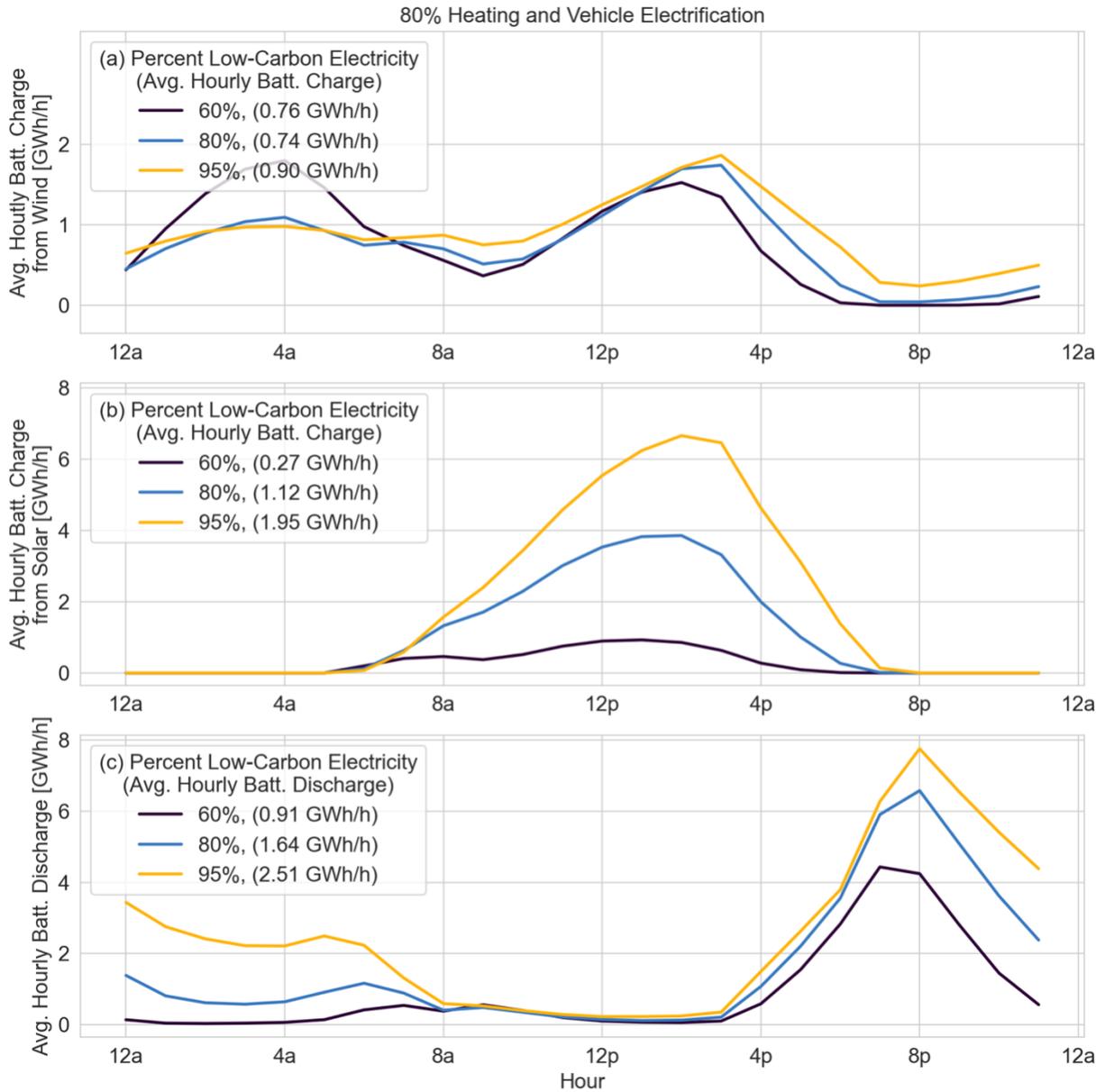

Supplementary Figure S10: *Average battery operation by hour for 60%, 80%, and 95% LCPs at 80% HVE. (a) Average hourly battery charging from wind (note y-axis scale is unique from (b) and (c)); (b) average hourly battery charging from solar; and (c) average battery discharge, all in GWh/h.*

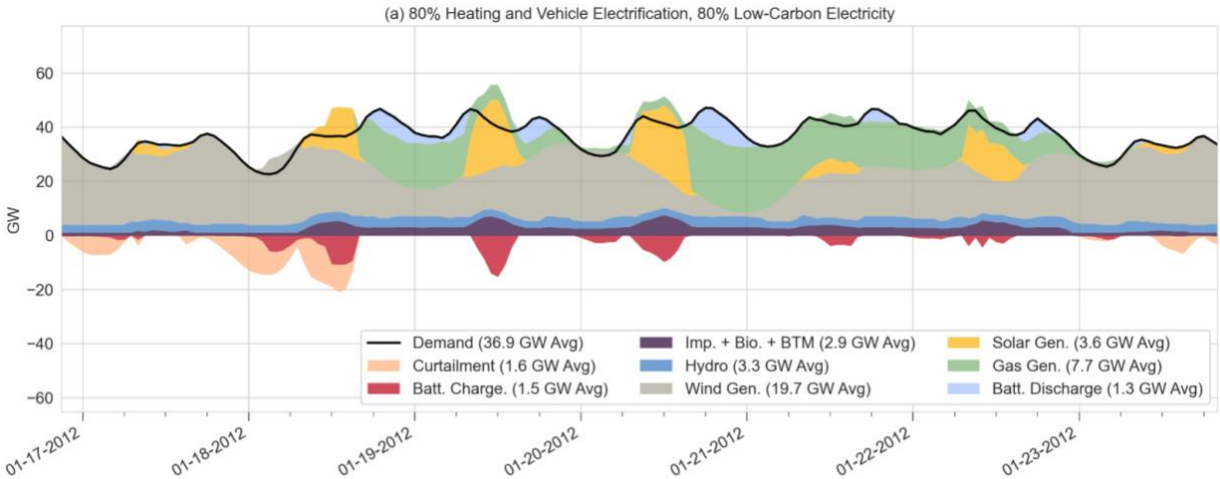

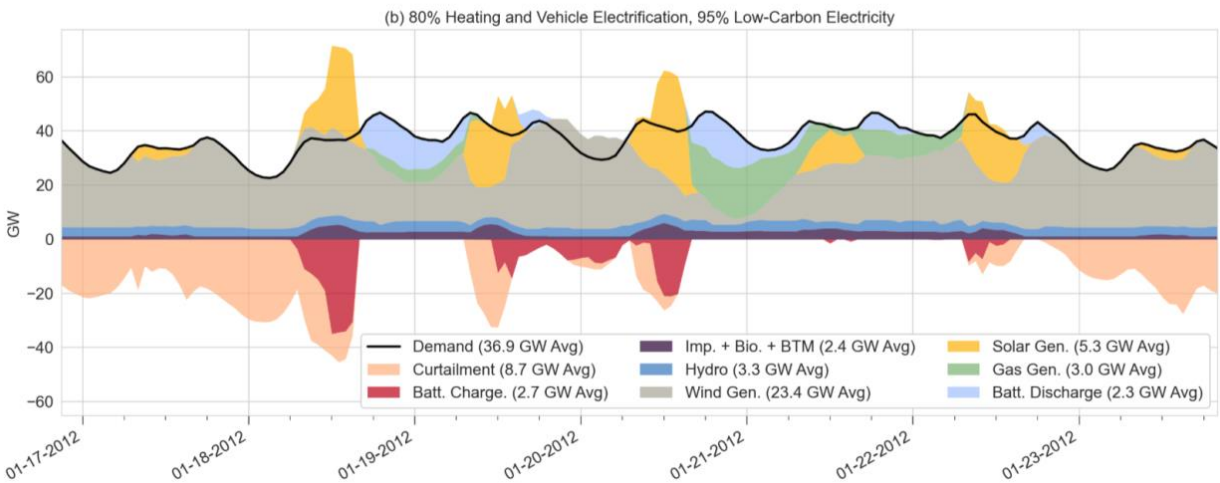

Supplementary Figure S11: Electricity generation and demand for a representative winter week with 80% HVE. (a) 80% LCP; (b) 95% LCP. 'Imp. + Bio. + BTM' represents the sum of imports, biofuel, and behind-the-meter solar generation. Average values reported in the legend are for the week shown.

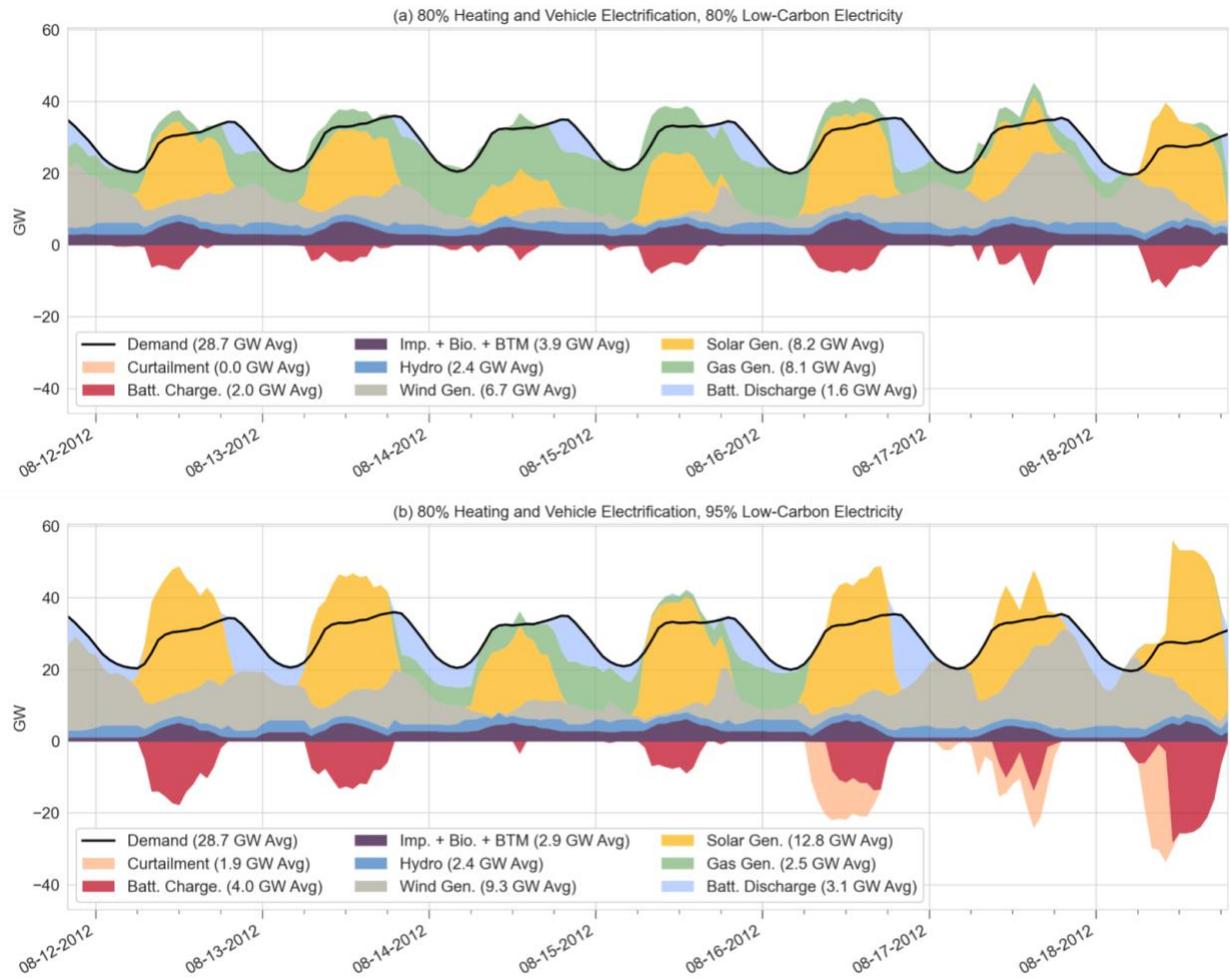

Supplementary Figure S12: Electricity generation and demand for a representative summer week with 80% HVE. (a) 80% LCP; (b) 95% LCP. 'Imp. + Bio. + BTM' represents the sum of imports, biofuel, and behind-the-meter solar generation. Average values reported in the legend are for the week shown.

## S4 References

[1] Waite M. Analyses of Energy Infrastructure Serving a Dense Urban Area: Opportunities and Challenges for Wind Power, Building Systems, and Distributed Generation 2016:1–282. https://academiccommons.columbia.edu/doi/10.7916/D8FT8M6H.

[2] US Energy Information Administration (EIA). Cost and Performance Characteristics of New Generating Technologies 2020:1–4. https://www.eia.gov/outlooks/aeo/assumptions/pdf/table_8.2.pdf.

[3] Bloom A, Townsend A, Palchak D, Novacheck J, King J, Barrows C, et al. Eastern Renewable Generation Integration Study 2016:TP-6A20-64. https://www.nrel.gov/grid/ergis.html.

[4] Conlon T, Waite M, Modi V. Assessing new transmission and energy storage in achieving increasing renewable generation targets in a regional grid. Appl Energy 2019;250:1085–98. https://doi.org/10.1016/j.apenergy.2019.05.066.

[5] Opalka W. Appeals Court Ratifies New York Capacity Zone. RTO Insid 2015. https://rtoinsider.com/rto/2d-cir-new-york-ferc-capacity-zone-14255/.

[6] New York Independent System Operator (NYISO). Market and Operational Data 2020. https://www.nyiso.com/energy-market-operational-data.

[7] Waite M, Modi V. Electricity Load Implications of Space Heating Decarbonization Pathways. Joule 2020;4:376–94. https://doi.org/10.1016/j.joule.2019.11.011.

[8] NOAA Centers for Environmental Information. Integrated Surface Dataset 2001. https://www.ncei.noaa.gov/products/land-based-station/integrated-surface-database.

[9] US Energy Information Administration (EIA). Updated Buildings Sector Appliance and Equipment Costs and Efficiencies 2018. https://www.eia.gov/analysis/studies/buildings/equipcosts/.

[10] Northeast Energy Efficiency Partnerships (NEEP). Northeast Energy Efficiency Partnerships Cold Climate Air Source Heat Pump Product Listing 2019. https://ashp.neep.org.

[11] Shapiro C, Puttagunta S. Field Performance of Heat Pump Water Heaters in the Northeast Consortium for Advanced Residential Buildings 2016. https://www.nrel.gov/docs/fy16osti/64904.pdf.

[12] US Energy Information Administration (EIA). State Energy Consumption Estimates 2018. https://www.eia.gov/state/seds/sep_use/notes/use_print.pdf.

[13] US Bureau of Transportation Statistics. Average Age of Automobiles and Trucks in Operation in the United States 2020. https://www.bts.gov/content/average-age-automobiles-and-trucks-operation-united-states.

[14] US Environmental Protection Agency (EPA). The 2020 EPA Automotive Trends Report: Greenhouse Gas Emissions, Fuel Economy, and Technology since 1975 2020:151. https://nepis.epa.gov/Exe/ZyPURL.cgi?Dockey=P1010U68.txt%0A.

[15] New York Open Data. New York State 2016 Gasoline Sales by County 2016. https://data.ny.gov/Energy-Environment/Estimated-Gasoline-Sales-Beginning-1995/cwrk-j5nn.

[16] National Renewable Energy Laboratory (NREL). EVI-Pro Lite API 2021.

https://developer.nrel.gov/docs/transportation/evi-pro-lite-v1/.

[17] New York Independent System Operator (NYISO). New York ISO Reliability Needs Assessment 2020. https://www.nyiso.com/documents/20142/2248793/2020-RNAReport-Nov2020.pdf/64053a7b-194e-17b0-20fb-f2489dec330d.

[18] Goldberg M, Keyser D. Transmission Line Jobs and Economic Development Impact (JEDI) Model User Reference Guide 2016. https://www.nrel.gov/analysis/jedi/transmission-line.html.

[19] MacDonald AE, Clack CTM, Alexander A, Dunbar A, Wilczak J, Xie Y. Future cost-competitive electricity systems and their impact on US CO2 emissions. Nat Clim Chang 2016. https://doi.org/10.1038/nclimate2921.

[20] New York Independent System Operator (NYISO). Western New York Public Policy Transmission Planning Report 2017.

[21] Transmission Developers Inc (TDI). Champlain Hudson Power Express: Project Development Portal 2019. http://ideas.mowerinteractive.com/clients/tdi/60394-website/site/index.php (accessed August 14, 2019).

[22] Starwood Energy Group. The Neptune Regional Transmission System 2007. http://starwoodenergygroup.com/wp-content/uploads/2014/06/6_NeptuneAnnouncement.pdf.

[23] US Energy Information Administration (EIA). Annual Energy Outlook 2020 with projections to 2050 2020. https://www.eia.gov/outlooks/aeo/pdf/aeo2020.pdf.

[24] U.S. Energy Information Administration (EIA). Electric Power Annual 2019 2020. https://www.eia.gov/electricity/annual/.

[25] New York Independent System Operator (NYISO). 2019 Load and Capacity Data: Gold Book 2019. https://www.nyiso.com/documents/20142/2226333/2019-Gold-Book-Final-Public.pdf/.

[26] New York Independent System Operator (NYISO). 2020 Load and Capacity Data: Gold Book 2020. https://www.nyiso.com/documents/20142/2226333/2020-Gold-Book-Final-Public.pdf/.

[27] US Energy Information Administration (EIA). Electric Power Annual Report 2019. Table 5.12. Accessed 10/21/2020. 2020. https://www.eia.gov/electricity/annual/.

[28] US Energy Information Administration (EIA). Electric Power Annual Report 2019. Table 3.11. Accessed 10/21/2020. 2020. https://www.eia.gov/electricity/annual/.

[29] New York Independent System Operator (NYISO). Locational Minimum Installed Capacity Requirements Study for the 2020-2021 Capability Year. 2020. https://www.nyiso.com/documents/20142/8583126/LCR2020-Report.pdf/4c9309b2-b13e-9b99-606a-7af426d93a47.

[30] New York State Energy Research and Development Authority (NYSERDA). New Efficiency: New York Analysis of Residential Heat Pump Potential and Economics 2019. https://www.nyserda.ny.gov/-/media/Files/Publications/PPSER/NYSERDA/18-44-HeatPump.ashx.

[31] Draxl C, Clifton A, Hodge BM, McCaa J. The Wind Integration National Dataset (WIND) Toolkit. Appl Energy 2015;151:355–66. https://doi.org/10.1016/j.apenergy.2015.03.121.

[32] Draxl C, Hodge B-M, Clifton A, McCaa J. Overview and Meteorological Validation of the Wind Integration National Dataset Toolkit. NREL 2015:87.

https://www.nrel.gov/docs/fy15osti/61740.pdf.
[33] Waite M, Modi V. Modeling wind power curtailment with increased capacity in a regional electricity grid supplying a dense urban demand. Appl Energy 2016. https://doi.org/10.1016/j.apenergy.2016.08.078.
[34] Musial W, Heimiller D, Beiter P, Scott G, Draxl C. 2016 Offshore Wind Energy Resource Assessment for the United States 2016. https://doi.org/NREL/TP-5000-66599.
[35] Wiser R, Bolinger M, Barbose G, Millstein D. 2016 Wind Technologies Market Report 2017. https://www.energy.gov/eere/wind/downloads/2016-wind-technologies-market-report.
[36] Wiser R, Bolinger M. 2017 Wind Technologies Market Report 2018:1–98. https://www.energy.gov/eere/wind/downloads/2017-wind-technologies-market-report.
[37] Wiser R, Bolinger M. 2018 Wind Technologies Market Report 2019:1–98. https://www.energy.gov/eere/wind/downloads/2018-wind-technologies-market-report.
[38] Wiser R, Jenni K, Seel J, Baker E, Hand M, Lantz E, et al. Expert elicitation survey on future wind energy costs. Nat Energy 2016;1. https://doi.org/10.1038/nenergy.2016.135.
[39] Bloomberg New Energy Finance (BNEF). 2018 Wind O&M Price Index. 2018.
[40] Stehly T, Beiter P, Heimiller D, Scott G, Stehly T, Beiter P, et al. 2017 Cost of Wind Energy Review 2018. https://www.nrel.gov/docs/fy18osti/72167.pdf.
[41] Walter M, Philipp B, Spitsen P, Nunemake J, Gevorgian V. 2018 Offshore Wind Technologies Market Report 2019:1–94. https://www.osti.gov/biblio/1572771-offshore-wind-technologies-market-report.
[42] Bosch J, Staffell I, Hawkes AD. Global levelised cost of electricity from offshore wind. Energy 2019;189:116357. https://doi.org/10.1016/j.energy.2019.116357.
[43] Hummon M, Ibanez E, Brinkman G, Lew D. Sub-Hour Solar Data for Power System Modeling From Static Spatial Variability Analysis. 2nd Int Work Integr Sol Power Power Syst 2012. https://www.nrel.gov/docs/fy13osti/56204.pdf.
[44] Blair N, Dobos AP, Freeman J, Neises T, Wagner M, Ferguson T, et al. System Advisor Model, 2014.1.14: General Description. Natl Renew Energy Lab 2014. http://www.nrel.gov/docs/fy14osti/61019.pdf.
[45] Brooks B, Xenergy K, Whitaker C. Guideline for the use of the Performance Test Protocol for Evaluating Inverters Used in Grid-Connected Photovoltaic Systems. Sandia Natl Lab 2005. https://www.energy.ca.gov/sites/default/files/2020-06/2004-11-22_Sandia_Test_Protocol_ada.pdf.
[46] Fu R, Feldman D, Margolis R. U.S. Solar Photovoltaic System Cost Benchmark: Q1 2018. NREL 2018:1–47. https://doi.org/10.7799/1325002.
[47] USDA National Agricultural Statistics Service. 2017 Census of Agriculture 2017. https://www.nass.usda.gov/Publications/AgCensus/2017/index.php.
[48] Ong S, Campbell C, Denholm P, Margolis R, Heath G. Land-Use Requirements for Solar Power Plants in the United States. 2013. https://doi.org/10.1016/j.rapm.2006.08.004.
[49] New York Independent System Operator (NYISO). Solar Impact on Grid Operations: An Initial Assessment 2016:1–57. https://www.transmissionhub.com/wp-content/uploads/2018/12/New-York-ISO-JUN-30-2016-Solar-Report.pdf.
[50] New York State Energy Research and Development Authority (NYSERDA). Solar Electric Programs Data (Accessed 06/01/2020) 2020. https://data.ny.gov/Energy-

Environment/Solar-Electric-Programs-Reported-by-NYSERDA-Beginn/3x8r-34rs.

[51] Sengupta M, Habte A, Gotseff P, Weekley A, Lopez A, Molling C, et al. A Physics-Based GOES Satellite Product for Use in NREL's National Solar Radiation Database. Natl Renew Energy Lab 2014. https://www.nrel.gov/docs/fy14osti/62237.pdf.

[52] US Energy Information Administration (EIA). Energy Mapping System 2020. https://www.eia.gov/state/maps.php.

[53] Cole W, Frazier AW. Cost Projections for Utility-Scale Battery Storage: 2020 Update. Natl Renew Energy Lab 2020. https://www.nrel.gov/docs/fy20osti/75385.pdf.

[54] Tesla. Powerpack System Specifications 2019. https://www.tesla.com/powerpack.

[55] Smith K, Saxon A, Keyser M, Lundstrom B, Cao Z, Roc A. Life prediction model for grid-connected Li-ion battery energy storage system. Proc Am Control Conf 2017:4062–8. https://doi.org/10.23919/ACC.2017.7963578.

[56] Aggreko. Aggreko delivers grid stability to New York State with 2MW/3.8MWh energy storage system for National Grid. GlobeNewswire 2019. https://www.globenewswire.com/news-release/2019/06/03/1863350/0/en/Aggreko-delivers-grid-stability-to-New-York-State-with-2MW-3-8MWh-energy-storage-system-for-National-Grid.html.

[57] Lockheed Martin. 1MWh GridStar® Lithium Energy Storage Installation in Syracuse, New York 2019. https://www.lockheedmartin.com/content/dam/lockheed-martin/mfc/documents/energy/energy-syracuses-project-summary.pdf.

[58] Key Capture Energy. KCE NY 1 Breaks Ground on 20 MW 2018. https://www.keycaptureenergy.com/kce-ny-1-breaks-ground-on-20-mw/.

[59] Clean Coalition. Long Island Community Microgrid Project (LICMP) 2019. https://clean-coalition.org/community-microgrids/long-island-community-microgrid-project/.

[60] Energy Storage Association. US Energy Storge: 2019 Year in Review 2019. https://energystorage.org/wp/wp-content/uploads/2020/04/ESA_AR_2020_FINAL.pdf.

[61] Guerra OJ, Zhang J, Eichman J, Denholm P, Kurtz J, Hodge BM. The value of seasonal energy storage technologies for the integration of wind and solar power. Energy Environ Sci 2020;13:1909–22. https://doi.org/10.1039/d0ee00771d.

[62] Steward D, Saur G, Penev M, Ramsden T, Steward D, Saur G, et al. Lifecycle Cost Analysis of Hydrogen Versus Other Technologies for Electrical Energy Storage. Natl Renew Energy Lab 2009. https://www.nrel.gov/docs/fy10osti/46719.pdf.

[63] Houchins C, James B. 2020 DOE Hydrogen and Fuel Cells Program Review 2020. https://www.hydrogen.energy.gov/pdfs/review20/st100_houchins_2020_o.pdf.

[64] Eichman J, Townsend A, Melaina M. Economic Assessment of Hydrogen Technologies Participating in California Electricity Markets. Natl Renew Energy Lab 2016. https://www.nrel.gov/docs/fy16osti/65856.pdf.

[65] Penev M, Rustagi N, Hunter C, Eichman J. Energy Storage: Days of Service Sensitivity Analysis 2019. https://www.nrel.gov/docs/fy19osti/73520.pdf.

[66] Walker SB, Mukherjee U, Fowler M, Elkamel A. Benchmarking and selection of Power-to-Gas utilizing electrolytic hydrogen as an energy storage alternative. Int J Hydrogen Energy 2016;41:7717–31. https://doi.org/10.1016/j.ijhydene.2015.09.008.

[67] Johnson S. New York's Indian Point nuclear power plant closes after 59 years of operation. US Energy Inf Adm Today Energy 2021.

https://www.eia.gov/todayinenergy/detail.php?id=47776#.

[68] New York State Energy Research and Development Authority (NYSERDA). Clean Energy Standard, 2020 Compliance Year: Load Serving Entity (LSE) Zero Emission Credit (ZEC) Rate. 2020. https://www.nyserda.ny.gov/All-Programs/Clean-Energy-Standard/LSE-Obligations/2020-Compliance-Year.

[69] US Energy Information Administration (EIA). Form EIA-923 detailed data with previous form data (EIA-906/920) 2020. https://www.eia.gov/electricity/data/eia923/.

[70] Mukerji R. NYISO Market Operations Report. New York Indep Syst Oper 2019. https://www.nyiso.com/documents/20142/8196990/Market+Operations+Report_+BIC_09.11.19.pdf.

[71] Whitney N. Amount of Capacity Qualified to Offer. New York Indep Syst Oper 2019. nyiso.com/documents/20142/3036383/4_Amt of Capacity Qualified to Offer.pdf/57f56a99-3293-d795-8584-21a70c495a5a.

[72] New York City Council. Local Laws of the City of New York for the Year 2019. No. 97. 2019. https://www1.nyc.gov/assets/buildings/local_laws/ll97of2019.pdf.

[73] NYSERDA. New York State Greenhouse Gas Inventory: 1990-2016 2019. https://www.nyserda.ny.gov/-/media/Files/EDPPP/Energy-Prices/Energy-Statistics/greenhouse-gas-inventory.pdf.

[74] New York State Senate. Climate Leadership and Community Protection Act — Final Bill Text 2019. https://www.nysenate.gov/legislation/bills/2019/s6599.

[75] Myrhe G, Shindell D, Huang J, Mendoza B, Daniel JS, Nielsen CJ, et al. Anthropogenic and natural radiative forcing. Clim Chang 2013 Phys Sci Basis Work Gr I Contrib to Fifth Assess Rep Intergov Panel Clim Chang 2013;9781107057:659–740. https://doi.org/10.1017/CBO9781107415324.018.

[76] Howarth RW. Methane Emissions and Greenhouse Gas Accounting: A Case Study of a New Approach Pioneered by the State of New York 2019:14. https://documents.dps.ny.gov/public/Common/ViewDoc.aspx?DocRefId=%7B3498AB82-B671-451E-A556-A917A61F939A%7D.

[77] Hayhoe K, Kheshgi HS, Jain AK, Wuebbles DJ. Substitution of natural gas for coal: Climatic effects of utility sector emissions. Clim Change 2002;54:107–39. https://doi.org/10.1023/A:1015737505552.

[78] Howarth RW, Santoro R, Ingraffea A. Methane and the greenhouse-gas footprint of natural gas from shale formations. Clim Change 2011;106:679–90. https://doi.org/10.1007/s10584-011-0061-5.

[79] Intergovernmental Panel on Climate Change (IPCC). Good Practice Guidance and Uncertainty Management in National Greenhouse Gas Inventories. Table 2, CH4 Emissions: Coal Mining and Handling. 1996. https://www.ipcc-nggip.iges.or.jp/public/gp/english/.

[80] National Energy Technology Laboratory (NETL). Petroleum-Based Fuel Life Cycle Greenhouse Gas Analysis - 2005 Baseline Model 2008. https://www.nata.aero/data/files/gia/environmental/bllcghg2005.pdf.

[81] Alvarez RA, Zavala-Araiza D, Lyon DR, Allen DT, Barkley ZR, Brandt AR, et al. Assessment of methane emissions from the U.S. oil and gas supply chain. Science (80- ) 2018;361:186–8. https://doi.org/10.1126/science.aar7204.

[82] Schneising O, Burrows JP, Dickerson RR, Buchwitz M, Reuter M, Bovensmann H. Remote sensing of fugitive methane emissions from oil and gas production in North American tight geologic formations. Earth's Futur 2014;2:548–58. https://doi.org/10.1002/2014ef000265.

[83] Engineering ToolBox. Fuels - Higher and Lower Calorific Values 2003. https://www.engineeringtoolbox.com/fuels-higher-calorific-values-d_169.html.

[84] EPA. Emission Factors for Greenhouse Gas Inventories 2019:1–5. https://www.ecfr.gov/current/title-40/chapter-I/subchapter-C/part-98#ap40.23.98_19.1.

[85] Hydro-Quebec. Hydro Quebec Annual Report 2019 2020. https://www.hydroquebec.com/data/documents-donnees/pdf/annual-report.pdf.